\documentclass[floatfix,nofootinbib,twocolumn,preprintnumbers,superscriptaddress,notitlepage]{revtex4-1}

\usepackage{times}
\usepackage{xcolor}
\usepackage{amsfonts}
\usepackage{amsmath}
\usepackage{amssymb}
\usepackage{bm}
\usepackage{dcolumn}
\usepackage{enumerate}
\usepackage{epsfig}
\usepackage{graphicx}
\usepackage{graphics}
\usepackage[latin1]{inputenc}
\usepackage{latexsym}
\usepackage{rotating}
\usepackage{hyperref}
\usepackage[caption=false]{subfig}
\usepackage[normalem]{ulem}

\newcommand{\<}{\begin{equation}}
\newcommand{\?}{\end{equation}}

\newcommand{\cD}{\mathcal{D}}
\newcommand{\cE}{\mathcal{E}}
\newcommand{\cN}{\mathcal{N}}
\newcommand{\cS}{\mathcal{S}}

\definecolor{mgreen}{rgb}{0.1,0.7,0.1}

\newcommand{\edit}[1]{{{#1}}}
\newcommand{\editII}[1]{{{#1}}}

\begin{document}

\title{Inferring spin tilts at formation from gravitational wave observations of binary black holes: Interfacing precession-averaged and orbit-averaged spin evolution}

\author{Nathan~K.~Johnson-McDaniel}
\affiliation{Department of Applied Mathematics and Theoretical Physics, Centre for Mathematical Sciences, University of Cambridge, Wilberforce Road, Cambridge, CB3 0WA, UK}
\affiliation{Department of Physics and Astronomy, The University of Mississippi, University, Mississippi 38677, USA}
\author{Sumeet Kulkarni}
\author{Anuradha Gupta}
\affiliation{Department of Physics and Astronomy, The University of Mississippi, University, Mississippi 38677, USA}

\date{\today}

\begin{abstract}
Two important parameters inferred from the gravitational wave signals of binaries of precessing black holes are the spin tilt angles, i.e., the angles at which the black holes' spin axes are inclined with respect to the binary's orbital angular momentum. The LIGO-Virgo parameter estimation analyses provide spin tilts at a fiducial reference frequency, often the lowest frequency used in the data analysis. However, the most astrophysically interesting quantities are the spin tilts when the binary was formed, which can be significantly different from those at the reference frequency for strongly precessing binaries. The spin tilts at formally infinite separation are a good approximation to the tilts at formation in many formation channels and can be computed efficiently for binary black holes using precession-averaged evolution.
Here, we present a new code for computing the tilts at infinity that combines the precession-averaged evolution with orbit-averaged evolution at high frequencies and illustrate its application to GW190521 and other binary black hole detections from \edit{O3}. We have empirically determined the transition frequency between the orbit-averaged and precession-averaged evolution to produce tilts at infinity with a given accuracy \edit{and find that using only the precession-averaged evolution can lead to errors in the cosines of the tilts at infinity of $>0.8$ for certain binary configurations}. \editII{However, the precession-averaged evolution alone is sufficient for good accuracy when obtaining the posterior distributions of the tilts at infinity for current detections.} We also have regularized the precession-averaged equations in order to obtain good accuracy for the very close-to-equal-mass binary parameters encountered in practice. This additionally allows us to investigate the singular equal-mass limit of the precession-averaged expressions, where we find \edit{that to a good approximation the results only depend on the orbital angular momentum $L$ through the combination $(1-q)L$, where $q$ is the mass ratio.}
\end{abstract}

\maketitle

\section{Introduction}
\label{sec:intro}

Gravitational wave observations of binary black holes have now revealed evidence for precessing spins, both mild evidence for GW190412~\cite{GW190412} and GW190521~\cite{GW190521_discovery,GW190521_implications} as well as stronger evidence for \editII{GW200129\_065458~\cite{Hannam:2021pit} and} the entire population~\edit{\cite{O3b_pop}}. Parameter inference of gravitational wave detections returns the spin angles, notably the misalignments with respect to the binary's Newtonian orbital angular momentum, or \emph{tilts}, at some fixed frequency ($20$~Hz for almost all the events in GWTC-2~\cite{GWTC-2_paper}).\footnote{\edit{For the new events in GWTC-3~\cite{GWTC-3_paper}, the tilts are quoted at infinite separation using the code introduced in this paper.}} However, in order to make contact with astrophysical formation scenarios, e.g., spin misalignments with the orbital angular momentum due to supernova kicks in the isolated binary channel (see, e.g., \cite{Kalogera:1999tq,OShaughnessy:2017eks,Gerosa:2018wbw}), one wants to know the spin tilts at formation, which are fairly well approximated by the tilts at formally infinite separation for most binaries. The differences between the values at infinity and a finite semimajor axis $a$ go roughly as $M^2/L \propto \sqrt{M/[a(1-e^2)]}$, where $M$ is the binary's total mass, $L$ is the magnitude of its orbital angular momentum, and $e$ is its eccentricity (using the Newtonian expression for the orbital angular momentum, since we are interested in cases where the binary's separation is large, so this is a good approximation). Alternatively, one can evolve population synthesis predictions forward to the reference frequency, as in~\cite{Gerosa:2018wbw}, but this evolution has to be performed statistically, since the precession-averaged evolution used \edit{in~\cite{Gerosa:2018wbw}} does not track the precessional phase, as discussed below.

In many standard formation scenarios, the orbital angular momentum at formation is large enough that the tilts at infinity approximate those at formation with an absolute accuracy of better than $\sim 10^{-2}$~rad. As we show in Sec.~\ref{sec:tilts_formation}, the uncertainties are (\emph{ceteris paribus}) larger for close to equal-mass binaries, with mass ratios differing from equal mass by $\lesssim 1\%$. Such binaries are present in population synthesis models, but are a relatively small fraction of even the detectable population of binary black holes (e.g., in the isolated binary calculations in~\cite{Belczynski:2017gds}). Thus, for most of the binary black holes we expect to detect with ground-based gravitational wave detectors, the tilts at infinity are a good approximation to those at formation. Nevertheless, this is a theoretical statement: Observationally, when analyzing a close-to-equal-mass binary, one will find samples that are very close to equal mass, for which the approximation of the tilts at formation by those at infinity will be much less accurate. However, this will only be relevant for constraining formation scenarios that predict nonnegligible numbers of binaries with such close-to-equal mass ratios. 

Additionally, quoting the tilts at infinity allows one to make direct comparisons for different binaries, while the fixed reference frequency used in LIGO-Virgo analyses corresponds to different points in the evolution of the binary, depending on the binary's parameters, particularly its total mass. Studies of the distribution of tilts for the entire population will thus also benefit from considering the tilts at infinity, \edit{as in~\cite{Mould:2021xst}}. The distribution of tilts is particularly important since it allows one to distinguish between different formation channels, most notably between the isotropic distribution expected for dynamical formation and the distribution favoring aligned spins expected for isolated formation (see, e.g.,~\cite{Stevenson:2017dlk,Talbot:2017yur,Farr:2017uvj,Tiwari:2018qch}). \edit{While current measurements of the tilts are still \editII{not very well constrained}, Ref.~\cite{Knee:2021noc} shows that we can expect to measure them (at the reference frequency) with a $90\%$ credible interval width of $\lesssim 0.2$~rad in some cases in the plus-detector era~\cite{Aasi:2013wya} (see Fig.~3 in~\cite{Knee:2021noc} in particular). While we are not aware of similar predictions for third-generation gravitational wave detectors (Cosmic Explorer~\cite{Reitze:2019iox,Hall:2020dps} and Einstein Telescope~\cite{Hild:2010id}), a na{\"\i}ve scaling based on their noise curves, which are a factor of $\sim 10$ more sensitive than those of the plus-era detectors, suggests that a similar network of detectors would be able to measure the tilts with $90\%$ credible interval widths of $\lesssim 0.02$~rad. Thus, it is important to understand and control the errors in the computation of tilts at infinity so that they remain below the statistical errors even with third-generation detectors. Moreover, as we will see, the errors in some computations of the tilts at infinity can be large enough to potentially even be important at current detector sensitivities.}

The precession-averaged post-Newtonian (PN) evolution introduced in~\cite{Kesden:2014sla,Gerosa:2015tea} (building on work in~\cite{Racine:2008qv}) and further developed in~\cite{Chatziioannou:2017tdw} provides an efficient way to compute the spin tilts at infinity. However, it is only accurate when the binary is sufficiently well separated, particularly since it is currently restricted to relatively low-order PN expressions for the precessional dynamics [$2$PN, which is next-to-leading order, in the conservative dynamics and only $1$PN in the dissipative dynamics, as discussed below Eq.~(38) in~\cite{Gerosa:2015tea}].\footnote{As usual, $n$PN refers to a term of relative order $(v/c)^{2n}$, where $v$ is the binary's orbital velocity and $c$ is the speed of light. The order counting usually starts at the leading order, though for the precessional equations it is instead counted so that the leading spin-orbit term is at $1.5$PN, the same order as the leading spin-orbit contribution to the binary's orbital dynamics.}
Additionally, it relies on the precession timescale $t_\text{prec}$ being much smaller than the radiation reaction timescale $t_\text{RR}$, and one has $t_\text{prec}/t_\text{RR} \sim (M/r)^{3/2}$ (see, e.g.,~\cite{Kesden:2014sla}), where $r$ is the binary's orbital separation. Thus, to obtain a good accuracy, one first needs to evolve the binary backwards in time to some transition point using orbit-averaged evolution (see, e.g.,~\cite{Apostolatos:1994mx,Kidder:1995zr}; the derivation of the precession-averaged evolution starts from the orbit-averaged equations), and then apply the precession-averaged evolution, which will now be sufficiently accurate, since the binary is sufficiently well separated. This was first appreciated in~\cite{Gerosa:2016sys} in the context of evolving forward in time. \edit{However, the accuracy of the precession-averaged evolution in obtaining the tilts at infinity has not been quantified, and thus it is not clear what transition point from orbit-averaged to precession-averaged evolution should be chosen to ensure a given accuracy in this calculation. Here we determine this transition point, as discussed below.} 

We use the SpinTaylorT5 orbit-averaged evolution~\cite{Ajith:2011ec} as implemented in LALSuite~\cite{LALSuite}, including the $3$PN spin-spin terms in the phasing and precession equations from~\cite{Bohe:2015ana} and the spin-orbit contributions to the orbital angular momentum in the precession equations (only needed through $1.5$PN to give $3$PN accurate contributions) from~\cite{Bohe:2012mr} in addition to the terms given in~\cite{Ajith:2011ec}, which are $3.5$PN accurate in the nonspinning phasing.\footnote{While the $3.5$PN spin-orbit terms were computed in~\cite{Bohe:2012mr}, they cannot be used in the orbit-averaged case, since there are uncomputed corrections at that order arising from the orbit averaging of the leading spin-spin terms at $2$PN, as discussed in~\cite{SpinTaylor_TechNote}.} \edit{Specifically, the $3$PN expressions include the first post-Newtonian corrections to the leading spin-orbit and spin-spin terms as well as the tail terms from the leading spin-orbit effects, while the $2$PN expressions used in the PRECESSION code~\cite{Gerosa:2016sys} are only the leading spin-orbit and spin-spin effects.} All these expressions assume quasicircular orbits, but we will see that the precession-averaged evolution does not need to be modified in eccentric cases for this application. The orbit-averaged evolution in eccentric cases can be carried out as in~\cite{Phukon:2019gfh}---we will consider this in the future, since most current analyses of LIGO-Virgo data do not include eccentricity (see~\cite{Romero-Shaw:2019itr,Romero-Shaw:2020thy,Romero-Shaw:2021ual} for analyses that measure eccentricity for binary black hole signals by reweighting analyses with a quasicircular waveform model, but only consider aligned spins, and~\cite{Wu:2020zwr} for analyses with a nonspinning, inspiral-only eccentric waveform model).

In this study we empirically determine the largest frequency at which one can switch from orbit-averaged to precession-averaged evolution to obtain a given accuracy (here $10^{-3}$~rad) in the spin tilts at infinity. (The analogous problem for evolving forward in time to obtain the remnant quantities has already been studied in~\cite{Reali:2020vkf}, albeit only with $2$PN precession equations for the orbit-averaged evolution.) Additionally, we regularize the expressions for the precession-averaged evolution from~\cite{Chatziioannou:2017tdw} to make the determination of the spin tilts at infinity numerically well-conditioned for mass ratios close to unity, where the equations become singular, due to a qualitatively different behavior for exactly equal masses, discussed in~\cite{Gerosa:2016aus}.\footnote{This regularization is different from that carried out in~\cite{Klein:2021jtd}, which instead changes variables. Additionally, our motivation for the regularization is slightly different, since we are particularly interested in regularizing a numerator in the expression for the tilts at infinity that vanishes in the equal-mass limit.} We also derive rigorous error bounds for simplifications that can be applied during various portions of the evolution to improve the robustness and speed of the method. We have implemented our method as part of the publicly available LALSuite package~\cite{LALSuite}; see~\cite{tilts_at_infinity}.

The aforementioned regularization is important for applications to LIGO-Virgo detections, where the parameter estimation produces samples with mass ratios quite close to $1$. For instance, one has mass ratios $q$ with $1 - q \simeq 1.5 \times 10^{-6}$ in the GW190521 and GW190929\_012149 samples from the LIGO-Virgo collaboration analysis~\cite{GWTC-2_PE}.\footnote{These very small values of $1 - q$ only occur in the samples without the reweighting to a prior uniform in comoving volume, since that reweighting significantly decreases the number of samples. In particular, the reweighting reduces the number of samples by factors of $\sim 6$ and $\sim 4$ for the two cases that give the smallest $1 - q$ values without the reweighting, viz., the GW190521 IMRPhenomPv3HM and GW190929\_012149 IMRPhenomPv2 samples, respectively.}
Mass ratios quite close to $1$ also are predicted in population synthesis calculations. For instance, of the $\sim 6\times 10^6$ precessing, unequal-mass binary black holes in the 2019 standard input physics (M30) isolated binary population synthesis results from~\cite{Belczynski:2017gds} that are detectable by LIGO and Virgo with their mid-high sensitivity with that paper's criterion of a single detector signal-to-noise ratio of at least $8$, $\sim 2\%$ have $1 - q < 10^{-2}$ and $\sim 0.2\%$ have $1 - q < 10^{-3}$ (we plot the distribution in Sec.~\ref{sec:tilts_formation}). While these close to equal mass ratio cases lead to the largest intrinsic uncertainties in the tilts at formation, as discussed above, it is still important to be able to calculate the tilts at infinity accurately for such cases, in order to use the tilts at infinity for comparisons between different binaries, also as discussed above.

As an example, we apply our hybrid orbit-averaged and precession-averaged evolution code to compute the distribution of tilts at infinity for some of the binaries with nonnegligible spins detected during \edit{O3}, including GW190521. We also use the regularized evolution to investigate the singular equal-mass limit of the precession-averaged evolution.

In Sec.~\ref{sec:prec_evol}, we introduce the standard precession-averaged evolution and computation of the spin tilts at infinity. We also discuss why these expressions are also applicable to most eccentric binaries when approximating the tilts at formation. We derive the regularized expressions in Sec.~\ref{sec:reg} and discuss their numerical implementation in Sec.~\ref{sec:numerics}. We then discuss the interface with the orbit-averaged evolution in Sec.~\ref{sec:interface}, while in Sec.~\ref{sec:tilts_formation} we discuss the uncertainties involved in approximating the tilts at formation by those at infinity. We apply the method to some of the binary black holes detected by LIGO and Virgo in Sec.~\ref{sec:appl}. We summarize and conclude in Sec.~\ref{sec:concl}. In Appendices~\ref{app:lin_err_bound} and~\ref{app:err_quadratic} we derive the error bounds for the simplifications to the precession-averaged evolution. In Appendix~\ref{app:prec_avg_internal_checks} we provide details about the internal checks in the precession-averaged evolution. Finally, in Appendix~\ref{app:examples} we give example uses of the code. We use $G = c = 1$ units throughout.

\section{Precession-averaged evolution and the spin tilts at infinity}
\label{sec:prec_evol}

The precession-averaged evolution introduced in~\cite{Kesden:2014sla,Gerosa:2015tea} allows one to evolve the spins of compact binaries over long timescales efficiently, even back to a formally infinite separation, at the cost of not tracking the precessional phase. This evolution is restricted to binary black holes, because it relies on the conservation of the effective spin (defined below), which is not conserved for binaries which contain an object that is not a black hole, due to the difference in spin-induced quadrupole moment~\cite{Racine:2008qv}.
While the expressions in~\cite{Kesden:2014sla,Gerosa:2015tea} are only applicable to quasicircular binaries, it turns out that they can be applied to eccentric binaries almost verbatim, as discussed below.

As pointed out by Gerosa~\emph{et al.}~\cite{Gerosa:2015tea}, the tilt angles at infinity are well defined, except for the case of an exactly equal-mass binary, where there is a qualitatively different behavior, as discussed in~\cite{Gerosa:2016aus} and Sec.~\ref{ssec:q1}. We now outline the computation of the tilts at infinity, which we refer to as $\theta_{1\infty}$ and $\theta_{2\infty}$, following the formulation of precession-averaged evolution from Chatziioannou~\emph{et al.}~\cite{Chatziioannou:2017tdw}. As in Chatziioannou~\emph{et al.}, we work in $M = 1$ units for the derivation, and denote the binary's mass ratio by $q := m_2/m_1 < 1$. We denote the binary's individual dimensionful spins by $\mathbf{S}_1$ and $\mathbf{S}_2$, while $\mathbf{S} := \mathbf{S}_1 + \mathbf{S}_2$ denotes the binary's total spin. Similarly, $\mathbf{J} = \mathbf{L} + \mathbf{S}$ denotes the binary's total angular momentum, where $\mathbf{L}$ is its orbital angular momentum. The non-boldface versions of all of these quantities denote their magnitudes, as usual.

Specifically, we want to compute [Eqs.~(45) in Gerosa~\emph{et al.}~\cite{Gerosa:2015tea}]
\begin{subequations}
\begin{align}
\cos\theta_{1\infty} &= \frac{-\xi + \kappa_\infty(1 + q^{-1})}{S_1(q^{-1} - q)},\\
\cos\theta_{2\infty} &= \frac{\xi - \kappa_\infty(1 + q)}{S_2(q^{-1} - q)}.
\end{align}
\end{subequations}
Here [Eq.~(12) in Gerosa~\emph{et al.}]
\<\label{eq:xi_def}
\xi := [(1 + q)\mathbf{S}_1 + (1 + q^{-1})\mathbf{S}_2]\cdot\hat{\mathbf{L}}
\?
is the effective spin (circumflexes denote unit vectors), which is a conserved quantity for the 2PN orbit-averaged evolution, and $\kappa_\infty$ is the value of [Eq.~(40) in Gerosa~\emph{et al.}]
\<\label{eq:kappa_def}
\kappa := \frac{J^2 - L^2}{2L} = \mathbf{S}\cdot\hat{\mathbf{L}} + \frac{S^2}{2L}
\?
at $L \to \infty$. The evolution of $\kappa$ as a function of $u := 1/(2L)$ is given by
\<\label{eq:kappa_eq}
\frac{d\kappa}{du} = \langle S^2\rangle_\text{pr}.
\?
This is an inline equation below Eq.~(44) in Gerosa~\emph{et al.}, which comes from that paper's Eq.~(41). Here $\langle S^2\rangle_\text{pr}$ is the precession average of $S^2$. Thus, $\kappa_\infty$ is given by the value of $\kappa$ at $u = 0$.

Now, from Eq.~(42) in Chatziioannou~\emph{et al.}~\cite{Chatziioannou:2017tdw} (rewritten slightly using the definition of $m$, defined below, to simplify it),
\<\label{eq:Ssq_pr}
\langle S^2\rangle_\text{pr} = S^2_+ + (S^2_+ - S^2_3)\left[\frac{E(m)}{K(m)} - 1\right],
\?
where $S^2_+ > S^2_- > S^2_3$ are the roots of the cubic equation (in $S^2$)
\<\label{eq:cubic}
S^6 + BS^4 + CS^2 + D = 0,
\?
whose coefficients are given in Eqs.~(B2--B4) in Chatziioannou~\emph{et al.}, $E$ and $K$ are complete elliptic integrals [whose definitions are given in Eqs.~\eqref{eq:ellip}], and [Eq.~(25) in Chatziioannou~\emph{et al.}]
\<
m := \frac{S^2_+ - S^2_-}{S^2_+ - S^2_3}.
\?

When performing this computation, we need to avoid catastrophic cancellations, which can occur in several places. One place involves the computation of the tilt angles themselves, where the denominators diverge as $q \nearrow 1$ (recall that we need to treat $1 - q$ as small as $\sim 1.5 \times 10^{-6}$ in the application to LIGO-Virgo detections). The other place involves the computation of $\langle S^2\rangle_\text{pr}$, where one obtains an $\infty\cdot 0$ indeterminate form as $S^2_3 \to \infty$, so $m \to 0$, since
\begin{subequations}
\begin{align}
E(m) &= \frac{\pi}{2}\left[1 - \frac{m}{4} + O(m^2)\right],\\
K(m) &= \frac{\pi}{2}\left[1 + \frac{m}{4} + O(m^2)\right].
\end{align}
\end{subequations}
In the next section, we describe how to rescale $\kappa$ to avoid the catastrophic cancellations for $q$ close to $1$. In Appendix~\ref{app:lin_err_bound}, we derive a bound on the error made in linearizing to obtain $\langle S^2\rangle_\text{pr} = (S_+^2 + S_-^2)/2 + O(m^2)$ to avoid the $\infty\cdot 0$ indeterminate form---see Eq.~\eqref{eq:m_bound} for the restriction on $m$ for linearization to lead to a given error in $\langle S^2\rangle_\text{pr}$ (in terms of the barred quantities introduced in the next section).
 
\subsection{The eccentric case}
\label{ssec:ecc}
 
As noted in Gerosa~\emph{et al.}~\cite{Gerosa:2015tea}, the generalization of the precession-averaged evolution to eccentric binaries is quite straightforward---this generalization was carried out by Yu~\emph{et al.}~\cite{Yu:2020iqj}. They found that the eccentricity only affects $dS/dt$ by an overall eccentricity-dependent factor---see their Eq.~(57). Thus, $\langle dJ/dL\rangle_\text{pr}$ [their Eq.~(59), with our notation for the precession average] does not depend on eccentricity, since the contribution from the explicit appearance of $dS/dt$ is cancelled by one from $\tau_\text{pre}$ [their Eq.~(58)]. Therefore, the expression for $d\kappa/du$ does not change in the eccentric case. The remainder of the calculation is purely geometric relations (see, e.g., the discussion in Sec.~III~A of Yu~\emph{et al.}). Therefore, when approximating the tilts at formation, the quasicircular expressions are equally applicable to eccentric binaries. Of course, the orbital angular momentum does not diverge at infinite separation for eccentric binaries like it does for quasicircular binaries---see Eq.~(5.11) in Peters~\cite{Peters:1964zz}, noting that $L \propto \sqrt{a(e)(1 - e^2)}$ remains finite as $e\nearrow 1$.
Thus, the tilts at infinite separation themselves are not well defined for eccentric binaries---they do not approach a single value. However, the tilts at infinite orbital angular momentum are still a good approximation to the tilts at formation in most eccentric cases, since this approximation only relies on the orbital angular momentum at formation being sufficiently large, as it is in standard formation scenarios (as discussed in Sec.~\ref{sec:tilts_formation}).
 
\section{Regularizing the precession-averaged equations for close-to-equal masses}
\label{sec:reg}

For exactly equal masses, the total spin is also conserved by the $2$PN orbit-averaged evolution, as discussed in~\cite{Gerosa:2016aus}. Thus, from Eq.~(2.6) in that paper, which says that
\<
\xi = \frac{J^2 - L^2 - S^2}{L} \qquad (q = 1)
\?
(in our $M = 1$ units), we have
\<
\kappa = \frac{S_0^2}{2L} + \frac{\xi}{2} \qquad (q = 1),
\?
where $S_0$ is the initial magnitude of the total spin. If we introduce $\epsilon := 1 - q > 0$ (where we are particularly interested in the method's accuracy for small $\epsilon$, but will obtain expressions valid for a general $\epsilon$) and take the ansatz\footnote{This ansatz is inspired by the special equal-mass case, though this is a singular limit, and we do not expect the $q < 1$ case to reduce to the $q = 1$ case in the limit $\epsilon \searrow 0$. In particular, as discussed in Sec.~\ref{ssec:q1}, the tilts at infinity are not well defined in the $q = 1$ case.} that
\<\label{eq:ansatz}
\kappa =  \frac{S_0^2}{2L} + \frac{\xi}{2} + \epsilon \kappa^{(\epsilon)} + O(\epsilon^2),
\?
we thus have
\begin{subequations}
\label{eq:tilts_ansatz}
\begin{align}
S_1\cos\theta_{1\infty} &= \frac{-(1 - \epsilon)\xi + (2 - \epsilon)\kappa_\infty}{2\epsilon} + O(\epsilon)\nonumber\\
&= \kappa^{(\epsilon)}_\infty - \xi/4 + O(\epsilon),\\
S_2\cos\theta_{2\infty} &= \frac{(1 - \epsilon)\xi - (2 - \epsilon)\kappa_\infty}{2\epsilon} + O(\epsilon)\nonumber\\
&= -\kappa^{(\epsilon)}_\infty + \xi/4 + O(\epsilon),
\end{align}
\end{subequations}
so that these are no longer singular in the limit $q \nearrow 1$ (i.e., $\epsilon \searrow 0$).

This motivates us to try to replace $\kappa$ with something like $\kappa^{(\epsilon)}$ as the variable being solved for. We thus note that if we know $\xi$ and $S_1\cos \theta_{1\infty}$, we can obtain $S_2\cos\theta_{2\infty}$ in a numerically stable way even for $q$ close to $1$ by using the definition of $\xi$ [Eq.~\eqref{eq:xi_def}] to obtain
\<
S_2\cos\theta_{2\infty} = \frac{\xi - (1 + q)S_1\cos \theta_{1\infty}}{1 + q^{-1}}.
\?
We therefore take our $\kappa^{(\epsilon)}$-like variable to reduce to $S_1\cos \theta_{1\infty}$ as $L\to\infty$. We also want to include the $S_0$ contribution from Eq.~\eqref{eq:ansatz}, which vanishes as $L \to \infty$, hence we define
\<\label{eq:kappaxiq_def}
\begin{split}
\kappa_{\xi q} &:= \frac{1}{1-q}\left(\kappa - \frac{S_0^2}{2L} - \frac{q\xi}{1 + q}\right)\\
&\;= S_1\cos\tilde{\theta}_1 + \frac{S^2 - S_0^2}{2L(1-q)},
\end{split}
\?
where the second equality comes from the second equality in Eq.~\eqref{eq:kappa_def}. Here we write $\tilde{\theta}_1$ to distinguish this tilt angle with respect to the direction of the full orbital angular momentum, including contributions from the black holes' spins [see, e.g., Eq.~(4.7) in~\cite{Bohe:2012mr}], with the tilt angles with respect to the direction of the binary's Newtonian orbital angular momentum (i.e., the normal to the orbital plane) used in gravitational wave data analysis~\cite{Farr:2014qka}. This distinction is only necessary when initializing the evolution, since the spin contributions to $\mathbf{L}$ only enter the precession equations at higher PN order than is used in the precession-averaged evolution. The term in the second equality involving $1/(1 - q)$ is well behaved as $q \nearrow 1$, since $dS/dt \propto 1 - q^2 = O(\epsilon)$ [see, e.g., Eq.~(8) in~\cite{Kesden:2014sla}], hence $S^2 - S_0^2$ is also $O(\epsilon)$.
Thus, we have
\begin{subequations}\label{eq:tilts_infty_reg}
\begin{align}
S_1\cos\theta_{1\infty} &= \kappa_{\xi q, \infty},\\
S_2\cos\theta_{2\infty} &= q\left(\frac{\xi}{1 + q} - \kappa_{\xi q, \infty}\right).
\end{align}
\end{subequations}
[The difference of $\xi/4$ compared to Eqs.~\eqref{eq:tilts_ansatz} is because we have $(1 - \epsilon)\xi/(2 - \epsilon)$ in the definition of $\kappa_{\xi q}$ as opposed to just $\xi/2$ in the definition of $\kappa^{(\epsilon)}$.]

The initial value for $\kappa_{\xi q}$ (e.g., the value at the reference frequency, in the application to gravitational wave observations) can be obtained from the second equality in Eq.~\eqref{eq:kappaxiq_def}, which gives
\<\label{eq:kappaxiq0}
\kappa_{\xi q}^{(0)} = S_1\cos\tilde{\theta}_1^{(0)}.
\?
Here we denote initial values by a superscript $(0)$.

We now consider the differential equation satisfied by $\kappa_{\xi q}$. Since $\xi$ is a conserved quantity, we have
\<\label{eq:ode}
\begin{split}
\frac{d\kappa_{\xi q}}{du} &= \frac{1}{1 - q}\left(\frac{d\kappa}{du} - S_0^2\right)\\
&= \bar{S}^2_+ + (\bar{S}^2_+ - \bar{S}^2_3)\left[\frac{E(m)}{K(m)} - 1\right]\\
&= \langle\bar{S}^2\rangle_\text{pr},
\end{split}
\?
where
\<\label{eq:Sbar_def}
\bar{S}^2_\star := \frac{S^2_\star - S_0^2}{1 - q}
\?
(denoting any of $+$, $-$, or $3$ by $\star$), so we have
\<
m = \frac{\bar{S}^2_+ - \bar{S}^2_-}{\bar{S}^2_+ - \bar{S}^2_3}
\?
and can obtain the equation to solve for the $\bar{S}^2_\star$ by substituting $S^2 = (1-q)\bar{S}^2 + S_0^2$ in Eq.~\eqref{eq:cubic}.\footnote{There are simpler forms of the equation in the special cases when $m = 0$ (i.e., $\bar{S}^2_+ = \bar{S}^2_-$), where $\langle\bar{S}^2\rangle_\text{pr} = \bar{S}^2_+$, and when $m = 1$ (i.e., $\bar{S}^2_3 = \bar{S}^2_-$), where $\langle\bar{S}^2\rangle_\text{pr} = \bar{S}^2_3$.} We then write the coefficients in terms of $\kappa_{\xi q}$, eliminating $J^2$, obtaining [after multiplying through by $q(1-q^2)u^2$ to regularize and simplify the coefficients]
\<\label{eq:cubic_barred}
q(1-q^2)u^2\bar{S}^6 + \bar{B}\bar{S}^4 + \bar{C}\bar{S}^2 + \bar{D} = 0,
\?
where
\begin{subequations}
\begin{align}
\bar{B} &= \frac{(1 - q)^2(1 + q)}{4} + q(1 - q)\left[\xi - 2(1+q)\kappa_{\xi q}\right]u\nonumber\\
&\quad  + \Upsilon u^2,\\
\bar{C} &= (1 - q)\left[(1 + q)\left(\Sigma + q\kappa_{\xi q}^2\right) - q\xi\kappa_{\xi q}\right]\nonumber\\ 
&\quad + 2\left(\zeta - \Upsilon\kappa_{\xi q}\right)u,\\
\bar{D} &= (1 + q)\left(\Sigma^2 - S_1^2S_2^2\right) + \frac{q^2 S_1^2\xi^2}{1 + q} - 2\zeta\kappa_{\xi q}  + \Upsilon\kappa_{\xi q}^2,
\end{align}
\end{subequations}
and
\begin{subequations}
\begin{align}
\Sigma &:= (S_0^2 - S_1^2 - S_2^2)/2,\\
\Upsilon &:= (1 + q)(2q\Sigma + q^2S_1^2 + S_2^2),\\
\zeta &:= q(\Sigma + qS_1^2)\xi.
\end{align}
\end{subequations}
We see that the singular nature of the $q \nearrow 1$ limit persists in this formulation through the $1 - q^2$ factor multiplying the $\bar{S}^6$ term in the cubic.

This version of the calculation of the tilts at infinity works well even for $1 - q \simeq 10^{-8}$, except for some very fine-tuned corner cases at or close to the endpoint of the up-down instability \edit{spin angles} obtained in~\cite{Mould:2020cgc} (see~\cite{Gerosa:2015hba} for more information about this instability), where the maximum mass ratio that can be evolved successfully is considerably smaller, as shown in Sec.~\ref{sec:numerics}.\footnote{\edit{The endpoint of the up-down instability computed in~\cite{Mould:2020cgc} corresponds to the (unphysical) limit of zero orbital angular momentum. Since we start our evolutions from a finite separation, the tilts at infinity corresponding to the up-down instability endpoint spin angles are not exactly $0$ and $\pi$, and can even differ considerably from these values for unequal mass ratios.}} The nonregularized version of the evolution implemented in PRECESSION~\cite{Gerosa:2016sys} runs in to difficulties in some of these corner cases for smaller mass ratios than the regularized evolution does, even in its updated development version~\cite{PRECESSION_GitHub_dev} that implements the method from Chatziioannou~\emph{et al.}~\cite{Chatziioannou:2017tdw}. We only compare with the more accurate development version in this paper. However, PRECESSION is able to evolve the up-down instability case for some mass ratios where the regularized evolution fails, though it loses accuracy for mass ratios close to $1$, as we illustrate in Sec.~\ref{sec:numerics}. It would be interesting to explore whether one can obtain more robust evolutions in this case using the alternative regularization in~\cite{Klein:2021jtd}, which replaces $S$ as the quantity to be evolved with a mass-weighted difference of spins.

Note that the error bound on linearizing in $m$ obtained in Appendix~\ref{app:lin_err_bound} still applies here with the substitution $S^2_\star \to \bar{S}^2_\star$. Additionally, for $q$ close to $1$ or $0$ and/or $u$ close to $0$, the coefficient of $\bar{S}^6$ in Eq.~\eqref{eq:cubic_barred} becomes quite small, and $\bar{S}^2_3$ becomes large, so we only need $\bar{S}^2_+ + \bar{S}^2_-$ in order to compute $d\kappa_{\xi q}/du$ to a good approximation. Additionally, we have $\bar{S}^2_+ + \bar{S}^2_- \simeq -\bar{C}/\bar{B}$ to a good approximation (obtained by setting the coefficient of $\bar{S}^6$ to $0$). We quantify the errors in this approximation in Appendix~\ref{app:err_quadratic}, specializing to the case where the coefficients of the cubic are all positive, which simplifies the analysis and is also the case encountered in practice.

One can also use the regularized expressions to obtain the range of tilt angles at some separation other than infinity. These are not a single value like they are at infinity, since the second term in Eq.~\eqref{eq:kappaxiq_def} does not vanish for finite $L$. However, one can obtain the upper and lower bounds on the tilt angles by evaluating that term using $S^2 \to S^2_\pm$. It is also possible to obtain an average value by evaluating it using $S^2 \to \langle S^2\rangle_\text{pr}$. Here $S^2_\pm$ and $\langle S^2\rangle_\text{pr}$ are calculated for the angular momentum corresponding to the desired semimajor axis and eccentricity. One can also write the additional $L$-dependent term directly in terms of the barred quantities in Eq.~\eqref{eq:Sbar_def}, so one has
\begin{subequations}
\label{eq:tilts_at_L}
\begin{gather}
\begin{align}
S_1\cos\theta_{1L}^\pm &= \kappa_{\xi q, L} - \frac{\bar{S}^2_{\pm, L}}{2L},\\
S_1\cos\theta_{1L}^\text{avg} &= \kappa_{\xi q, L} - \frac{\langle\bar{S}^2\rangle_{\text{pr},L}}{2L},
\end{align}
\end{gather}
and
\<
S_2\cos\theta_{2L}^{\pm,\text{avg}} = q\left(\frac{\xi}{1 + q} - S_1\cos\theta_{1L}^{\pm,\text{avg}}\right).
\?
\end{subequations}
Here the subscript $L$s denote the value when the binary's orbital angular momentum has the magnitude $L$.
We also have $\theta^-_{1L} \leq \theta_{1L}(t) \leq \theta^+_{1L}$ and $\theta^+_{2L} \leq \theta_{2L}(t) \leq \theta^-_{2L}$, where $\theta_{1,2L}(t)$ denotes the (time dependent) tilt angles when the binary's orbital angular momentum has the magnitude $L$ (letting the binary evolve conservatively with fixed magnitude of $L$). We use these expressions to assess the uncertainties in approximating the tilts at formation by those at infinity in Sec.~\ref{sec:tilts_formation}.

\subsection{The $q = 1$ case}
\label{ssec:q1}

We now discuss the $q = 1$ case of determining the tilts at infinity. As mentioned previously, the tilts at infinity are not well defined for $q = 1$. In that case, one can describe the precessional motion by the angle $\varphi'$ between the projection of $\mathbf{S}_1$ orthogonal to the total spin and a reference direction, and obtains [Eqs.~(2.15) and~(2.16) in~\cite{Gerosa:2016aus}]
\begin{subequations}
\label{eq:tilts_q1}
\begin{align}
\cos\theta_1 &= \frac{1}{4S_1S_0^2}\left[\xi(S_0^2 + S_1^2 - S_2^2) + \cS^3\cos\varphi'\right],\\
\cos\theta_2 &= \frac{1}{4S_2S_0^2}\left[\xi(S_0^2 + S_2^2 - S_1^2) - \cS^3\cos\varphi'\right],
\end{align}
\end{subequations}
where
\<
\cS^6 := (\xi^2 - 4S_0^2)[(S_1 - S_2)^2 - S_0^2][(S_1 + S_2)^2 - S_0^2]
\?
(defined so that $\cS$ has units of angular momentum), and we have converted to the $M = 1$ units we use, as well as replaced $S^2 \to S_0^2$, since it is constant for $q = 1$. The tilt angles depend on $L$ through $\cos\varphi'$ and $d\varphi'/dL \propto S_0 L$ (at leading PN order), from Eq.~(2.9) in~\cite{Gerosa:2016aus} and Eq.~(36) in~\cite{Gerosa:2015tea}, recalling that $L \propto r^{1/2}$. Thus, except in the special case $S_0 = 0$, where $\varphi'$ is constant, the tilt angles continue to oscillate between the bounds given by substituting $\cos\varphi' \to \pm 1$ in Eqs.~\eqref{eq:tilts_q1} [given explicitly in Eqs.~(2.19) and~(2.20) of~\cite{Gerosa:2016aus}] as $L \to\infty$ without approaching a limit.

Nevertheless, the regularized expressions in Sec.~\ref{sec:reg} are well behaved for $q \nearrow 1$, so it is interesting to see what they give in this case. Noting that the cubic equation~\eqref{eq:cubic_barred} degenerates into a quadratic in this case and $m \to 0$, we have
\<
\begin{split}
\frac{d\kappa_{\xi q}}{du} &= \langle\bar{S}^2\rangle_\text{pr}\\
&= \frac{\bar{S}^2_+ + \bar{S}^2_-}{2}\\
&= -\frac{\bar{C}}{2\bar{B}}\\
&= \left[\kappa_{\xi q} - \frac{S_0^2 + S_1^2 - S_2^2}{4S_0^2}\xi\right]\frac{1}{u}.
\end{split}
\?
This has a solution of
\<
\kappa_{\xi q} = \frac{S_0^2 + S_1^2 - S_2^2}{4S_0^2}\xi + \alpha u,
\?
where $\alpha$ is a constant that is fixed by the initial conditions [cf.\ Eq.~\eqref{eq:kappaxiq0}] and we thus have
\<
\kappa_{\xi q,\infty} = \frac{S_0^2 + S_1^2 - S_2^2}{4S_0^2}\xi,
\?
so, from Eqs.~\eqref{eq:tilts_infty_reg},
\begin{subequations}
\begin{align}
\cos\theta_1 &= \frac{S_0^2 + S_1^2 - S_2^2}{4S_1S_0^2}\xi,\\
\cos\theta_2 &= \frac{S_0^2 + S_2^2 - S_1^2}{4S_2S_0^2}\xi,
\end{align}
\end{subequations}
which are the average values of $\cos\theta_1$ and $\cos\theta_2$ over a precession cycle (i.e., their values for $\cos\varphi' = 0$) from Eqs.~\eqref{eq:tilts_q1}.

\begin{figure}
\includegraphics[width=0.48\textwidth]{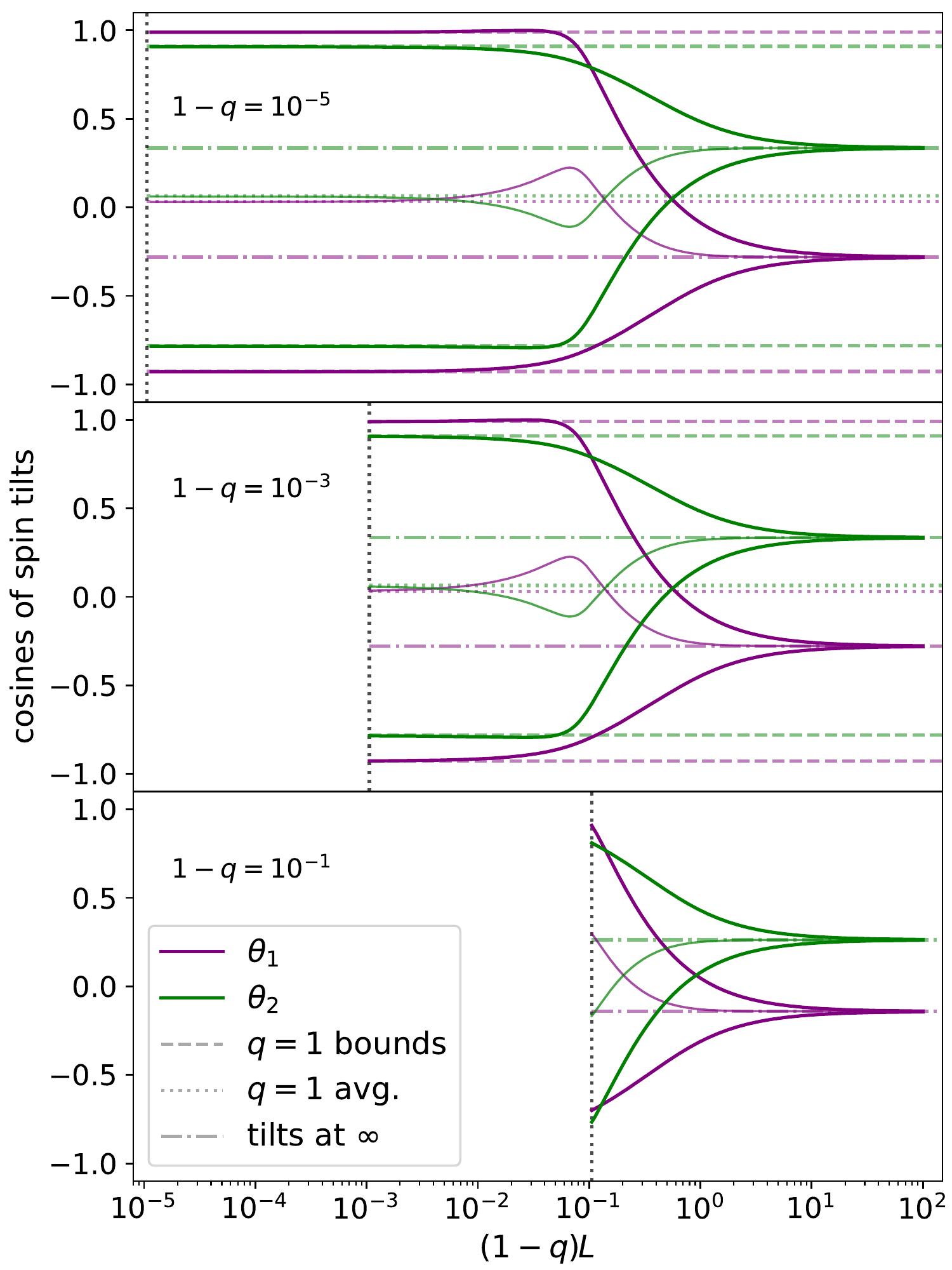}
\caption{\label{fig:tilts_vs_L} The maximum and minimum (thick lines) and average (thin lines) \edit{cosines of the} tilts as a function of $L$ given by the precession-averaged expressions for a significantly precessing binary, with parameters given in the text and mass ratios $q$ with $1 - q = 10^{-5}, 10^{-3}, 10^{-1}$ (the binary's total mass is fixed). We also plot the maximum, minimum, and average \edit{cosines of the} tilts for $q = 1$ in the upper two plots (the same in both plots) as dashed and dotted horizontal lines, respectively. The \edit{cosines of the} tilts at infinity for each mass ratio are shown as horizontal dash-dotted lines. We also show the initial value of $L$ for each mass ratio as a vertical dotted line. We plot versus $(1 - q)L$ to illustrate the approximate scaling of these quantities with $1/(1 - q)$. We do not necessarily expect the precession-averaged expressions to be very accurate for the smallest $L$ values plotted, and just show results for those values to illustrate these expressions' behavior in this regime.}
\end{figure}

Numerical experiments indicate that these are not the $q \nearrow 1$ limit of the $q < 1$ tilts at infinity. This is expected, since that limit is singular. As illustrated in Fig.~\ref{fig:tilts_vs_L}, for $q$ close to $1$ but not exactly equal to it, these tilts agree well with the average tilts obtained from Eqs.~\eqref{eq:tilts_at_L} for small $L$ values, where the maximum and minimum tilts [also obtained from Eqs.~\eqref{eq:tilts_at_L}] are also very close to those one obtains for $q = 1$ from Eqs.~\eqref{eq:tilts_q1} when substituting $\cos\varphi' \to \pm 1$. However, for large $L$, the average, maximum, and minimum tilts all approach their values at infinity, as expected. The magnitude of $L$ necessary to transition from the equal-mass-like tilts to the tilts at infinity grows as $1/(1 - q)$, thus illustrating why the $q \nearrow 1$ limit is singular. Indeed, all the behavior of the maximum, minimum, and average tilts as a function of $L$ scales very well with $1/(1 - q)$ for $q$ close to $1$, as illustrated in Fig.~\ref{fig:tilts_vs_L}.

The binary considered in Fig.~\ref{fig:tilts_vs_L} was chosen to be significantly precessing, to provide a good illustration, but otherwise had its parameters selected quasi-randomly; all the other binaries we tried had qualitatively similar behavior for the scaling with $1/(1 - q)$ and the relation to the equal-mass and $L\to\infty$ limits.
These parameters are a total mass of $60M_\odot$ and dimensionless spins of $\chi_1 = 0.82$, $\chi_2 = 0.93$, with spin angles of $\theta_1 = 1.2$~rad, $\theta_2 = 1.8$~rad, $\phi_{12} = 2.3$~rad at a dominant gravitational-wave frequency of $f_0 = 20$~Hz. Here $\theta_{1,2}$ denotes the spins' tilt angles (with respect to the direction of the binary's Newtonian orbital angular momentum, i.e., the normal to the orbital plane), and $\phi_{12}$ denotes the angle between the components of spin~1 and spin~2 in the binary's orbital plane. These spin angles are the ones used in gravitational wave data analysis~\cite{Farr:2014qka} [though the description of $\phi_{12}$ in~\cite{Farr:2014qka} is a bit confusing, referring to the angle of the difference of spin vectors in the text below their Eq.~(2.10), when the associated footnote makes it clear that the difference of azimuthal spin angles is intended].

\section{Numerical implementation of precession-averaged evolution}
\label{sec:numerics}

We have implemented the regularized equations and simplifications detailed above in a Python code \verb|calc_tilts_prec_avg_regularized|, using the numpy~\cite{Harris:2020xlr} and scipy~\cite{Virtanen:2019joe} packages for the default evolution, as well as the mpmath~\cite{mpmath} arbitrary precision package in order to evolve difficult cases, e.g., as a fallback. Given the binary's masses, spin magnitudes, and spin angles $\theta_1$, $\theta_2$, $\phi_{12}$ (defined above) at a dominant mode gravitational wave reference frequency $f_0$,\footnote{We use $f_0$ for the reference frequency for the precession-averaged evolution to distinguish it from $f_\text{ref}$, the reference frequency used in the parameter estimation, which is also the reference frequency for the orbit-averaged evolution when applying our hybrid evolution to posterior samples.} the function \verb|prec_avg_tilt_comp| outputs either the tilts at infinity or the bounds and average tilts at a finite separation, the latter also requiring the input of a final orbital angular momentum magnitude $L_\text{f}$. 

Following the initial implementation of the precession-averaged evolution in the PRECESSION code~\cite{Gerosa:2016sys}, we solve Eq.~\eqref{eq:ode} using the LSODA integrator~\cite{ODEPACK,Petzold}. However, we use the implementation in \verb|scipy.integrate.ode| rather than the one in \verb|scipy.integrate.odeint| used in the original version of PRECESSION. This gives us access to a number of other integrators, though we found that LSODA indeed appears to be the best choice for this problem of the integrators available through that function. We also included the option to use the LSODA implementation in \verb|scipy.integrate.solve_ivp|, which is the interface the development version of the PRECESSION code~\cite{PRECESSION_GitHub_dev} uses, though with a Runge-Kutta integrator instead of LSODA. While we found that this version of LSODA is faster (not needing to be applied in the sequence of steps described below) and allows for tighter tolerance settings than the \verb|scipy.integrate.ode| one, we also found that it apparently hangs for some more difficult cases, which is why we do not use it as the default. Future work will investigate whether it is possible to avoid these hangs with appropriate settings.

We apply the integrator in a sequence of small steps with a fixed step size to improve accuracy, where the default step size is $\Delta u = 10^{-3}$. We set the integrator's absolute  tolerance to a default value of $\delta_\text{abs} = \max[\delta_\text{abs}^\text{base}\min(\chi_1\sin\theta_1,\chi_2\sin\theta_2),\delta_\text{abs}^\text{floor}]$, with $\delta_\text{abs}^\text{base} = 10^{-8}$ and $\delta_\text{abs}^\text{floor} = 10^{-13}$, and the relative tolerance to a default value of $\delta_\text{rel} = 0.1\delta_\text{abs}$. The dependence of the tolerance on the in-plane spins was set after noticing that a fixed tolerance led to larger errors for cases with small in-plane spins. The floor on the tolerance is because LSODA produces errors for some binaries for smaller values of the tolerance, e.g., $10^{-14}$. By default, we set the linearization tolerance (see Appendix~\ref{app:lin_err_bound}) to be the same as the integrator's absolute tolerance, i.e., $\delta_\text{lin} = \delta_\text{abs}$. For comparison, the development version of PRECESSION uses $10^{-8}$ for both $\delta_\text{abs}$ and $\delta_\text{rel}$.

\begin{figure*}
\includegraphics[width=0.98\textwidth]{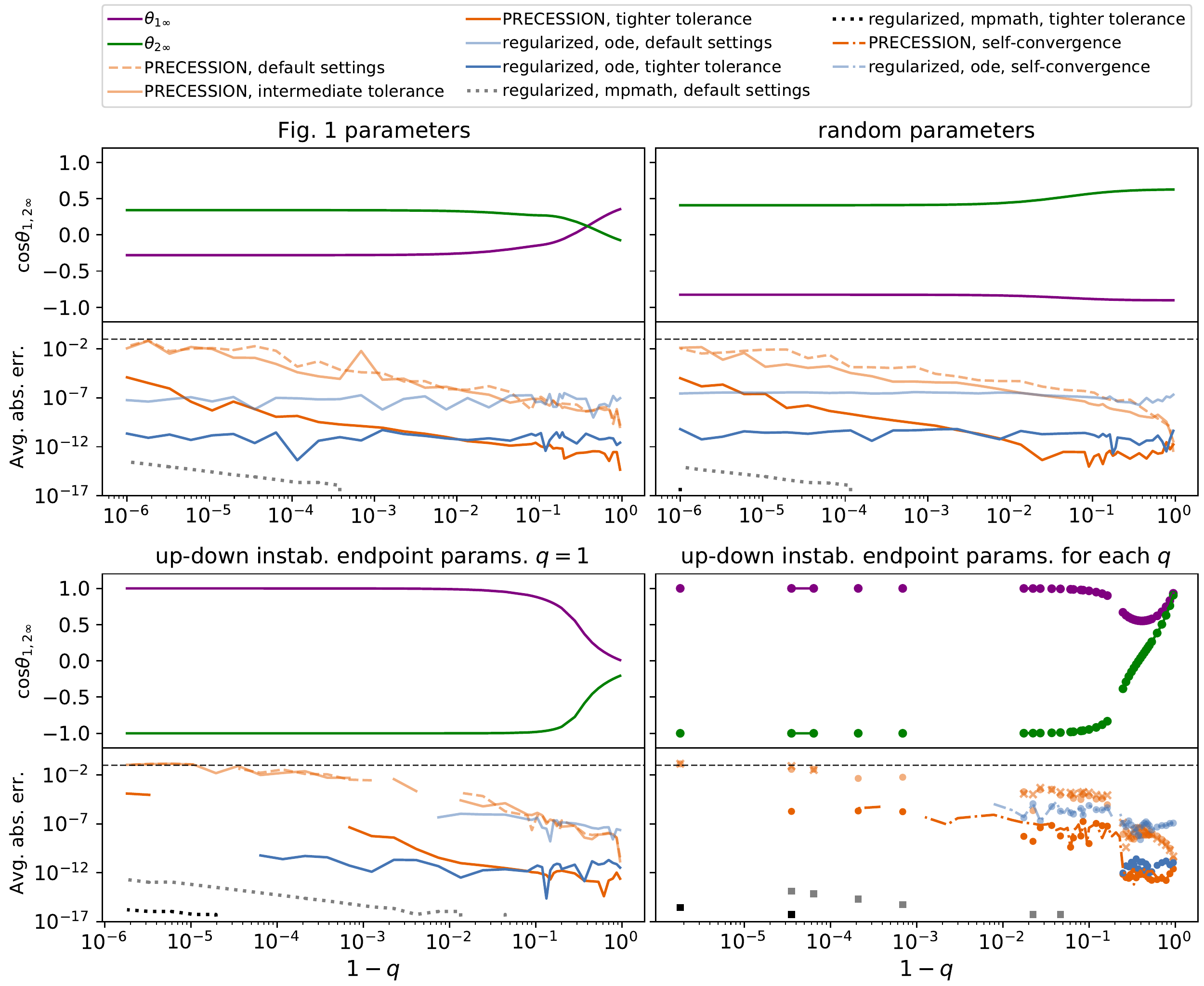}
\caption{\label{fig:accuracy} The \edit{cosines of the} tilts at infinity and their average errors as a function of mass ratio (expressed as $1 - q$ to highlight the mass ratios close to $1$) for four binary configurations, detailed in the text.
As discussed in the text, all errors are computed with respect to an mpmath evolution with strict tolerance settings, which also gives the results for the tilts at infinity that we plot. The text also gives the specifics of the different accuracy results and discusses the issue with the mpmath evolution that causes us to plot most results in the lower right-hand panel as symbols (defined in the text). In the lower right-hand panel we also plot the mpmath results for the tilts with circles and show the results from PRECESSION with the tighter tolerance settings as lines. The dashed line marks an error of \edit{$0.1$}. Above the lower bound of $1 - q = 10^{-6}$, the portions of the plot for which there is no trace for a given error are the ones for which the evolution(s) required to compute that trace did not succeed. This is also true for the large gaps between symbols in the lower right-hand panel. While the mass ratios are unevenly sampled, the places in the lower right-hand panel where there are very large gaps are those where the mpmath evolution does not succeed. All plots extend to a minimum value of $q$ of $0.05$ (i.e., $1 - q = 0.95$).
}
\end{figure*}

We also provide a fallback evolution (enabled by default) for cases where the primary evolution fails, e.g., if it encounters complex roots of the cubic equation~\eqref{eq:cubic}, since for corner cases where this occurs, the evolution can often succeed with more stringent (and thus more time-consuming) settings. The default fallback settings are $\Delta u = 10^{-5}$ and $\delta_\text{abs} = 10^{-13}$ (with all other settings being the same). The fallback settings are necessary for evolving the cases with the endpoint of up-down instability spin angles from~\cite{Mould:2020cgc} and slight perturbations thereof for binaries with mass ratios close to $1$. Specifically, those angles are
\begin{subequations}
\begin{align}
\cos\theta_1 = \cos\theta_2 &= \frac{\chi_1 - q\chi_2}{\chi_1 + q\chi_2},\\
\phi_{12} &= 0.
\end{align}
\end{subequations}
Binaries with these angles are the most difficult cases to evolve that we have encountered. For instance, for the up-down instability endpoint case considered in Fig.~\ref{fig:accuracy}, one is not able to evolve some mass ratios greater than $0.997$ with any evolution method we have tried.\footnote{Perturbing the tilts slightly makes this case much easier to evolve. For instance, if one uses the default $2$PN computation for the initial orbital angular momentum instead of the $0$PN computation used in Fig.~\ref{fig:accuracy} for comparison with the PRECESSION code, so that the orbital angular momentum receives contributions from the spins and is thus no longer parallel to the Newtonian orbital angular momentum used to define the tilts, then one can evolve this case with $1 - q$ as small as $10^{-5}$ with the first fallback evolution.}

We also have a second fallback evolution for even more difficult-to-evolve cases using the mpmath package's Taylor series-based differential equation integrator, its polynomial root finder (as opposed to the numpy polynomial root finder), and implementation of the elliptic integrals, as opposed to scipy's (as well as the final arccosines used to obtain the tilts instead of numpy's). This is necessary to evolve particularly difficult cases (mostly close to equal mass cases near the up-down instability endpoint) without errors, but is much slower than the LSODA integration. See the bottom two panels in Fig.~\ref{fig:accuracy} for some cases where this fallback evolution is triggered. The default fallback settings are $30$~digits, an integrator tolerance of $10^{-15}$, and $50$ digits of extra precision in the root finder; we otherwise use the default settings for the integrator and root finder. However, this second fallback evolution is optional and the code will work without mpmath being installed, though the second fallback evolution is enabled by default if mpmath is present. The code also has several internal checks that the evolution is proceeding well and the roots of the cubic equation~\eqref{eq:cubic} are being obtained accurately, as discussed in Appendix~\ref{app:prec_avg_internal_checks}.

To initialize the evolution, the code needs to convert the reference frequency $f_0$ into the binary's initial orbital angular momentum. It does this using the same PN expression used in the orbit-averaged evolution, viz., Eq.~(4.7) in~\cite{Bohe:2012mr} [see also Eq.~(4) in~\cite{SpinTaylor_TechNote}], with the orbit averaging discussed in, e.g., Sec.~II of~\cite{SpinTaylor_TechNote}. This expression contains orbit-averaged contributions from the black holes' spins by default. To allow evolutions to a given final separation, the code also provides the ability to convert an orbital separation and eccentricity to a magnitude of the orbital angular momentum. Since there are several different PN eccentricity parameters, the code only implements the Newtonian expression, for simplicity---see, e.g.,~\cite{Mora:2003wt,Memmesheimer:2004cv} for PN corrections.

\subsection{Illustrations of accuracy}

We illustrate the accuracy of the method as a function of mass ratio for four binary configurations in Fig.~\ref{fig:accuracy}, all with $f_0 = 20$~Hz: The first configuration is the same one used in Fig.~\ref{fig:tilts_vs_L}, \edit{and the magnitude of the initial orbital angular momentum, $L_0$, ranges from $0.17M^2$ to $0.94M^2$ for the range of mass ratios we consider. Here and in following comparisons, we use the Newtonian orbital angular momentum expression to compute the initial orbital angular momentum from the binary's parameters and $f_0$ for the regularized evolution, in order to facilitate the comparison with PRECESSION.} The second is completely random parameters: $\chi_1 = 0.485$, $\chi_2 = 0.171$, $\theta_1 = 2.699$~rad, $\theta_2 = 0.893$~rad, and $\phi_{12} = 4.746$~rad, with $M = 63.957M_\odot$, \edit{so $L_0$ ranges from $0.17M^2$ to $0.92M^2$ for the range of mass ratios we consider}. The third is the up-down instability endpoint angles for equal masses and equal dimensionless spins, here $\chi_1 = \chi_2 = 0.95$, i.e., $\theta_1 = \theta_2 = \pi/2$~rad, and $\phi_{12} = 0$, with $M = 30M_\odot$, \edit{so $L_0$ ranges from $0.22M^2$ to $1.19M^2$ for the range of mass ratios we consider}. The fourth is the up-down instability endpoint angles for each mass ratio with $\chi_1 = 0.7$, $\chi_2 = 0.5$, and $M = 30M_\odot$ \edit{and thus the same range of $L_0$ as the previous configuration}. 

We calculate the accuracy by comparing with the results computed with the mpmath evolution with $70$~digits, an integrator tolerance of $10^{-21}$, and $110$ extra digits of precision in the root finder as a sufficiently accurate approximation for the exact result. We verify that this is indeed very accurate by comparing with the mpmath evolution with its default second fallback settings (some intermediate settings) of $30$ ($50$) digits, an integrator tolerance of $10^{-15}$ ($10^{-18}$), and $50$ ($80$) extra digits. There we find errors that are orders of magnitude smaller than the ones with the scipy evolution (using the default ode function implementation of the LSODA integrator). We show the average of the absolute values of the errors of the two tilt angles, since their errors are generally very similar (only differing by more than a factor of $10$ for a few points in all cases except for the lower left-hand plot, and even there only a few points differ by more than a factor of $100$).

For the scipy evolution, we consider the default settings of $\Delta u = 10^{-3}$ and the default in-plane-spin-dependent $\delta_\text{abs}$ with the default values of $\delta_\text{abs}^\text{base} = 10^{-8}$ and $\delta_\text{abs}^\text{floor} = 10^{-13}$ (here the fallback evolution is disabled). We also consider the first fallback settings of $\Delta u = 10^{-5}$ and $\delta_\text{abs} = 10^{-13}$ (with the second fallback disabled). Additionally, we compare with the accuracy of the development version of PRECESSION~\cite{PRECESSION_GitHub_dev}, which uses a Runge-Kutta integrator, both with the default settings and with a slight modification to tighten the integrator tolerance to the same values used in the regularized integration.

For the $q = 1$ up-down instability endpoint case, only the mpmath evolution is able to evolve with no failures for all the mass ratios we consider. Moreover, both the PRECESSION evolution with the default settings and the one with the same tolerance settings as the default regularized evolution have $\sim 50\%$ errors for the mass ratios closest to $1$ that we consider. For the up-down instability endpoint for each $q$ case, none of the evolutions we consider is able to evolve all the mass ratios we consider. PRECESSION is able to evolve the most mass ratios, since it implements a case to deal with systems close to resonances. However, it is not able to evolve mass ratios too close to $1$. The regularized evolution with the mpmath integration is also not able to evolve many cases with mass ratios close to $1$, just isolated cases, \edit{as illustrated in Fig.~\ref{fig:accuracy}.}

The following accuracy results are plotted in Fig.~\ref{fig:accuracy}, all comparing with the mpmath evolution with the largest number of digits we consider, except the final two, and always using the development version of PRECESSION. All but the final two of these are plotted with symbols with the same color as the lines in the lower right-hand panel, due to the issues with the mpmath evolution in that case, discussed above. We thus note the symbols used for these, as well:
\begin{itemize}
\item PRECESSION, default settings (also plotted as crosses): The PRECESSION evolution with its default settings
\item PRECESSION, intermediate tolerance (also plotted as circles): The PRECESSION evolution with the same tolerance settings as the regularized evolution defaults
\item PRECESSION, tighter tolerance (also plotted as circles): The PRECESSION evolution with $\delta_\text{abs} = 10^{-13}$ and $\delta_\text{rel} = 10^{-14}$
\item regularized, ode, default settings (also plotted as circles): The regularized evolution with its default settings, including using the ode LSODA integrator
\item regularized, ode, tighter tolerance (also plotted as circles): The regularized evolution with $\delta_\text{abs} = 10^{-13}$ and $\delta_\text{rel} = 10^{-14}$, still using the ode LSODA integrator
\item regularized, mpmath, default settings (also plotted as squares): The regularized evolution using mpmath and its default second fallback settings
\item regularized, mpmath, tighter tolerance (also plotted as squares): The regularized evolution using mpmath and the intermediate settings
\item PRECESSION, self-convergence: Comparison of the PRECESSION evolution with $\delta_\text{abs} = 10^{-13}$ and $\delta_\text{rel} = 10^{-14}$ with the evolution with $\delta_\text{abs}^\text{base} = 10^{-12}$ and $\delta_\text{rel} = 0.1\delta_\text{abs}^\text{base}$ (and $\delta_\text{abs}^\text{floor} = 10^{-13}$)
\item regularized, ode, self-convergence: Comparison of the regularized evolution using the ode LSODA integrator with the default settings and with tolerances of $10^{-2}$ times the default settings
\end{itemize}

We find that for mass ratios close to $1$, the regularized evolution with LSODA is more accurate than PRECESSION for a given value of $\delta_\text{abs}^\text{base}$ (with the other accuracy settings fixed to their default values), as expected, while the PRECESSION evolution's errors increase roughly like a positive power of $1/(1 - q)$ for $q$ close to $1$. The regularized evolution's errors with LSODA only increase as $q\nearrow 1$ in the up-down instability endpoint cases, where they also increase roughly like a positive power of $1/(1 - q)$. The errors for the regularized evolution with the mpmath integration also increase roughly like a positive power of $1/(1 - q)$ in all cases.\footnote{The feature in the mpmath errors seen for mass ratios close to $1$ in the equal-mass up-down instability endpoint case is caused by $\theta_{2\infty}$ equaling $\pi$ to all $18$ decimal places to which we output the data.}

\begin{figure*}
\centering
\subfloat{
\includegraphics[width=0.95\textwidth]{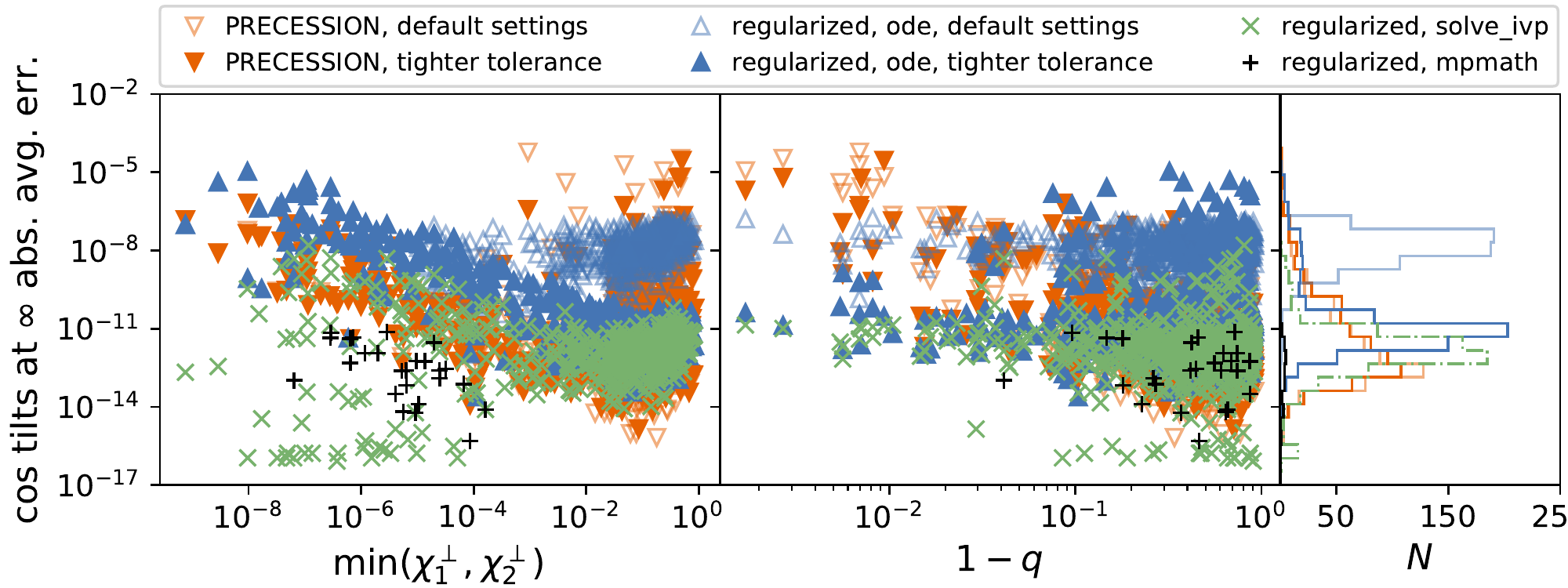}
}
\caption{\label{fig:accuracy_random} The accuracy of the \edit{cosines of the} tilts at infinity for randomly sampled binaries with the development version of PRECESSION and the regularized evolution with both the scipy ode and solve\_ivp LSODA integrators and the mpmath integrator, computed as discussed in the text. We plot the average of the absolute values of the errors of the \edit{cosines of the} two tilts versus the minimum of the magnitudes of the in-plane dimensionless spins of the two black holes ($\chi_1^\perp, \chi_2^\perp$) as well as versus $1 - q$ to show the scalings with both of these quantities. For PRECESSION, we show both the evolution with the default settings (the unfilled triangles and light histogram) and the one with the same evolution settings as the defaults for the regularized evolution (the filled triangles and heavy histogram). For the regularized evolution with the ode interface, the unfilled triangles and light histogram give the results with the default settings, while the filled triangles and heavy histogram give the results with $\delta_\text{abs}^\text{base} = 10^{-12}$. The small number of points plotted for the mpmath evolution are the only ones for which the error is larger than numpy's minimum resolvable difference of $\sim 10^{-16}$.}
\end{figure*}

In Fig.~\ref{fig:accuracy_random}, we compare the accuracy of the development version of PRECESSION and the regularized evolution with the default settings for the $500$ random binaries (starting from the spin tilts and transition frequencies from the orbit-averaged evolution determined there) used in Sec.~\ref{ssec:v_trans_validation} to validate the expression for the transition frequencies. We augment this set by $150$ additional binaries sampled from those $500$ but with the spins scaled randomly to extend to smaller values, to check the accuracy of the evolution in such cases. There are spin magnitudes as small as $\sim 3\times 10^{-8}$ for the GW190630\_185205 samples from the LIGO-Virgo collaboration analysis~\cite{GWTC-2_PE}. (The case with the smallest spin magnitudes also has the smallest in-plane spin magnitudes.) We compute the accuracy by comparing with the mpmath evolution using the intermediate settings given above, also checking the accuracy of the mpmath evolution with the default second fallback settings. As above, we consider the PRECESSION evolution both with the default settings and with the same integrator tolerance settings as the default regularized evolution. We also show the results for the regularized evolution with the solve\_ivp interface, still using the LSODA integrator, with $\delta_\text{abs}^\text{base} = 10^{-12}$ and also taking $\delta_\text{abs}^\text{floor} = 10^{-20}$, which is possible with this integrator and gives increased accuracy for cases with small in-plane spins.

We find that the development version of PRECESSION with default settings gives better accuracy for many systems than the regularized evolution with its default accuracy or even the much tighter tolerance. This is presumably because the right-hand side of the differential equation for the regularized evolution [Eq.~\eqref{eq:ode}] becomes orders of magnitude smaller than the non-regularized version [Eq.~\eqref{eq:kappa_eq}] that PRECESSION uses as $q \searrow 0$. Thus, in future work we may consider the optimal mass ratio to switch between the regularized and non-regularized evolution or an alternative $\kappa_{\xi q}$ expression that is better behaved for small mass ratios. However, as expected, the regularized evolution is more accurate for close-to-equal-mass systems. The errors increase for smaller minimum in-plane spins, though the solve\_ivp regularized evolution is able to keep them from growing too large, since it is able to use a much smaller setting for $\delta_\text{abs}^\text{floor}$, so the scaling of the tolerance with the in-plane spins is effective for much smaller in-plane spins (which require much tighter tolerances).

\section{Interface with orbit-averaged evolution}
\label{sec:interface}

We now present \verb|hybrid_spin_evolution|, a code that performs a hybrid orbit-averaged and precession-averaged
evolution of the spin angles of a binary from a given reference frequency $f_{\rm ref}$  to infinite separation. 
The evolution is carried out by the function \verb|calc_tilts_at_infty_hybrid_evolve| and proceeds in two stages.
In the first stage, we use orbit-averaged evolution \edit{as implemented in the SpinTaylor code} in LALSimulation~\cite{LALSuite}, which
is more accurate at higher frequencies (i.e., at smaller separations), until an empirically determined 
transition frequency $f_{\rm trans}$. At this point, the code switches to the second stage, 
which involves precession-averaged evolution to infinite separation using the \verb|prec_avg_tilt_comp|
function discussed in the previous section.
The transition frequency is set to a value that ensures that the tilts at infinity are accurate, 
with estimated absolute errors \edit{in their cosines} less than $10^{-3}$, 
while the evolution from $f_{\rm ref}$ to $f_{\rm trans}$ is still computationally efficient. 
In what follows, we describe how we determined $f_{\rm trans}$ and
the tests we performed to validate it.

\subsection{Determining the point of transition}

\begin{figure}[h]
\centering
\subfloat{
\includegraphics[width=0.45\textwidth]{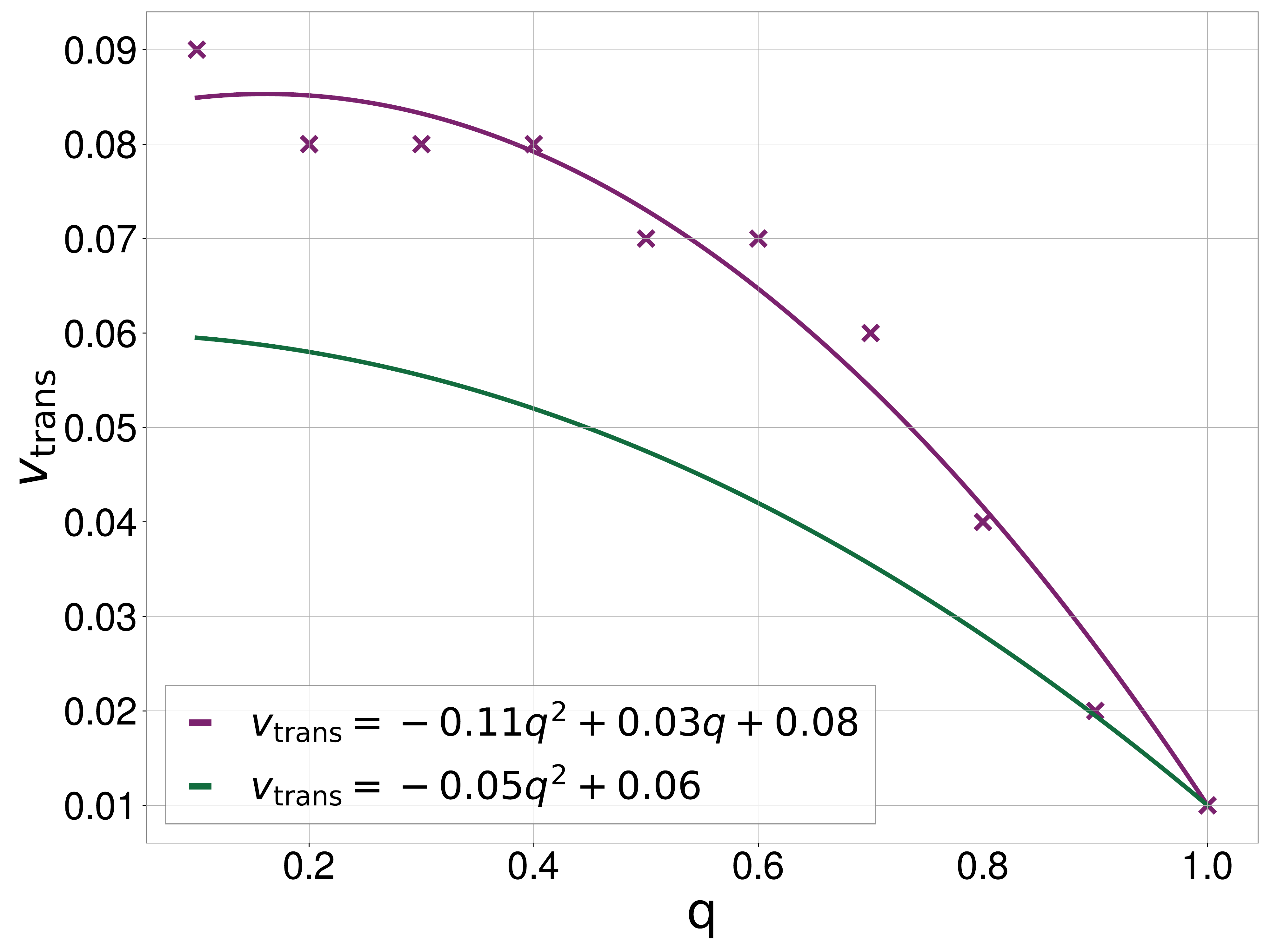}
}
\caption{\label{fig:v_trans_fit} The maximum transition orbital speed $v_{\rm trans}$ as a function of mass ratio $q$ that leads to estimated absolute errors 
in the tilts at infinity to be less than or equal to $10^{-3}$ rad.
The $v_{\rm trans}$ values (purple crosses) are given for each mass-ratio bin.
The purple line is the fit we get by using the purple $v_{\rm trans}$ vs $q$ data points. 
However, we choose a simple conservative $v_{\rm trans}$ vs $q$ curve (green line) to ensure that we are not above any of the data points. 
}
\end{figure}

In order to determine the transition frequency to switch from orbit-averaged to precession-averaged 
evolution, we work with a related dimensionless quantity, the transition orbital speed, $v_{\rm trans} = (\pi M f_{\rm trans})^{1/3}$,
where $M$ is the total mass of the binary. 

The gravitational wave frequency (or orbital speed) at which the precession-averaged equations become 
accurate enough for binary evolution depends upon the binary's parameters. We found that the dominant dependence is on the mass ratio.
To find $v_{\rm trans}$ as a function of $q$ we did a convergence test using a population of $1000$ binaries having their mass-ratios 
divided into $10$ bins, which are equally spaced, except for the one closest to $q = 1$. The first $9$ bins have mass ratios $q$ between $0.1$ and $0.9$ in steps of $0.1$, and are populated by $100$ binaries with mass ratios distributed uniformly within the bounds $q \pm 0.02$. The tenth bin has its $100$ binaries distributed uniformly in mass ratios between $0.98$ and $1$. The resultant $1000$ binaries
have their total mass and spin parameters at $f_{\rm ref}= 10$~Hz distributed uniformly as following: $M \in [10,200]M_{\odot}$; 
$\chi_1, \chi_2 \in [0, 1]$; $\theta_1, \theta_2 \in [0, \pi]$~rad; $\phi_{12} \in [0, 2\pi]$~rad. \edit{Thus, the magnitude of the initial Newtonian orbital angular momentum ranges from $0.14M^2$ to $1.72M^2$ (giving this for comparison with previous examples).}
We evolve the binaries using the orbit-averaged equations back to frequencies corresponding to orbital
speeds $v$ between $0.12$ and $0.01$ in steps of $0.01$ and then 
compute tilts at infinity using \verb|prec_avg_tilt_comp|. We then compute the absolute difference 
between the tilts at infinity $\Delta\theta_{i\infty}$ ($i \in \{1,2\}$) 
for $v$ and $v + 0.01$ and identify the largest $v$ value below which $\max(\Delta \theta_{i\infty})\leq10^{-3}$ rad
for all binaries in a given mass-ratio bin---this is our $v_{\rm trans}$ for a given mass-ratio bin. \edit{While we consider the cosines of the tilts at infinity (as the more physically relevant quantity) in the rest of the paper, here we are more conservative and set the accuracy in terms of the tilts themselves, noting that $|\cos(\theta + \Delta\theta) - \cos\theta| = |(1 - \cos\Delta\theta)\cos\theta + \sin\Delta\theta\sin\theta| \leq (2 - 2\cos\Delta\theta)^{1/2} \leq |\Delta\theta|$, using the Cauchy-Schwarz inequality and the standard trigonometric inequality $1 - \cos\alpha \leq \alpha^2/2$.}

Figure~\ref{fig:v_trans_fit} shows the $v_{\rm trans}$ values for which the tilts at infinity have an absolute error less than $10^{-3}$ rad for each mass ratio bin. 
We see that for close to equal mass binaries, the transition frequency needs to be very small, with $v_{\rm trans} = 0.01$.
We find that the dependence of $v_{\rm trans}$ on $q$ is approximately quadratic, with a best fit of $v_{\rm trans} =  -0.11q^2 + 0.03q + 0.08$.
However, there is significant scatter. Thus we choose a conservative
quadratic relation between $v_{\rm trans}$ and $q$ that ensures that this $v_{\rm trans}$ is not above any of the data points. 
This quadratic expression for $v_{\rm trans}$ is given by
\begin{equation}
\label{eq:v_trans}
v_{\rm trans} = -0.05 q^{2} + 0.06.
\end{equation}
For mass-ratios much smaller than $1$, this expression gives a $v_{\rm trans}$ that is considerably 
lower than the one given by curve fitting. However, these cases are not computationally 
expensive to evolve, so that is not a significant concern (see Sec.~\ref{ssec:hybrid_evol_code}).  

For the simulations performed for Fig.~\ref{fig:v_trans_fit}, we used the orbit-averaged evolution with 
the $3.5$PN accurate binding energy and flux, the $2.5$PN orbit-averaged precession equations 
including the leading order spin contributions to the orbital angular momentum, and the SpinTaylorT5
approximant (see~\cite{SpinTaylor_TechNote} for details). At the time when we performed the calculations for Fig.~\ref{fig:v_trans_fit}
the $3$PN spin-spin terms in the orbit-averaged precession equations were not available in \edit{the LALSimulation SpinTaylor code. We also only had the less accurate v1 evolution, which performed the orbit-averaged evolution stepwise in order to use the existing SpinTaylor code. (We updated the SpinTaylor code in v2.)} Since the calculations
to obtain the purple crosses in Fig.~\ref{fig:v_trans_fit} are rather computationally expensive, we thus chose to verify that the final expression in Eq.~\eqref{eq:v_trans}
is also appropriate for the $3$PN equations \edit{and with v2 of the evolution}, as well, rather than redoing these calculations with the $3$PN equations \edit{and v2 of the evolution}. In fact, we have verified that the $v_{\rm trans}$
expression given in Eq.~(\ref{eq:v_trans}) is robust for other orders of the precession equations and different SpinTaylor approximants, as discussed in Sec.~\ref{ssec:v_trans_validation}. \edit{We also find that the v2 evolution gives results that easily satisfy the accuracy requirements we placed.}
Thus, all the hybrid evolution results we present henceforth were obtained using the $3$PN order SpinTaylorT5 orbit-averaged precession equations, unless specified
otherwise, \edit{and all use v2 of the evolution}.

\begin{figure}[h]
\centering
\subfloat{
\includegraphics[width=0.47\textwidth]{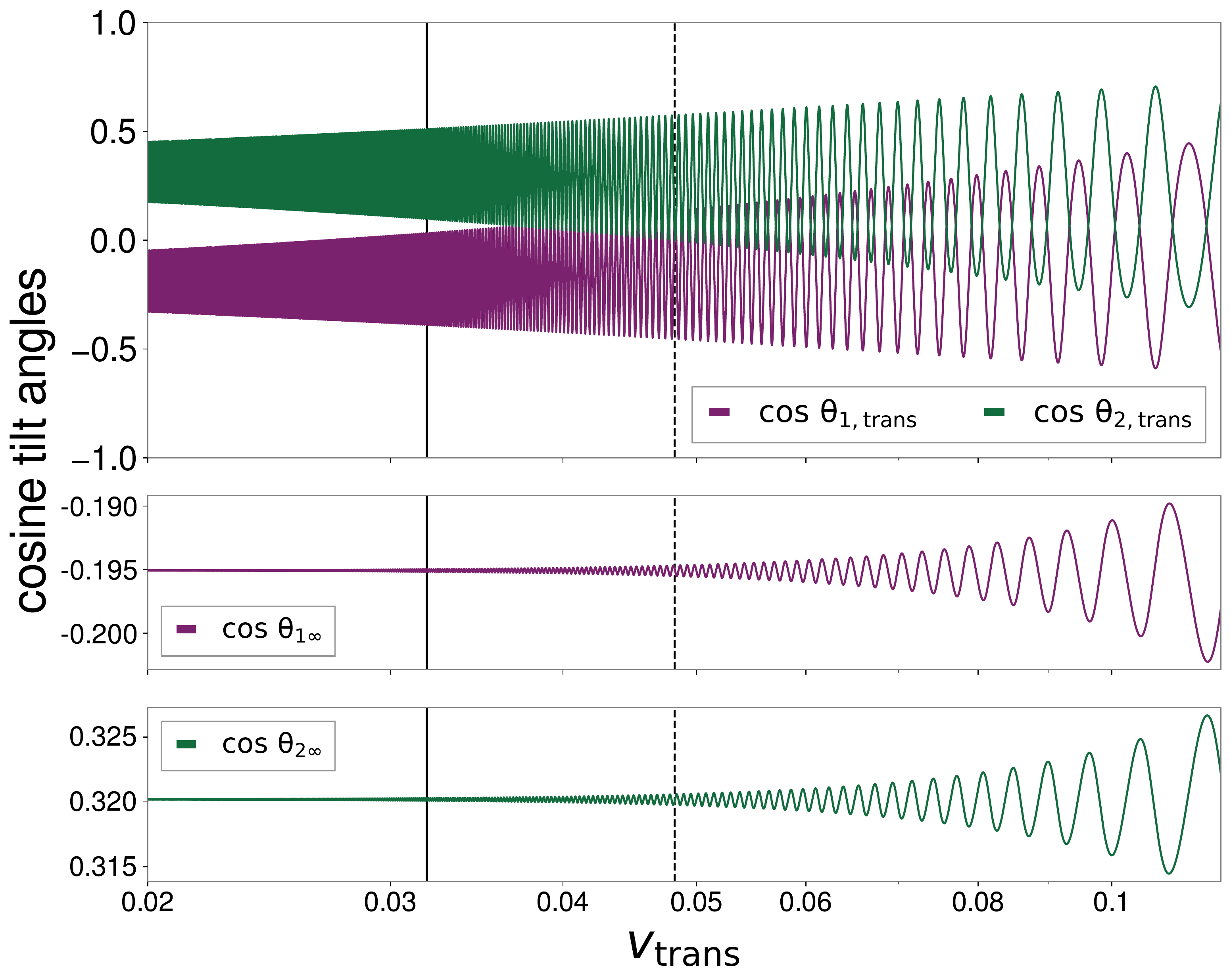}
}
\caption{\label{fig:tilts_vs_vtrans} The evolution of \edit{the cosines of the} tilt angles as a function of the orbital speed, in the regime where we select our transition orbital speed values. The upper panel shows the tilts at the given $v_\text{trans}$ value and the lower panel shows the tilts at infinity obtained using that $v_\text{trans}$ value in the hybrid evolution.
This is illustrated for the same binary parameters as in Fig.~\ref{fig:tilts_vs_L} with $q = 0.75$. \edit{The solid vertical lines represent $v_\text{trans}$ for this binary given by Eq.~\eqref{eq:v_trans}, while the dashed vertical lines display the value of $v_\text{trans}$ given by the purple curve fitting our data in Fig.~\ref{fig:v_trans_fit}.}
}
\end{figure}

Figure~\ref{fig:tilts_vs_vtrans} shows the \edit{cosines of the} tilt angles computed at infinity as well as 
intermediate transition orbital speeds ($v_{\rm trans}$), as a function of $v_{\rm trans}$, 
for the same binary parameters as in Fig.~\ref{fig:tilts_vs_L} with $q = 0.75$. The top panel shows the oscillation of the \edit{cosines of the} tilt angles with orbital 
speed over the course of the orbit-averaged evolution. These are the usual oscillations over a precessional cycle of the binary.
The bottom two panels show that \edit{$\cos \theta_{1 \infty}$} and \edit{$\cos \theta_{2 \infty}$} also oscillate with $v_{\rm trans}$ but they converge to the
$v_\text{trans} \searrow 0$ limit much faster than the tilts at the transition frequency do. 
The oscillations in \edit{$\cos \theta_{i \infty}$ presumably arise because the precession-averaged evolution does not track the evolution of the spins over
a precessional cycle. The tilts at infinity one obtains in the limit $v_\text{trans} \searrow 0$ are apparently the same as the ones one would obtain starting from
a finite $v_\text{trans}$ and using the average values of $S^2$ over a precessional cycle (when using the same PN order for the orbit-averaged and precession-averaged evolution).
Thus, the oscillations in $\cos \theta_{i \infty}$ are due to starting the precession-averaged evolution from $S^2$ values at different points in the precessional cycle.}
The $v_{\rm trans}$ for this binary is $0.03$ [from Eq.~\eqref{eq:v_trans}], and it is evident that the tilts at infinity
computed using the hybrid evolution with this value of $v_{\rm trans}$ are very close to those that would be computed in the limit $v_\text{trans} \searrow 0$.

\subsection{Validation of $v_{\rm trans}$}
\label{ssec:v_trans_validation}

\begin{figure}[h]
\centering
\subfloat{
\includegraphics[width=0.47\textwidth]{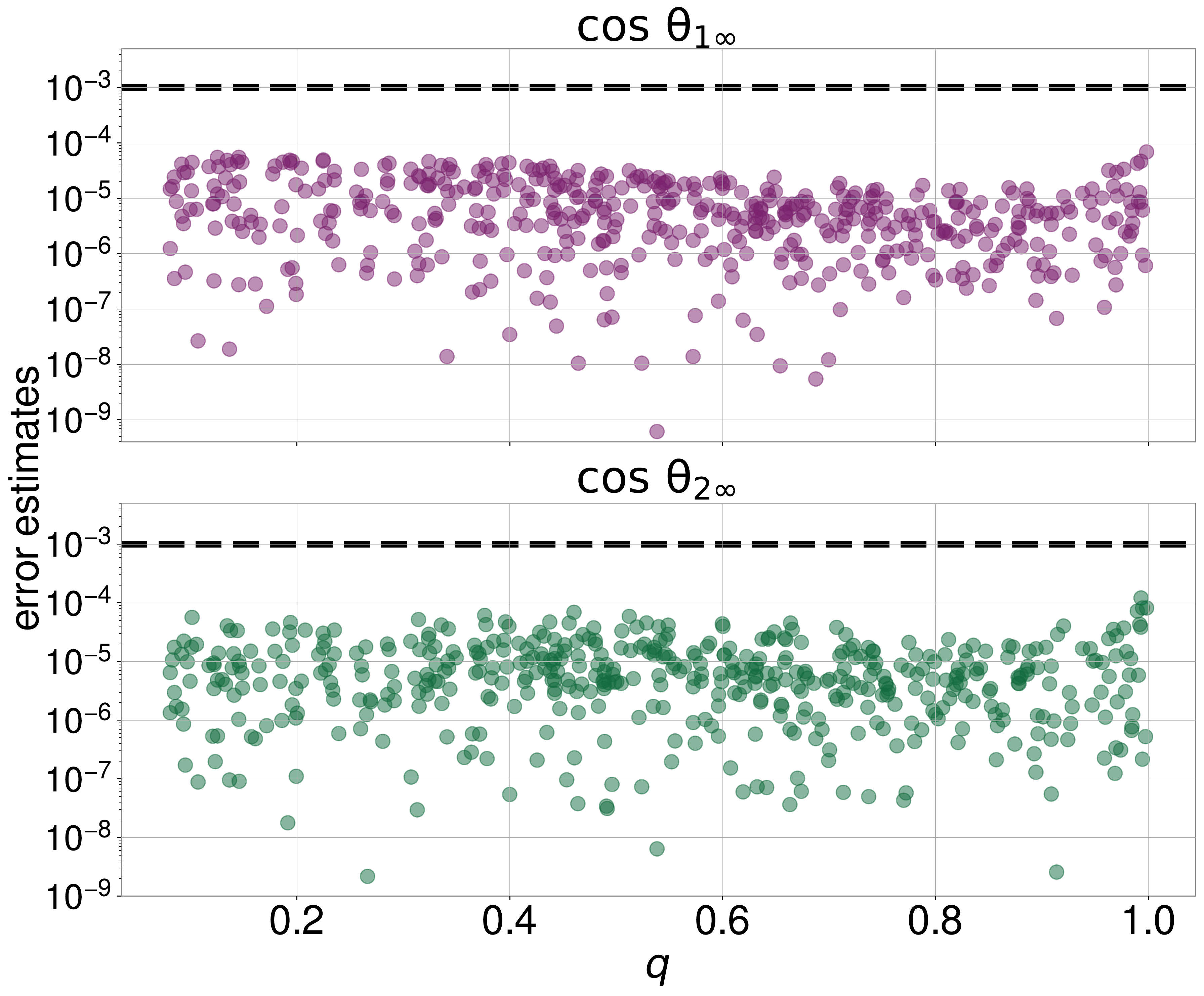}
}
\caption{\label{fig:randomset_validation} Plots showing the estimated error in \edit{$\cos\theta_{i\infty}$} (computed as discussed in the text)
for a random dataset of $500$ binaries using a transition orbital speed given by Eq.~\eqref{eq:v_trans} and a lower transition speed value of $0.6v_{\rm trans}$.
}
\end{figure}


\begin{figure*}
\centering
\subfloat{
\includegraphics[width=\textwidth]{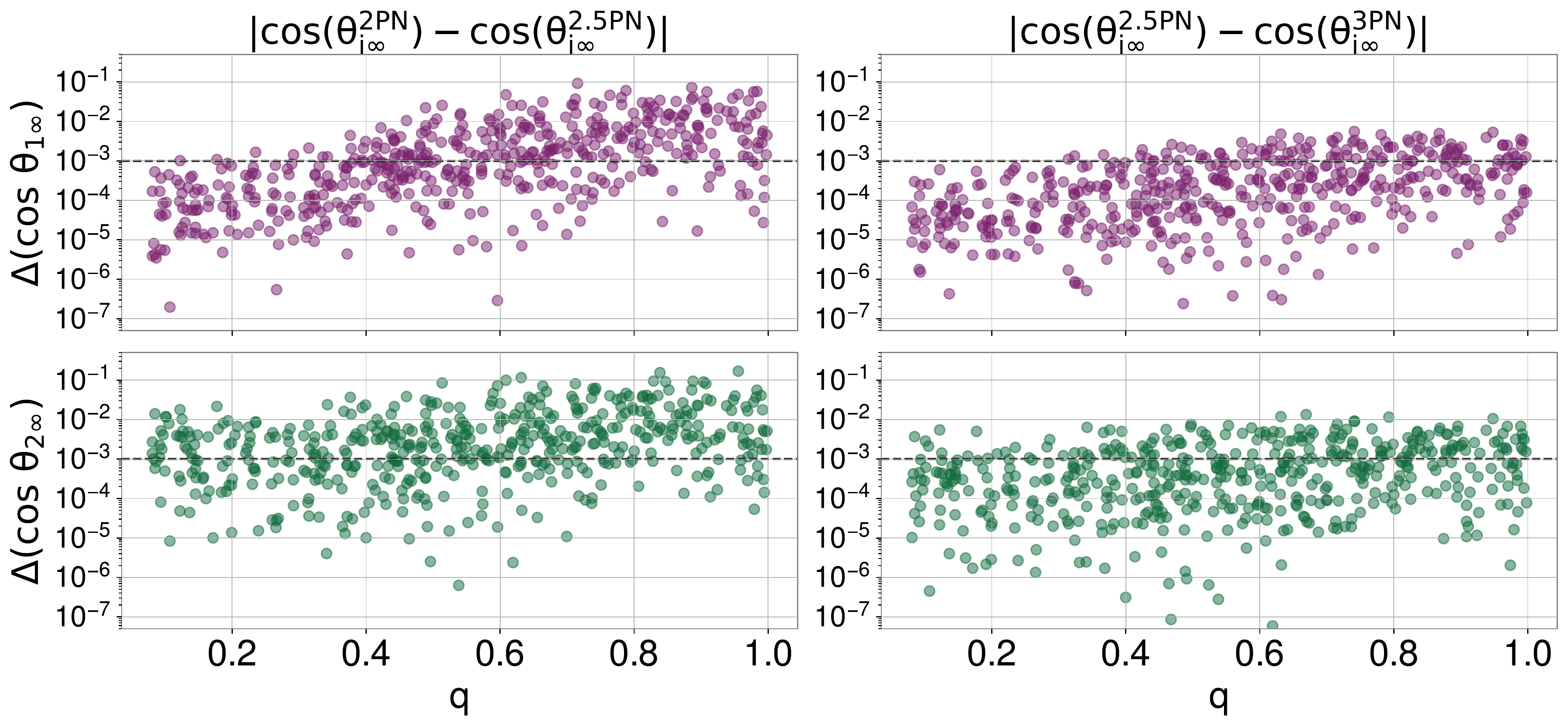}
}
\caption{\label{fig:spinO4_t1} Comparison of \edit{cosines of tilt angles at infinity} while using 2PN and 2.5PN accurate (left) and 2.5PN and 3PN accurate (right) orbit-averaged precession equations in the hybrid evolution code.
}
\end{figure*}

To check whether the quadratic expression for $v_{\rm trans}$ given in Eq.~\eqref{eq:v_trans} gives
satisfactory results for binaries outside the training dataset, we employ a population of 500 random binaries
with the following parameters:
$m_{1}, m_{2} \in [5,100] M_{\odot}$; $q \in [0,1)$;
$\chi _{1}, \chi_{2} \in [0,1]$;
$\theta_{1}, \theta_{2} \in [0,\pi]$~rad; $\phi_{12} \in [0, 2\pi]$~rad; $f_{\rm ref} = 20$~Hz. \edit{This gives the same range of orbital angular momentum magnitudes as for the training set.}
For each of these 500 binaries, we estimate the error in tilts at infinity by
computing the absolute difference between \edit{$\cos\theta_{i\infty}$} at two transition orbital speeds:
the $v_{\rm trans}$ given by Eq.~\eqref{eq:v_trans} and \edit{$0.6v_{\rm trans}$. We do not use the obvious choice of $0.5v_{\rm trans}$ since it is significantly more computationally expensive for the close-to-equal-mass binaries.} 

Figure~\ref{fig:randomset_validation} shows the estimated errors in \edit{$\cos\theta_{i\infty}$} for all $500$ binaries.
As expected, all errors are less than $10^{-3}$, which validates the expression for $v_{\rm trans}$ in Eq.~\eqref{eq:v_trans}.
The results shown in Fig.~\ref{fig:randomset_validation} were obtained with the $3$PN order SpinTaylorT5 orbit-averaged precession equations. The same upper bound on the
errors holds with SpinTaylorT1 and SpinTaylorT4. If one instead considers the \edit{$2$PN and $2.5$PN orbit-averaged precession equations
with SpinTaylorT5, the upper bound on the estimated errors is also less than $10^{-3}$.} This suggests that the $v_{\rm trans}$ given by Eq.~\eqref{eq:v_trans},
is valid for all PN orders of the orbit-averaged precession equations and SpinTaylor approximants.
Moreover, as expected, we found the estimated errors for $2$PN orbit-averaged precession equations to be in general smaller compared to
those for the $3$PN order equations, particularly for smaller mass ratios and $\theta_{1\infty}$. This is because the $2$PN order matches the PN order used in the
precession-averaged evolution.
\edit{We also compared $\cos \theta_{i\infty}$ for $500$ random binaries while using various
PN approximants: SpinTaylorT1, SpinTaylorT4, and SpinTaylorT5 and found that the tilts at infinity are
not significantly different for different approximants. The differences in the cosines of tilts at infinity are mostly below $10^{-3}$ with a maximum difference of $0.003$  
between SpinTaylorT1 and SpinTaylorT4.} 

\begin{figure}[h]
\centering
\subfloat{
\includegraphics[width=0.47\textwidth]{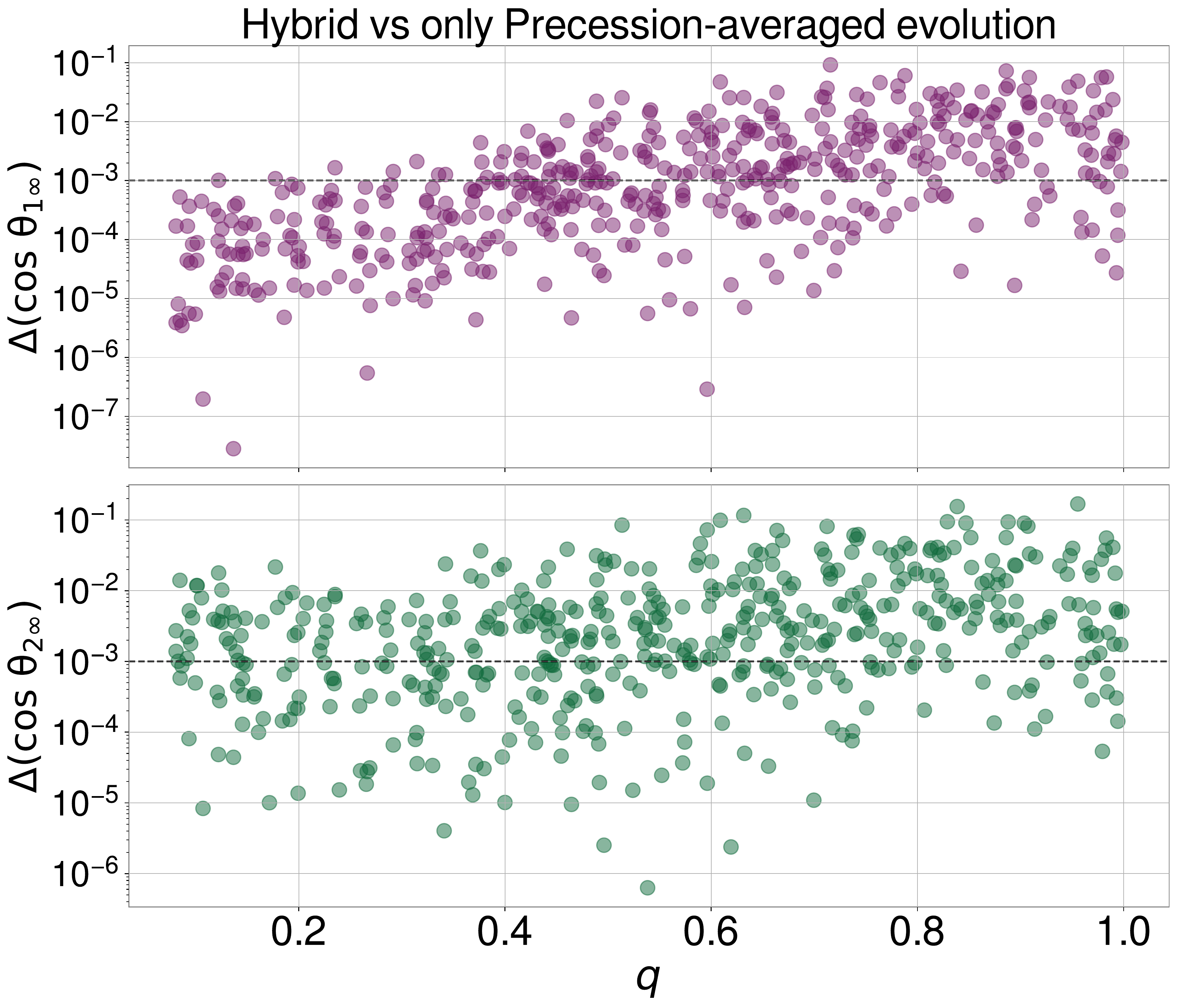}
}
\caption{\label{fig:hyb_vs_prec_randomset} Comparison of \edit{the cosines of the} tilts at infinity computed using the hybrid evolution 
and purely precession-averaged evolution for the random dataset of 500 binaries.
}
\end{figure}

We also compare the results using different PN orders directly. In Fig.~\ref{fig:spinO4_t1}, we compare \edit{$\cos \theta_{i\infty}$}, for the same $500$ binaries as in
Fig.~\ref{fig:randomset_validation}, 
using different PN orders for the orbit-averaged precession equations in the hybrid 
evolution code. Specifically, we compare the results between $2$PN, $2.5$PN, and $3$PN accurate orbit-averaged 
precession equations.  
We find that the absolute differences in $\cos \theta_{i\infty}$ between $2$PN and $2.5$PN are quite noticeable (most of them are above $10^{-3}$). 
However, the differences between the $2.5$PN and $3$PN results are smaller than those between the $2$PN and $2.5$PN results. This gives us
confidence that the $3$PN results we are using give good accuracy, particularly since comparisons of PN precessional
dynamics with numerical relativity in~\cite{Ossokine:2015vda} (particularly Sec.~III~C) find the $3$PN results to be more accurate than lower orders.

\begin{figure}[h]
\centering
\subfloat{
\includegraphics[width=0.5\textwidth]{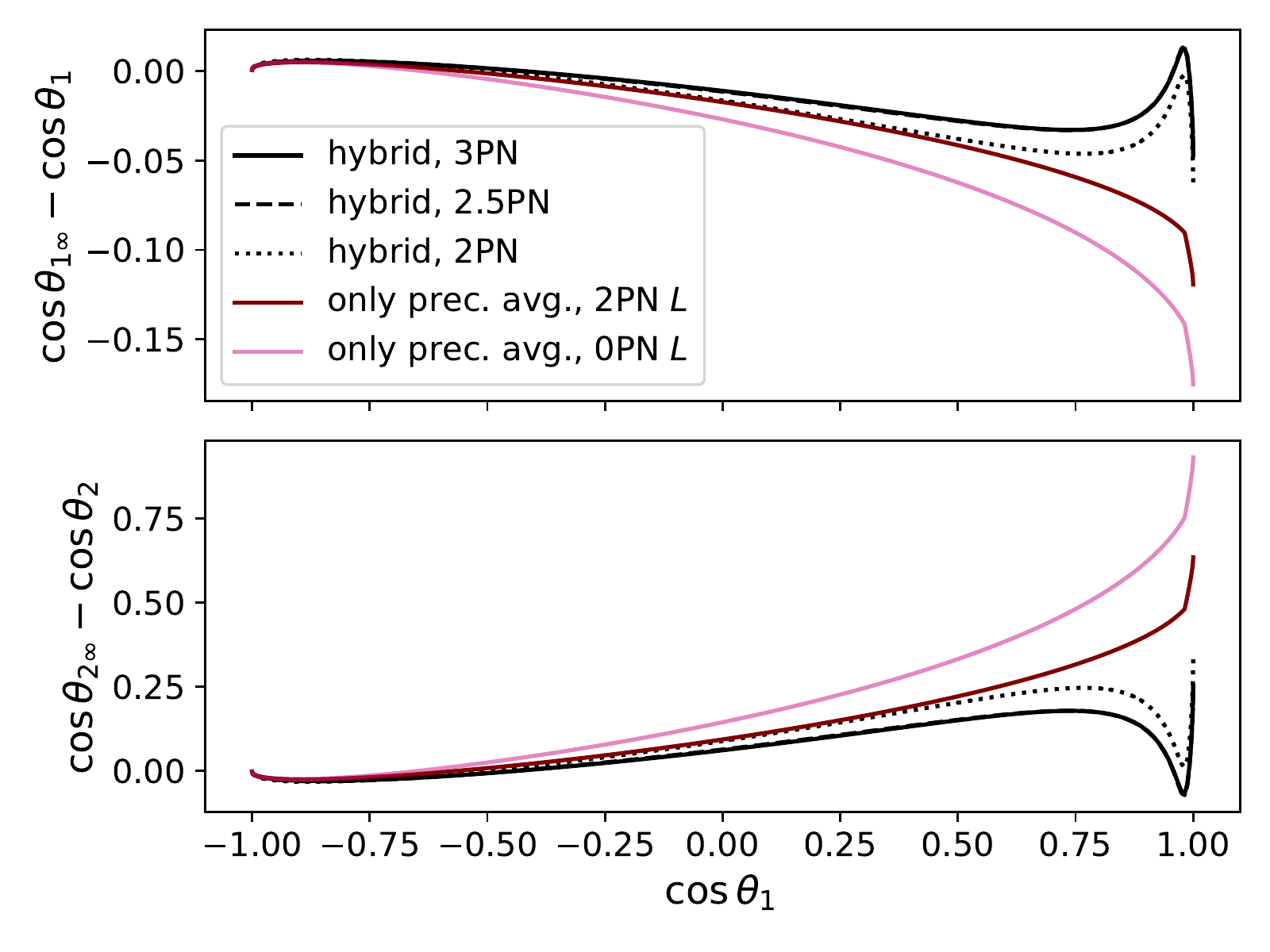}
}
\caption{\label{fig:hyb_and_prec_vs_tilt1} \edit{Comparison of the difference in the cosines of the tilts at infinity with those at the reference frequency as a function of tilt~$1$ at the reference frequency. This is for an example binary described in the text that shows a significant difference between the hybrid and purely precession-averaged evolution. The $2$PN purely precession-averaged evolution does not include the contributions of the spins to the orbital angular momentum and the $2.5$PN hybrid evolution curve is almost indistinguishable from the $3$PN curve. The kinks in the purely precession-averaged curves close to $\cos\theta_1 = 1$ are not numerical artifacts (e.g., they remain when plotting using a finer mesh and also when using different integrators and PRECESSION), but we are not sure why they occur. 
}
}
\end{figure}

In Fig.~\ref{fig:hyb_vs_prec_randomset} we compare the tilts at infinity computed with the hybrid evolution with those computed using only precession-averaged evolution. Here we consider the same $500$ binaries as before. In the only precession-averaged case, we initialize the precession-averaged evolution using the $2$PN expression for the orbital angular momentum without the spin contributions, so it is parallel to the Newtonian orbital angular momentum which is used to define the tilt angles in the LIGO-Virgo analysis~\cite{Farr:2014qka}. This is the recommended setup for obtaining quick results for the tilts at infinity. We find that there are relatively significant differences in the cosine tilts at infinity in many cases, though there are still a significant fraction of binaries for which the difference is less than $10^{-3}$.

\edit{We also find that there are certain parts of parameter space that lead to even larger differences between the two evolutions, in particular if we initialize the purely precession-averaged evolution with the Newtonian expression for the orbital angular momentum, as is done in PRECESSION. While we reserve a complete exploration of the portions of parameter space that lead to the largest to differences to future work, we found that certain parameters lead to much larger differences in the tilts at infinity than found in our set of $500$ random binaries. We obtained these parameters by considering the sample from the analysis of GW190521 with the SEOBNRv4PHM waveform model (as discussed in Sec.~\ref{ssec:GW190521}) that gives the largest difference in $\cos\theta_{2\infty}$ when using the Newtonian orbital angular momentum in this comparison and varying $\theta_1$. We also scaled the masses to a reference frequency of $20$~Hz, to illustrate that this case is included in the range of parameters used to construct the $500$ random binaries.}

\edit{Specifically, we found that for $m_1 = 88.4M_\odot$, $m_2 = 66.6M_\odot$, $\chi_1 = 0.721$, $\chi_2 = 0.180$, $\cos\theta_1 \in [0.9,1]$, $\cos\theta_2 = -0.90$, $\phi_{12} = 5.22$~rad, at $f_\text{ref} = 20$~Hz (so the magnitude of the initial Newtonian orbital angular momentum is $0.67M^2$), there are differences in magnitude of $>0.50$ ($> 0.28$) and a maximum of $0.83$ ($0.56$) in $\cos\theta_{2\infty}$ when comparing the hybrid evolution with the purely precession-averaged evolution initialized with the Newtonian ($2$PN) orbital angular momentum. The magnitudes of the differences for $\cos\theta_{1\infty}$ are $\sim 0.15$ ($\sim 0.10$) for the two comparisons. We illustrate the difference in the tilts at infinity calculated with the hybrid evolution and purely precession-averaged evolution for this binary in Fig.~\ref{fig:hyb_and_prec_vs_tilt1}, showing qualitatively different behavior in how the results depend on tilt~1 at the reference frequency.}

\subsection{Code runtimes}
\label{ssec:hybrid_evol_code}

\begin{figure}[h]
\centering
\subfloat{
\includegraphics[width=0.47\textwidth]{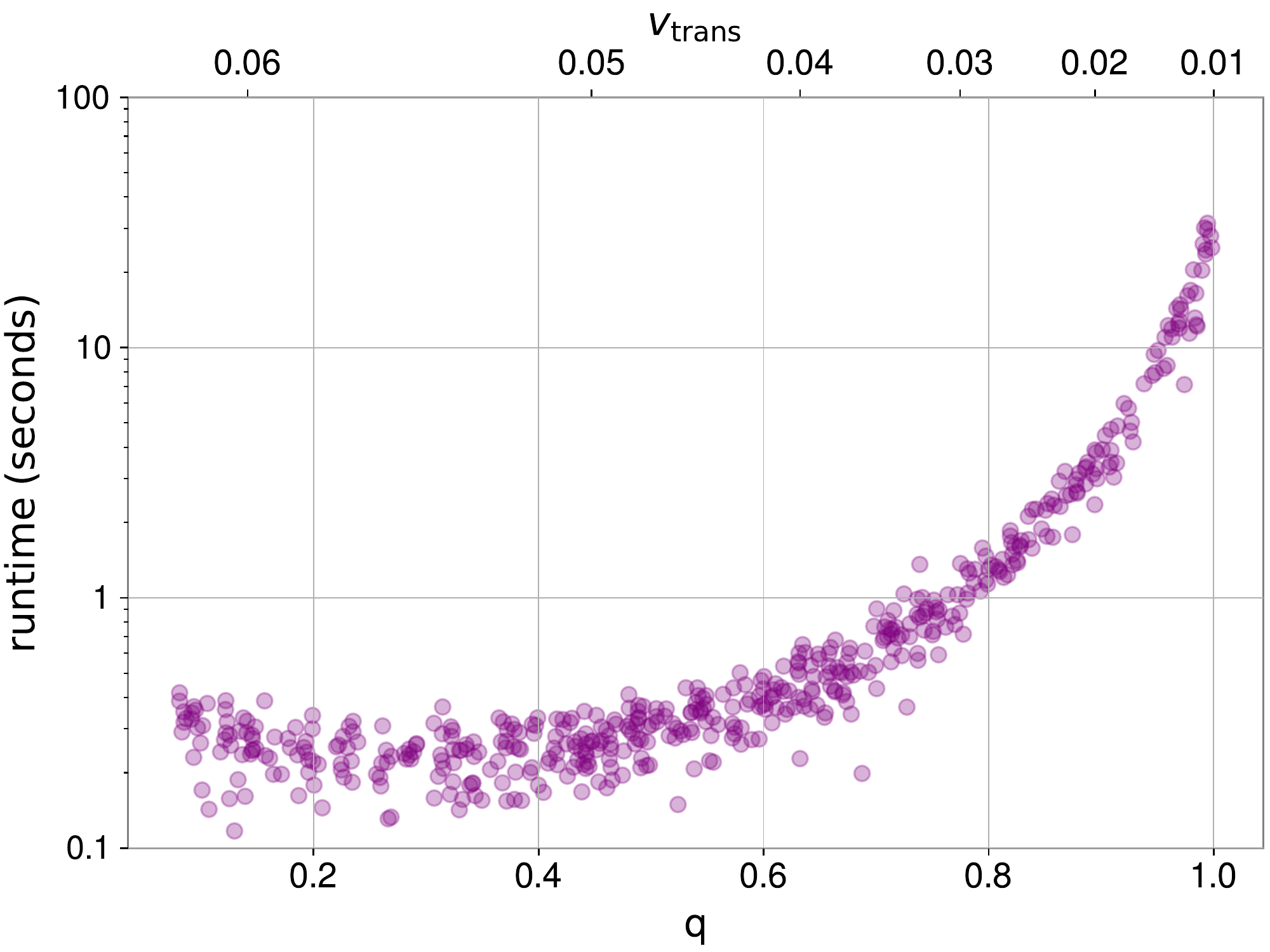}
}
\caption{\label{fig:hybrid_evol_runtimes}
Run time of hybrid evolution for binaries having different mass ratios. These are the same $500$ binaries as in
Figs.~\ref{fig:randomset_validation} and \ref{fig:spinO4_t1}.
}
\end{figure}

Figure~\ref{fig:hybrid_evol_runtimes} shows the run times of the hybrid evolution code for the same
dataset of $500$ random binaries. The run times are higher for close-to-equal-mass binaries, taking about $40$ seconds to complete. 
These timing results were obtained using a  $2.6$~GHz Intel Xeon E5-2680-v3 (12C) with 256~GB RAM. 

For comparison, with just the precession-averaged evolution, the average speed for these $500$ random binaries from the transition frequencies is $\sim 7 \text{ ms}/\text{binary}$ for the development version of PRECESSION either with its default settings or with the same integrator tolerance settings as the default for the regularized evolution, and $\sim 15 \text{ ms}/\text{binary}$ for the default regularized evolution, but only $\sim 7 \text{ ms}/\text{binary}$ when using solve\_ivp. With the fallback tolerance settings, we have speeds of $\sim 0.014 \text{ s}/\text{binary}$ for PRECESSION and $\sim 0.16$, $\sim 0.014$, and $\sim 2 \text{ s}/\text{binary}$ for the regularized evolution with the scipy ode, scipy solve\_ivp, and mpmath integrators, respectively. From $20 \text{ Hz}$, the default regularized evolution has an average speed of $\sim 30 \text{ ms}/\text{binary}$. These timing results were obtained using a $2.2$~GHz Intel Core i7.
\section{Uncertainties in approximating the tilts at formation by those at infinity}
\label{sec:tilts_formation}

\begin{figure}
\includegraphics[width=0.5\textwidth]{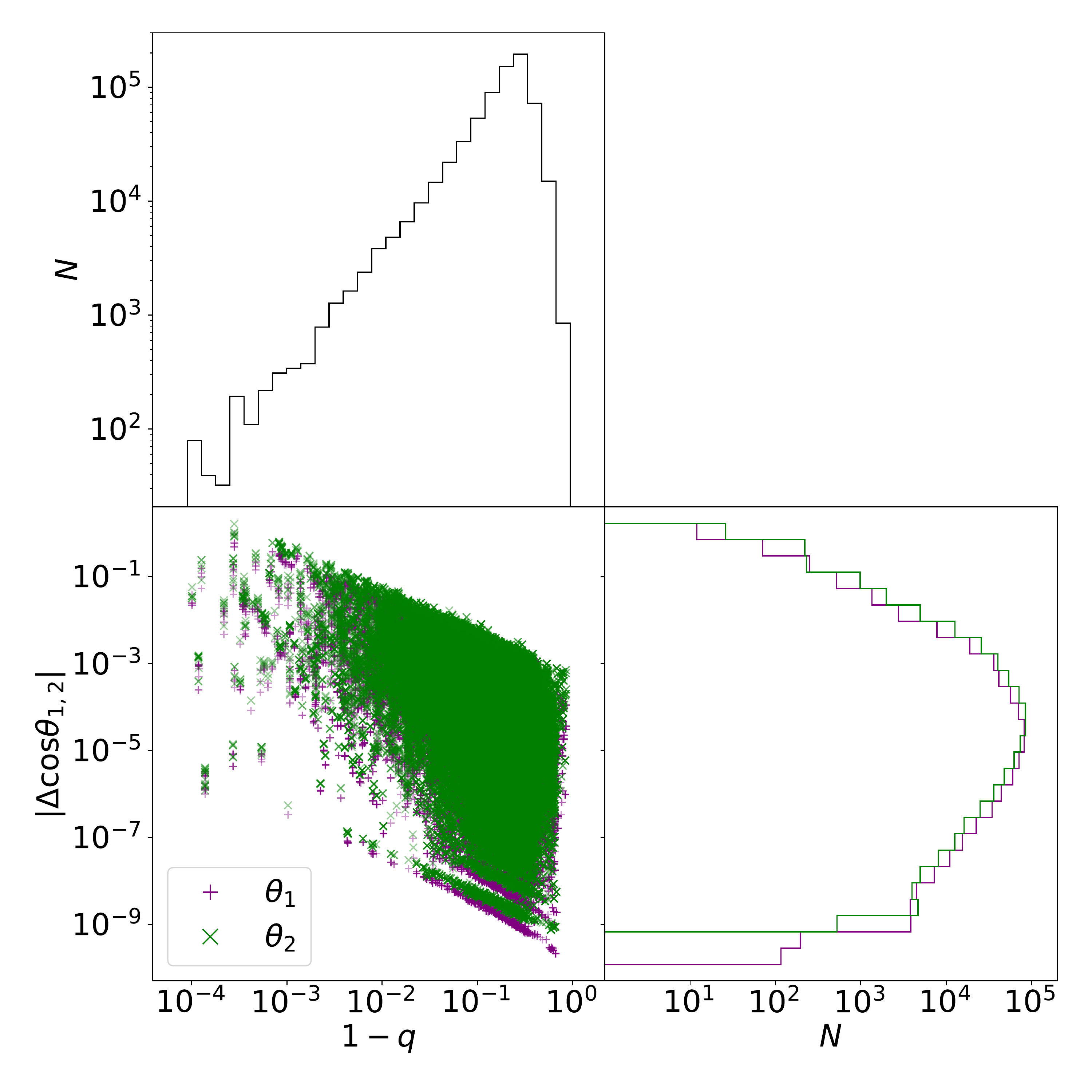}
\caption{\label{fig:pop_syn_tilt_uncertainties} The uncertainties in inferring the \edit{cosines of the} tilts at formation due to the amplitude of precession at that point in the binary's evolution. This plots the difference between the \edit{cosines of the} maximum and minimum tilts given in Eqs.~\eqref{eq:tilts_at_L} for a selection of the binary black hole mergers from the M30 population synthesis model from~\cite{Belczynski:2017gds} that are detectable by LIGO and Virgo by that paper's criterion, neglecting aligned-spin cases or binaries that are equal mass to the accuracy of the data provided. The value of $\phi_{12}$ is chosen randomly for each binary. The histograms show the results for the first $2 \times 10^6$ binaries in the data set (giving $\sim 6.8 \times 10^5$ unequal mass precessing binary black holes)---the computation is expensive enough that we do not consider more binaries---while the scatter plot shows a random selection of $2 \times 10^5$ binaries (from the full set we consider), for clarity. We also do not show \edit{five} outliers \edit{(for each tilt)} with uncertainties in \edit{both $\cos\theta_1$ and $\cos\theta_2$} that are smaller than \edit{$5 \times 10^{-11}$} in the histograms.}
\end{figure}

Here we assess the uncertainties in approximating the tilts at formation by those at infinity. We first note that the orbital angular momentum at formation is relatively large in most cases.
For instance, in the isolated binary and dynamical formation scenarios for GW150914 given in Fig.~1 of both~\cite{Belczynski:2016obo} and~\cite{Rodriguez:2016avt}, respectively, and the isolated binary formation scenario for GW170104 given in Fig.~8 of~\cite{Belczynski:2017gds}, the orbital angular momentum at the time of the formation of a black hole binary is $\sim 130M^2$, so one expects that the \edit{cosines of the} tilts at infinity approximate those at formation with an accuracy of better than a percent. However, the angular momentum at binary black hole formation can be a factor of a few less in some cases, e.g., it is $\sim 50M^2$ in the isolated binary formation scenarios for GW170729, GW190412, and GW190521 given in Fig.~9 of~\cite{Belczynski:2017gds}, Fig.~1 of~\cite{Olejak:2020oel}, and Fig.~2 of~\cite{Belczynski:2020bca}, respectively. Additionally, it is $36 M^2$ for the dynamical formation scenario for GW190814 shown in Fig.~2 of~\cite{ArcaSedda:2021zmm}, due to its high eccentricity ($0.97$) and unequal mass ratio. Nevertheless, even in those cases the accuracy of the approximation is still expected to be a few percent. In fact, for \edit{the analyses of GW190521 considered in Sec.~\ref{sec:appl}}, the uncertainties in the distributions of the \edit{cosines of the} tilts at an orbital angular momentum of $50M^2$ are \edit{only $\pm 0.06$}.

In fact, $35M^2$ seems to be a reasonable lower bound on the expected orbital angular momentum at formation for isolated binary formation: Considering the precessing, unequal-mass binary black holes in the detectable systems of the synthesized population with 2019 standard input physics (M30) from~\cite{Belczynski:2017gds} (also considered in the mass ratio discussion in Sec.~\ref{sec:intro}), only $9$ systems out of $\sim 6 \times 10^6$ have an orbital angular momentum of $35M^2$ or less, though the smallest value is $8M^2$.

The orbital angular momentum at formation is relatively large for other formation scenarios, e.g., the smallest orbital angular momentum for the hierarchical second-generation merger scenario for GW190814 with parameters given in Fig.~3 of~\cite{Lu:2020gfh} is $\sim 90M^2$. The situation for formation in the disks of active galactic nuclei (e.g.,~\cite{Bartos:2016dgn,Stone:2016wzz,McKernan:2017umu}) is more complicated, since the binary interacts with the disk during much of its evolution (see, e.g.,~\cite{Grobner:2020drr}). Thus, our results are not applicable to this case.

The only exception to the accuracy estimates given above is for very close to equal mass cases, with $1 - q \lesssim 10^{-3}$, since the intrinsic uncertainties (due to the amplitude of precession) at a fixed $L$ increase roughly as a positive power of $1/(1 - q)$ until they reach \edit{$2$}---see the illustration in Fig.~\ref{fig:pop_syn_tilt_uncertainties}. However, while such close-to-equal-mass systems are predicted by population synthesis calculations, they are relatively uncommon. For instance, considering the same selection from the synthesized population from~\cite{Belczynski:2017gds} plotted in the histograms in Fig.~\ref{fig:pop_syn_tilt_uncertainties}, the uncertainty in either of the \edit{cosines of the} tilts at formation is $> 10^{-2}$ ($10^{-1}$) for only \edit{$\sim 1\%$} (\edit{$\sim 0.1\%$}) of these. (This comparison excludes the binaries that are exactly equal mass to the three decimal places used for the data, since the tilts at infinity are not well defined in exactly equal-mass cases, but these are only $0.1\%$ of the total.)

This estimate only accounts for the intrinsic uncertainty in the tilts due to the amplitude of spin precession at a finite separation, not the added uncertainty in approximating the tilts at finite separation by those at infinity if the tilts at infinity are not contained between the bounds on the tilts at finite separation. Such a scenario can occur, particularly near the up-down instability parameters (see, e.g.,~\cite{Mould:2020cgc}), though it is not generic---see Fig.~11 in~\cite{Gerosa:2015tea}---and just occurs for much smaller separations than one expects for most formation scenarios, except for quite close to equal mass binaries, $1 - q \lesssim 10^{-2}$. In particular, of the selection of the population of binary black holes from~\cite{Belczynski:2017gds} plotted in the histograms in Fig.~\ref{fig:pop_syn_tilt_uncertainties}, only $\sim 0.1\%$ of them have either of the tilts at infinity outside the bounds on the tilts at formation, though in about \edit{$25\%$} (\edit{$5\%$}) of these cases, at least one of the \edit{cosines of the} tilts at infinity is $> 10^{-2}$ ($10^{-1}$) different from the range \edit{for the} tilts at formation, at most \edit{$\sim 0.24$}; all of those binaries have $1 - q < \edit{3 \times 10^{-4}}$. As one would expect, the cases with a larger difference between the range of tilts at formation and those at infinity have \edit{more misaligned} tilts at formation, \edit{with cosines of the tilts as small as $\sim -0.97$} (the \edit{smallest cosines of} tilts in the selection of binary black holes plotted in the histograms in Fig.~\ref{fig:pop_syn_tilt_uncertainties} are \edit{$\sim -0.9999$}).

\section{Application to GW events}
\label{sec:appl}

We applied the hybrid evolution code to compute the tilts at infinity for selected binary black 
hole detections from the LIGO-Virgo catalog \edit{GWTC-3} \edit{\cite{GWTC-3_paper}} using the publicly available posterior samples~\cite{GWTC-2_PE, PE_GWTC-3}. 
\edit{Specifically, we considered the events that had evidence for misaligned spins and found the three events that show the largest differences between the tilts at the reference frequency and at infinity: GW190521, GW191109\_010717, and GW200129\_065458.}

\begin{figure*}[h]
\centering
\subfloat{
\includegraphics[width=\linewidth]{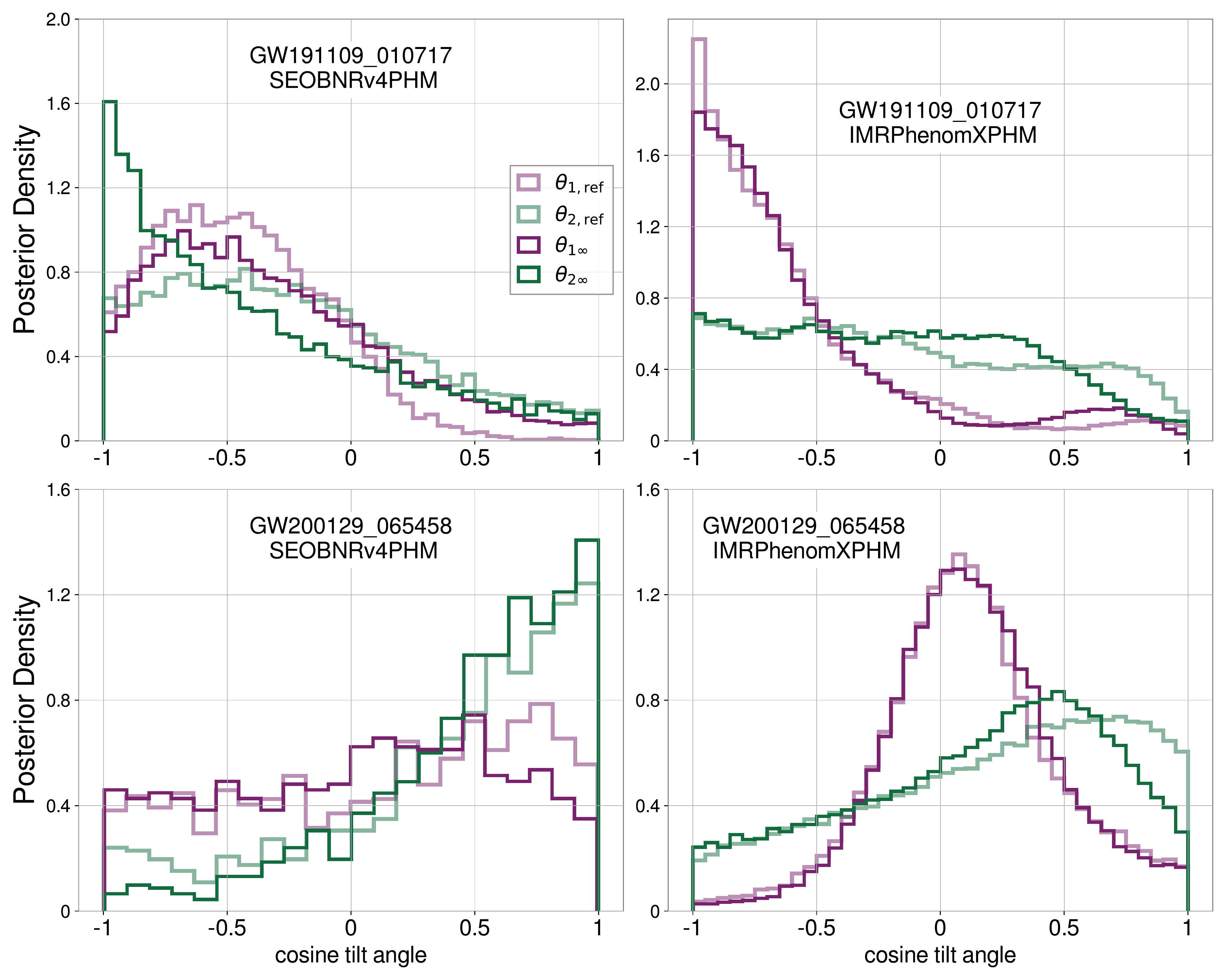}
}
\caption{\label{fig:tilts_dist_PEsamples_condensed} \edit{Posterior distributions for the cosines of spin tilt angles at
$f_{\rm ref}$ and at infinity computed using the hybrid evolution code for the two O3b events with the largest differences between spin tilts at $f_{\rm ref}$ and at infinity, with two different waveform models.
The light colored curves give the distributions at $f_\text{ref}$ and the dark ones the distributions at infinity.}
}
\end{figure*}

\edit{Figure~\ref{fig:tilts_dist_PEsamples_condensed} shows the posterior distributions of the cosines of the tilts
at $f_{\rm ref} = 20$~Hz and at infinity for GW191109\_010717 and GW200129\_065458. We consider the results with both the \textsc{SEOBNRv4PHM}~\cite{Ossokine:2020kjp,Babak:2016tgq,Pan:2013rra} and \textsc{IMRPhenomXPHM}~\cite{Pratten:2020ceb} waveform models (both of which include higher modes in addition to precession), to illustrate the qualitatively different behavior.\footnote{\editII{For GW200129\_065458, Ref.~\cite{Hannam:2021pit} points out that neither of these waveform models is sufficiently accurate to analyze the signal. They instead carry out an analysis with the numerical relativity surrogate model NRSur7dq4~\cite{Varma:2019csw}. Unfortunately neither they nor Ref.~\cite{Varma:2022pld}, which also carries out an analysis of this signal with NRSur7dq4, have released their samples. However, Ref.~\cite{Hannam:2021pit} finds that IMRPhenomXPHM agrees better with NRSur7dq4 for signals like GW200129\_065458 than SEOBNRv4PHM does.}} In order to compare the 1d marginal distributions of the tilts, we use the quantity $\max \Delta Q$, which represents the maximum of the absolute value of the difference between the $5\%$, $50\%$, and $95\%$ quantiles of the two distributions (i.e., the differences in the median and the $90\%$ credible interval around it).} 

\edit{The most interesting case is likely GW191109\_010717 in the \textsc{SEOBNRv4PHM} analysis, where there is significantly more support for tilt~$2$ being close to antialigned at infinity, with a $15\%$ probability of $\cos\theta_2 < -0.9$, compared to only $6\%$ at the reference frequency. However, in the \textsc{SEOBNRv4PHM} analysis of GW191109\_010717, it is tilt~$1$ that is close to being antialigned and there is slightly less support for this at infinity. In the GW200129\_065458 \textsc{IMRPhenomXPHM} analysis, tilt~$2$ has more support for being misaligned at infinity. We can also consider the $\max \Delta Q$ values of the distributions, the largest two values of which are $0.39$ ($0.38$) for $\cos\theta_{1}$ ($\cos\theta_{2}$) in the GW191109\_010717 (GW200129\_065458) \textsc{SEOBNRv4PHM} analysis.}

\begingroup
\setlength{\tabcolsep}{11pt}
\renewcommand{\arraystretch}{1.25}
\begin{table}[]
\caption{\label{tab:Median-GW190521}
$\max \Delta Q$ values for GW190521 between the distributions of cosine tilt angles at $f_{\rm ref}$ and infinity computed using the hybrid evolution code, for different waveform models.
}
\begin{tabular}{ccc}
\hline\hline
Approximant            & $\cos\theta_{1}$          & $\cos\theta_{2}$          \\
\hline
SEOBNRv4PHM  & $0.12$ & $0.16$ \\
IMRPhenomTPHM & $0.14$ & $0.12$ \\
IMRPhenomPv3HM  & $0.06$ &  $0.05$ \\
NRSur7dq4  & $0.05$ & $0.04$ \\
\hline\hline
\end{tabular}
\end{table}
\endgroup


\begin{figure*}[h]
\centering
\subfloat{
\includegraphics[width=\linewidth]{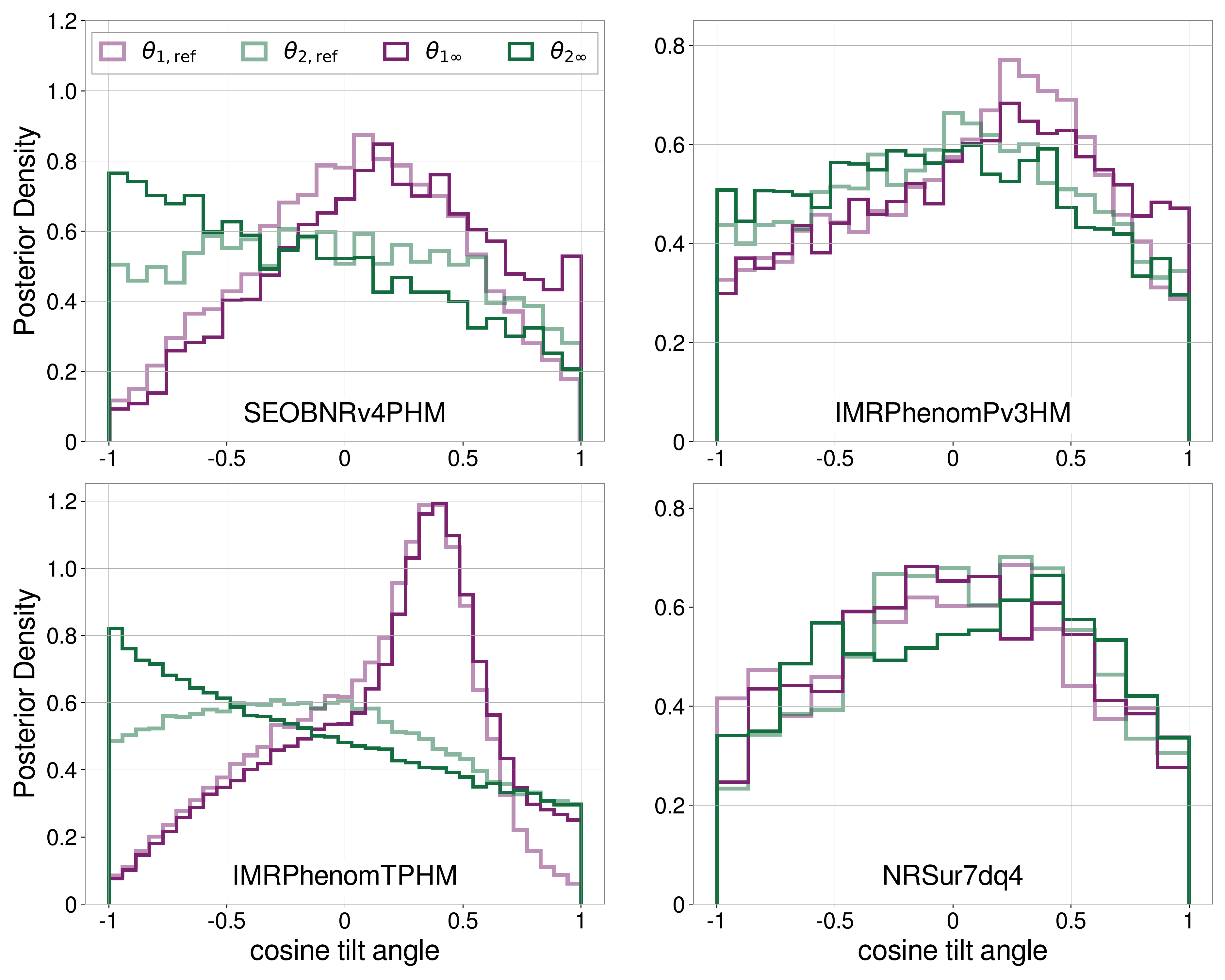}
}
\caption{\label{fig:tilts_dist_GW190521} The \edit{distributions of the cosines of the spin tilts} at $f_{\rm ref}$ ($4.4$~Hz for \textsc{IMRPhenomTPHM} and $11$~Hz for other cases) and at infinite
separation for GW190521 computed using the hybrid evolution code starting from the results obtained using different waveform models.
}
\end{figure*}

\subsection{GW190521}
\label{ssec:GW190521}

\edit{We now focus on GW190521, since we have results with a variety of waveform models and it is an intrinsically interesting event, due to its very high mass. Figure~\ref{fig:tilts_dist_GW190521}} shows the posterior \edit{distributions} for tilts at $f_{\rm ref}$
and at infinity for different waveform models \edit{that incorporate precession and higher harmonics and are used in~\cite{GWTC-2_paper}: \textsc{SEOBNRv4PHM}~\cite{Ossokine:2020kjp,Babak:2016tgq,Pan:2013rra}, \textsc{IMRPhenomPv3HM} \cite{Khan:2019kot}, 
and \textsc{NRSur7dq4}~\cite{Varma:2019csw} (see Sec.~V~A of~\cite{GWTC-2_paper} 
for details about these waveform models).} We also consider the results obtained with \textsc{IMRPhenomTPHM} 
in~\cite{Estelles:2021jnz} (we use the \textsc{LALInference} samples).
\textsc{IMRPhenomTPHM} is a new phenomenological time-domain waveform model that also includes a description of precession
and higher harmonics \cite{Estelles:2020osj,Estelles:2020twz,Estelles:2021gvs}.
The analysis with \textsc{IMRPhenomTPHM} in~\cite{Estelles:2021jnz} obtains multimodal mass posteriors, which had also been found using an older
version of the phenomenological  frequency-domain \textsc{IMRPhenomXPHM} waveform model~\cite{Pratten:2020ceb} in~\cite{Nitz:2020mga}.
However, the posteriors differ, particularly in the weights of the modes. While~\cite{Estelles:2021jnz} also presents results using the updated version of
\textsc{IMRPhenomXPHM}, we only give results for \textsc{IMRPhenomTPHM}, due to its more accurate treatment of precession.
\textsc{SEOBNRv4PHM} gives the largest difference 
between the tilts at $f_{\rm ref}$ and infinity, followed by \textsc{IMRPhenomTPHM}. \textsc{IMRPhenomPv3HM} and \textrm{NRSur7dq4} have the least 
difference in tilts at $f_{\rm ref}$ and at infinity. Here $f_\text{ref} = 4.4$~Hz for \textsc{IMRPhenomTPHM} and $11$~Hz for the other models.
Table~\ref{tab:Median-GW190521} \edit{gives the $\max \Delta Q$ values comparing the distributions at infinity and the reference frequency} for these four models.


\edit{In all cases except for NRSur7dq4, the tilts at infinity give a primary spin that it closer to being aligned and a secondary spin that is closer to being antialigned. This reduces the support for an isolated formation channel, since it is more difficult to obtain the significant spin misalignments in such cases.}


One important check of the accuracy of our results is that they are not very sensitive to the 
reference point employed in the analysis.\footnote{Here we refer to a more general reference point instead of a
reference frequency, since we consider setting a reference time in this comparison, which
corresponds to a different reference frequency for each sample.} It is simple to check this for the
\textrm{NRSur7dq4} waveform model, which allows one to evolve its spins to an arbitrary 
point within the length of time covered by the model. For this check, 
we evolved the spins backwards to a time of $-4200M$ before the peak of the waveform (close to the start of the surrogate model~\cite{Varma:2019csw}) using
the \textrm{NRSur7dq4} surrogate's spin evolution before applying the hybrid evolution. 
Here we compute the reference frequency for each sample at that time using the \textrm{NRSur7dq4} orbital frequency. 
\edit{We found that the distributions of tilts at infinity we obtain starting from the reference 
frequencies obtained this way and from a reference frequency of $11$~Hz are quite 
similar, with $\max \Delta Q \simeq 0.02$, while the distributions of 
tilts at the two reference points differ more substantially, with $\max \Delta Q \simeq 0.06$. 
However, we found that the individual tilts at infinity samples can be quite different when evolved from these different reference points.
For instance, the maximum absolute values of the differences for $\cos \theta_{1\infty}$ and $\cos \theta_{2\infty}$ when evolved from the reference frequency of $11$~Hz and a
time of $-4200M$ are $0.64$ and $1.04$, respectively.
Similarly, these maximum differences are larger at the two reference points, with values of $1.40$ and $1.33$ for  
$\cos \theta_1$ and $\cos \theta_2$, respectively.} 

\begin{figure}[h]
\centering
\subfloat{
\includegraphics[width=0.47\textwidth]{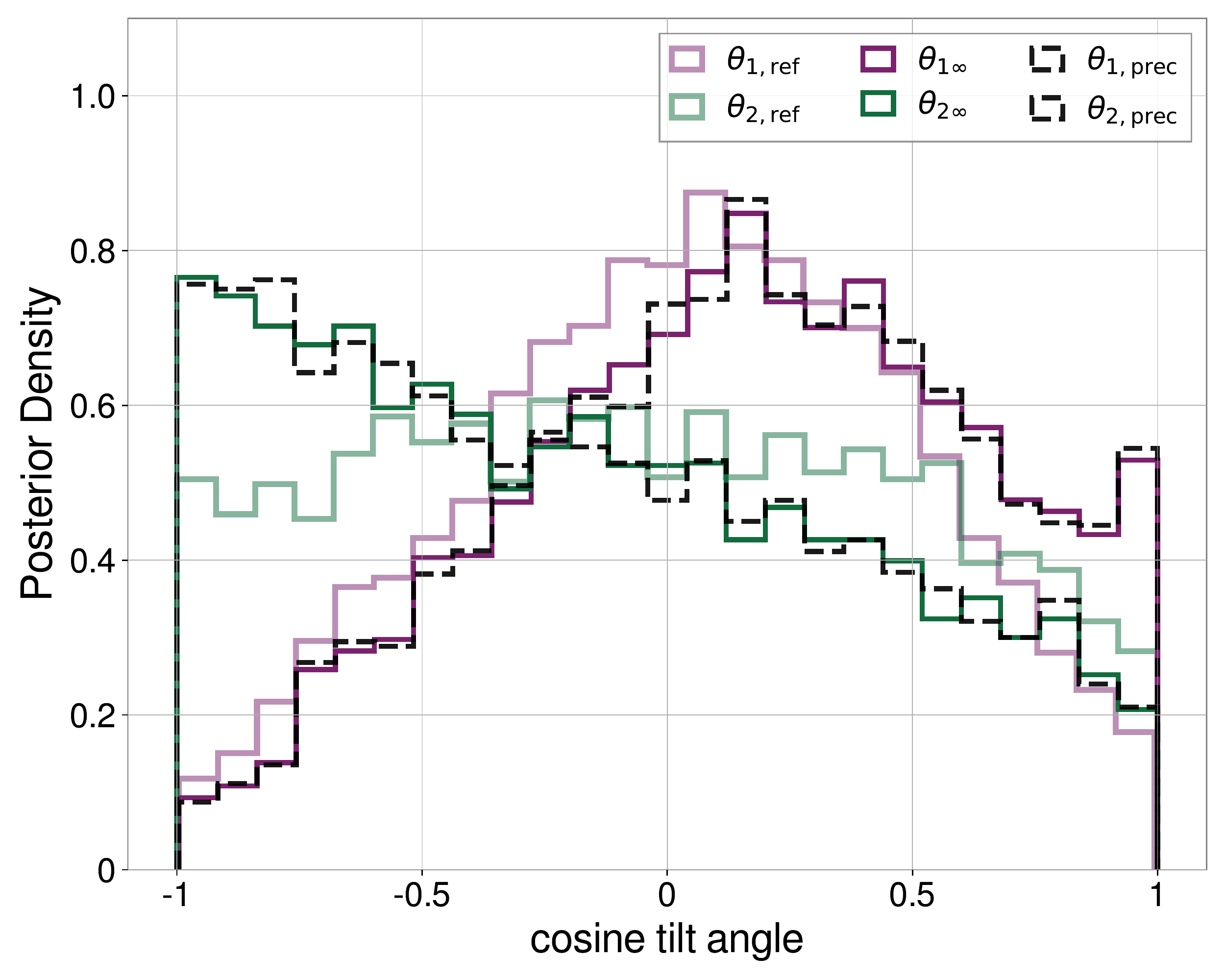}
}
\caption{\label{fig:GW190521_prec_only_compare} Comparison of tilts at infinity for GW190521 computed using the hybrid evolution 
and purely precession-averaged evolution using \textsc{SEOBNRv4PHM} samples. 
The tilts at $f_{\rm ref}$ are also shown for comparison.
}
\end{figure}

Additionally, it is useful to see how much difference the hybrid evolution makes in computing these distributions. Figure~\ref{fig:GW190521_prec_only_compare} shows a
comparison of tilts at $f_{\rm ref}$ and infinity computed using only precession-averaged evolution and the hybrid orbit-averaged and precession-averaged evolution,
for GW190521 using \textsc{SEOBNRv4PHM} samples. We initialize the precession-averaged evolution the same way as in Fig.~\ref{fig:hyb_vs_prec_randomset}.
\edit{Comparing the tilts at infinity computed using the hybrid evolution and purely precession-averaged evolution, we see no significant differences, 
with a $\max \Delta Q$ of $2 \times 10^{-3}$ for the $\cos \theta_{2\infty}$ distributions. Thus, for this event the precession-averaged evolution alone would be sufficient to compute the
distribution of tilts at infinity with good accuracy. The maximum difference between individual samples is $0.37$ for $\cos \theta_{1\infty}$, though the $90\%$ upper bound on this difference is only $0.03$.}

Finally, we consider the effects of the evolution to infinity on the effective spin $\xi$ [defined in Eq.~\eqref{eq:xi_def} and often denoted $\chi_\text{eff}$
in gravitational wave astronomy] and the
effective precession spin parameter $\chi_{\rm p}$ \cite{Hannam:2013oca,Schmidt:2014iyl}. In particular, we verify that the effective spin is
approximately conserved by our spin evolution to infinity. It is exactly conserved for the precession-averaged evolution and is conserved by
the orbit-averaged evolution through $2$PN order. For GW190521, we find that the
posterior distributions for $\xi$ at $f_{\rm ref}$ and at infinity are nearly
identical\edit{: The largest $\max \Delta Q$ between the posterior distributions at $f_{\rm ref}$ and at infinity  
among the four approximants is $8 \times 10^{-4}$ for the \textsc{NRSur7dq4} samples.}
Moreover, the largest \edit{differences} between individual $\xi$ 
samples at $f_{\rm ref}$ and infinity are all $< 10^{-2}$. 

\begin{figure*}[h]
\centering
\subfloat{
\includegraphics[width=\linewidth]{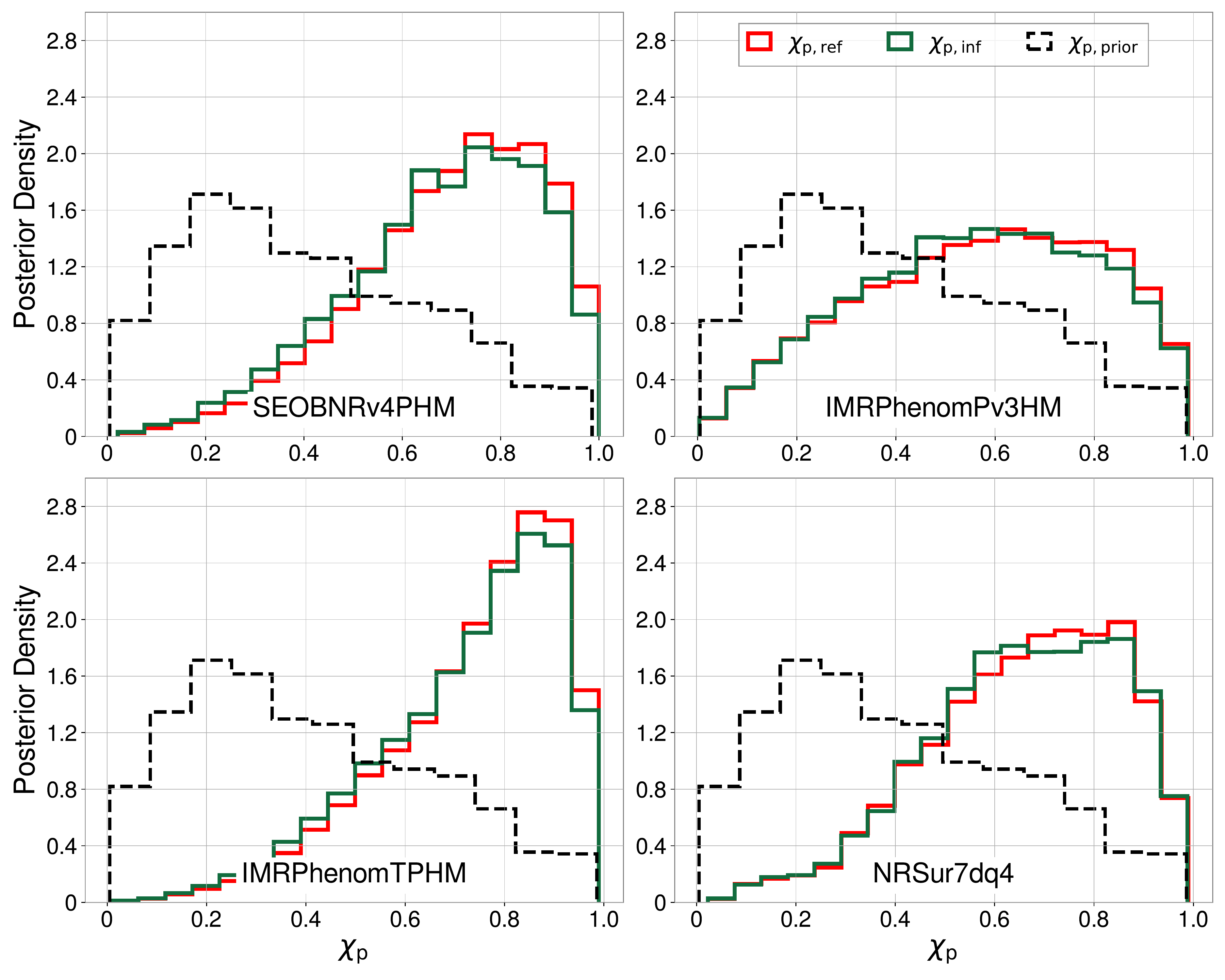}
}
\caption{\label{fig:chi_p_dist_PEsamples_condensed} The distribution of the effective precession spin parameter $\chi_\text{p}$ for GW190521 at $f_{\rm ref}$ and at infinity computed using the hybrid evolution code, as well as the prior distribution of $\chi_\text{p}$ conditioned on the $\xi$ posterior, which is the same at both $f_{\rm ref}$ and infinity.
}

\end{figure*}

The effective precession spin parameter $\chi_{\rm p}$ depends on the projection of the component
spin vectors into the orbital plane [see Eq.~(3.4) in~\cite{Schmidt:2014iyl} for the definition]. It is used to assess the evidence for precession both for individual events (e.g.,~\cite{GWTC-2_paper}) and the population as a whole (e.g.,~\cite{O3a_pop}). There is a recently introduced improved version of $\chi_{\rm p}$ that removes an inconsistency in the definition~\cite{Gerosa:2020aiw}. However, we consider the original version here, to show how it is affected by the spin evolution. We compare the posterior distributions of $\chi_{\rm p}$ at $f_{\rm ref}$ and infinity.
Figure~\ref{fig:chi_p_dist_PEsamples_condensed} shows the posterior distribution for GW190521 using all four waveform models. As expected,
the largest difference is for \textsc{SEOBNRv4PHM} and \textsc{IMRPhenomTPHM} while \textsc{IMRPhenomPv3HM} and \textsc{NRSur7dq4}
show smaller differences in $\chi_{\rm p}$ distribution at $f_{\rm ref}$ and at infinity. 
\edit{The $\max \Delta Q$ values between the $\chi_{\rm p}$ distributions at $f_{\rm ref}$ and infinity are 
$0.03$, $0.03$, $0.02$, and $0.01$
for \textsc{SEOBNRv4PHM},  \textsc{IMRPhenomTPHM}, \textsc{IMRPhenomPv3HM}, 
and \textsc{NRSur7dq4}, respectively.} 

In Fig.~\ref{fig:chi_p_dist_PEsamples_condensed}, we also show
the prior distribution of $\chi_{\rm p}$ at $f_{\rm ref}$ and infinity conditioned on the $\xi$ posterior.
Conditioning the $\chi_{\rm p}$ prior on the $\xi$ posterior accounts for the correlated
prior between $\chi_{\rm p}$ and $\xi$ \edit{with the standard spin prior choices} and helps us identify events 
for which \edit{the} data are informative about precession, as in~\edit{\cite{GWTC-2_paper, GWTC-3_paper}} (see in particular \edit{Figs.~11 and 10, respectively}).
Since this prior comes from isotropic priors on the individual spins, we find that it is the same at the reference
frequency and at infinity, in agreement with the finding that an isotropic distribution evolves to an isotropic
distribution for both the precession-averaged~\cite{Gerosa:2015tea} and orbit-averaged~\cite{Bogdanovic:2007hp} evolution
(though~\cite{Bogdanovic:2007hp} only uses the 2PN precession equations without the contribution of the black holes'
spin-induced quadrupoles).

The posteriors on $\chi_\text{p}$ at infinity all prefer slightly smaller values than the ones at the reference frequency. Thus, the evidence for precession given by the Kullback-Leibler (KL) divergence~\cite{Kullback:1951zyt} from the prior to the posterior will be slightly smaller at infinity. Indeed, these KL divergences at $\{f_{\rm ref}, \infty\}$ are $\{0.51,0.45\}$, $\{0.54,0.49\}$, $\{0.16,0.15\}$, and $\{0.44,0.42\}$, for \textsc{SEOBNRv4PHM},  \textsc{IMRPhenomTPHM}, \textsc{IMRPhenomPv3HM},
and \textsc{NRSur7dq4}, respectively.

\section{Conclusions}
\label{sec:concl}

We have developed a new code that combines together orbit-averaged and precession-averaged evolution to compute the tilts at infinity, which is available as part of LALSuite~\cite{LALSuite} in the tilts\_at\_infinity module~\cite{tilts_at_infinity} in LALSimulation. It is also implemented as an option (calling the LALSuite implementation) in the PESummary postprocessing code~\cite{Hoy:2020vys}, either to evolve the spins with the fast but less accurate precession-averaged evolution or the slower but more accurate hybrid evolution. \edit{This implementation has been used in the analysis of the new events in GWTC-3~\cite{GWTC-3_paper}.} The precession-averaged evolution implements the regularized equations we have derived. These are much more accurate than the standard equations for close-to-equal-mass cases, which one encounters when applying the method to gravitational wave detections. In particular, our implementation of the precession-averaged evolution does not lose accuracy for close-to-equal-mass cases as the implementation in PRECESSION~\cite{Gerosa:2016sys,PRECESSION_GitHub_dev} does.

There are various extensions to the code that we may consider in future work. The most direct extension would be to use PN evolution without orbit averaging for high frequencies, determining the empirical transition frequency between this evolution and the orbit-averaged evolution using the same method we used here to determine the transition frequency between the orbit-averaged and precession-averaged evolution. One can also use the spin evolution from the waveform model itself---this is particularly straightforward to do with the surrogate models and we have already checked that it does not make a significant difference for GW190521 with \textrm{NRSur7dq4}.

It may also be possible to make the orbit-averaged evolution more efficient through code optimization and/or through using, e.g., the techniques introduced in~\cite{Yang:2019oqm}, though creating a surrogate model for this mapping, similar to the surrogate models for the final state of precessing binaries in~\cite{Varma:2018aht,Varma:2019csw}, is likely to provide the largest speed-up. The results in Fig.~\ref{fig:tilts_vs_vtrans} also suggest that it might be possible to obtain the hybrid evolution results to good accuracy by averaging the tilts at infinity over a cycle of the oscillations in terms of $v_\text{trans}$ and/or some extrapolation without needing to evolve backwards to such low frequencies using the orbit-averaged evolution. We will investigate if this can provide substantial reductions in computational cost. \edit{A simpler way to obtain a faster code at the cost of less accuracy (since the current accuracy is overly stringent for many applications) will be to allow the user to specify the desired tolerance and set the transition frequency accordingly, since a looser tolerance will necessitate less of the more computationally expensive orbit-averaged evolution.}

If one can obtain a very significant reduction in computational cost, likely requiring a surrogate model, one could consider sampling directly on the tilts at infinity in the inference on gravitational wave signals. Here one would parameterize the four spin angular degrees of freedom by the two tilts at infinity and two phases at transition points (one precessional phase and one orbital phase), or their equivalent, giving the information about the binary's evolution that is not captured by the precession-averaged and orbit-averaged evolutions, respectively. Specifically, one would first evolve forward to a transition frequency with precession-averaged evolution, which does not need the two phases, and then evolve with the orbit-averaged evolution, initializing this evolution by augmenting the output of the precession-averaged evolution with the precessional phase. Finally, one would augment the output of the orbit-averaged evolution with the orbital phase to perform the final evolution with no averaging, which will give the output necessary for waveform generation, or to evaluate the waveform model directly.

Additional extensions include using eccentric orbit-averaged evolution, which can be performed following~\cite{Phukon:2019gfh}, though this will not be relevant for interpreting gravitational wave detections until eccentric precessing waveform models become available. However, eccentric orbit-averaged evolution will be relevant for the problem of evolving forward in time to obtain the remnant properties (final mass, spin, and recoil velocity) using a combination of precession-averaged, orbit-averaged, and instantaneous evolutions. This problem has been studied for quasicircular binaries using $2$PN orbit-averaged evolution in~\cite{Reali:2020vkf}---it will be useful to extend this to higher PN orders and to include instantaneous evolution and eccentricity. While there are also not fits for remnant properties for eccentric, precessing binary black holes, in cases where the eccentricity close to merger is small, one can apply the quasicircular fits, even if the eccentricity is not small earlier in the evolution.

There are also improvements that can be made to the precession-averaged evolution. These include small improvements such as determining how to use the faster solve\_ivp evolution without it leading to apparent hangs in difficult cases and determining the optimal mass ratio at which to switch from the standard to the regularized version of the equations to increase the accuracy of the results or how best to modify the definition of $\kappa_{\xi q}$ to avoid needing to do this. However, there are also much more significant improvements that may be possible, notably extending the precession-averaging to higher PN order and allowing for non-black hole spin-induced quadrupoles (so it can be applied to systems involving neutron stars). Neither of these is straightforward, since the current precession-averaged evolution relies on the effective spin being conserved, and this conservation is lost at higher PN order or with non-black hole spin-induced quadrupoles, though there is a proposal for a method to include these effects perturbatively in~\cite{Klein:2021jtd}. Further development is also needed to be able to evolve the up-down instability endpoint cases accurately for all mass ratios close to $1$. There the alternative regularization given in~\cite{Klein:2021jtd} may also be relevant.

Nevertheless, while there are plenty of improvements that are possible, the current implementation is already well suited to compute the tilts at infinity for current and future binary black hole detections.

\acknowledgments

We thank Christopher Berry and Richard O'Shaughnessy for suggesting this project and Hang Yu for useful discussions. We also thank Charlie Hoy for the implementation in PESummary and Sylvain Marsat, Marta Colleoni, Lucy Thomas, and Eleanor Hamilton for reviewing the implementation in LALSuite and providing many useful comments on the code. Additionally, we thank Riccardo Sturani for implementing the $3$PN SpinTaylor precession equations and Frank Ohme and Serguei Ossokine for useful suggestions. Finally, we thank Davide Gerosa and Sylvain Marsat for a careful reading of the paper \edit{and the anonymous referee and Leo Stein for useful comments}.
N.~K.~J.-M.\ acknowledges support from STFC Consolidator Grant 
No.~ST/L000636/1. Also, this work has received funding from the European Union's Horizon 
2020 research and innovation programme under the Marie Sk{\l}odowska-Curie grant agreement 
No.~690904. This research was supported in part by Perimeter Institute for Theoretical 
Physics. Research at Perimeter Institute is supported by the Government of Canada through 
Industry Canada and by the Province of Ontario through the Ministry of Economic Development 
\& Innovation. N.~K.~J.-M.\  also acknowledges support from the AIRBUS Group Corporate Foundation through a chair in
``Mathematics of Complex Systems'' at the International Centre for Theoretical Sciences, Tata Institute of Fundamental Research during initial work on this project.

This material is based upon work supported by NSF's LIGO Laboratory which is a major facility fully funded by the National Science Foundation. This research has made use of data obtained from the Gravitational Wave Open Science Center (www.gw-openscience.org), a service of LIGO Laboratory, the LIGO Scientific Collaboration and the Virgo Collaboration. LIGO is funded by the US National Science Foundation (NSF). Virgo is funded by the French Centre National de Recherche Scientifique (CNRS), the Italian Istituto Nazionale della Fisica Nucleare (INFN) and the Dutch Nikhef, with contributions by Polish and Hungarian institutes.

The authors are grateful for computational resources provided by the Leonard E Parker Center for Gravitation, Cosmology and Astrophysics at the University of Wisconsin-Milwaukee and the LIGO Laboratory and supported by National Science Foundation Grants PHY-1626190, PHY-1700765, PHY-0757058, and PHY-0823459. Additionally, we acknowledge the use of IUCAA LDG cluster Sarathi for the computational/numerical work.

This study used the Python software packages matplotlib~\cite{Hunter:2007ouj}, mpmath~\cite{mpmath}, numpy~\cite{Harris:2020xlr}, PESummary~\cite{Hoy:2020vys},
PRECESSION~\cite{Gerosa:2016sys,PRECESSION_GitHub_dev}, and scipy~\cite{Virtanen:2019joe}.

This is LIGO document number P2100029.

\appendix

\section{Linearization error bound}
\label{app:lin_err_bound}

When $m$ is small (notably when $L$ is large, so $S^2_3$ is also large and negative), we can linearize $\langle S^2\rangle_\text{pr}$ [Eq.~\eqref{eq:Ssq_pr}] in $m$, obtaining
\<\label{eq:m_lin}
\langle S^2\rangle_\text{pr} = \frac{1}{2}(S^2_+ + S^2_-) + O(m^2).
\?
This is especially convenient numerically, since when $S^2_3$ is large and negative, the residual one obtains when inserting the numerically determined value into the cubic is also large, making it untrustworthy, but it turns out that one does not need to compute it at all.

However, in order for us to be able to use this linearization only when it is really warranted, for a desired accuracy, we need a strict bound on the error incurred. Fortunately, such a bound is relatively straightforward to obtain.
We start by considering the bounds on the linearization of the complete elliptic integrals $E$ and $K$. These functions are defined by the following integrals (see, e.g., Sec.~17.3 in~\cite{AS})
\begin{subequations}
\label{eq:ellip}
\begin{align}
E(m) &:= \int_0^{\pi/2}\sqrt{1 - m\sin^2\theta}\,d\theta,\\
K(m) &:= \int_0^{\pi/2}\frac{d\theta}{\sqrt{1 - m\sin^2\theta}}
\end{align}
\end{subequations}
(with $m \in [0,1]$).
Thus, as mentioned previously, we have linearized versions of these functions of
\begin{subequations}
\begin{align}
E_\text{lin}(m) &= \frac{\pi}{2}\left(1 - \frac{m}{4}\right),\\
K_\text{lin}(m) &= \frac{\pi}{2}\left(1 + \frac{m}{4}\right).
\end{align}
\end{subequations}
Now, using Taylor's theorem with Lagrange remainder, we have the following expressions for the error incurred by linearization:
\begin{subequations}
\begin{align}
E(m) - E_\text{lin}(m) &= -\frac{m^2}{8}\int_0^{\pi/2}\frac{\sin^4\theta}{(1 - \bar{m}\sin^2\theta)^{3/2}}\,d\theta,\\
K(m) - K_\text{lin}(m) &= \frac{3m^2}{8}\int_0^{\pi/2}\frac{\sin^4\theta}{(1 - \bar{\bar{m}}\sin^2\theta)^{5/2}}\,d\theta,
\end{align}
\end{subequations}
for some $\bar{m}, \bar{\bar{m}} \in (0, m)$. Since the integrands are nonnegative, increasing functions of $\bar{m}, \bar{\bar{m}}$, we thus have
\begin{subequations}
\begin{align}
-\frac{m^2}{8}\int_0^{\pi/2}\frac{\sin^4\theta}{(1 - m\sin^2\theta)^{3/2}}\,d\theta \leq E(m) - E_\text{lin}(m) \leq 0,\\
0 \leq K(m) - K_\text{lin}(m) \leq \frac{3m^2}{8}\int_0^{\pi/2}\frac{\sin^4\theta}{(1 - m\sin^2\theta)^{5/2}}\,d\theta.
\end{align}
\end{subequations}
We can then use $(1 - m\sin^2\theta)^{-\alpha} \leq (1 - m)^{-\alpha}$ for $\alpha > 0$ (since $m > 0$) and note that $\int_0^{\pi/2}\sin^4\theta\,d\theta = 3\pi/16$ to obtain
\begin{subequations}
\label{eq:ellip_lin_bounds}
\begin{align}
-\frac{3\pi m^2}{128(1-m)^{3/2}} &\leq E(m) - E_\text{lin}(m) \leq 0,\\
0 &\leq K(m) - K_\text{lin}(m) \leq \frac{9\pi m^2}{128(1-m)^{5/2}}.
\end{align}
\end{subequations}

We now want to bound the difference between $E(m)/K(m)$ and its linearization, viz.,
\<\label{eq:ellip_lin_diff}
\left|\frac{E(m)}{K(m)} - 1 + \frac{m}{2}\right|.
\?
We will do this by bounding $|E/K - E_\text{lin}/K_\text{lin}|$ and noting that
\<
\label{eq:ellip_lin_express}
\frac{E_\text{lin}(m)}{K_\text{lin}(m)} - 1 + \frac{m}{2} = \frac{m^2}{8 + 2m},
\?
so we can combine this with the bound on $|E/K - E_\text{lin}/K_\text{lin}|$ to obtain the desired bound.

To obtain the bound on $|E/K - E_\text{lin}/K_\text{lin}|$, we first rewrite $E/K$ using a convenient zero as
\<\label{eq:ND_expr}
\frac{\cN + \epsilon_\cN}{\cD + \epsilon_\cD} = \frac{\cN}{\cD} + \frac{\cD\epsilon_\cN - \cN\epsilon_\cD}{\cD(\cD + \epsilon_\cD)},
\?
where (using $\cN$ for the numerator and $\cD$ for the denominator)
\begin{subequations}
\begin{align}
\cN &= E_\text{lin}(m), \qquad \epsilon_\cN = E(m) - E_\text{lin}(m),\\
\cD &= K_\text{lin}(m), \qquad \epsilon_\cD = K(m) - K_\text{lin}(m).
\end{align}
\end{subequations}
Thus $\cN, \cD, \epsilon_\cD \geq 0$ and $\epsilon_\cN \leq 0$ and we have
\<
\begin{split}
\left|\frac{E(m)}{K(m)} - \frac{E_\text{lin}(m)}{K_\text{lin}(m)}\right| &= \left|\frac{\cN + \epsilon_\cN}{\cD + \epsilon_\cD} - \frac{\cN}{\cD}\right|\\
&= \frac{\cD|\epsilon_\cN| + \cN\epsilon_\cD}{\cD(\cD + \epsilon_\cD)}\\
&\leq \frac{|\epsilon_\cN|}{\cD} + \frac{\cN}{\cD^2}\epsilon_\cD.
\end{split}
\?
Now, from the above and Eqs.~\eqref{eq:ellip_lin_bounds}, we have
\begin{widetext}
\<
\begin{split}
\left|\frac{E(m)}{K(m)} - \frac{E_\text{lin}(m)}{K_\text{lin}(m)}\right| &\leq \frac{|E(m) - E_\text{lin}(m)|}{K_\text{lin}(m)} + \frac{E_\text{lin}(m)}{K^2_\text{lin}(m)}[K(m) - K_\text{lin}(m)]\\
&\leq \frac{3m^2}{(64+16m)(1-m)^{3/2}}\left(1 + \frac{3}{1-m}\frac{4 - m}{4 + m}\right).
\end{split}
\?

Thus, noting that $E/K - E_\text{lin}/K_\text{lin} \leq 0$, $E_\text{lin}/K_\text{lin} - 1 + m/2 \geq 0$, and $|E/K - E_\text{lin}/K_\text{lin}| \geq E_\text{lin}/K_\text{lin} - 1 + m/2$ (which can be checked numerically), we have
\<\label{eq:basic_bound}
\begin{split}
\left|\frac{E(m)}{K(m)} - 1 + \frac{m}{2}\right| &= \left|\frac{E(m)}{K(m)} - \frac{E_\text{lin}(m)}{K_\text{lin}(m)} + \frac{E_\text{lin}(m)}{K_\text{lin}(m)} - 1 + \frac{m}{2}\right|\\
&\leq \left[\frac{3}{16(1-m)^{3/2}}\left(1 + \frac{3}{1-m}\frac{4 - m}{4 + m}\right) - \frac{1}{2}\right]\frac{m^2}{4 + m}\\
&=: m^2\cE_\text{ratio}(m),
\end{split}
\?
\end{widetext}
where we have introduced $\cE_\text{ratio}(m)$ (which is an increasing function of $m$) to represent the overall scaling and the subdominant corrections to the dominant $m^2$ dependence of the error. This is a strong bound for the small values of $m$ we are interested in when applying it, but becomes quite weak as $m \nearrow 1$.

Thus, the difference between $\langle S^2\rangle_\text{pr}$ and its linearization is
\<
\left|\frac{S^2_- - S^2_+}{m}\left[\frac{E(m)}{K(m)} - 1 + \frac{m}{2}\right]\right| \leq (S^2_+ - S^2_-)m\cE_\text{ratio}(m).
\?
This translates immediately into a bound in terms of the barred quantities introduced in Sec.~\ref{sec:reg}, which is what is implemented in the code.
Specifically, we linearize in $m$ when
 \<\label{eq:m_bound}
 m\cE_\text{ratio}(m) \leq \delta_\text{lin}\max\left(1,\frac{1}{\bar{S}^2_+ - \bar{S}^2_-}\right),
 \?
where $\delta_\text{lin}$ sets the tolerance. In particular, $\delta_\text{lin}$ is roughly the relative tolerance in Eq.~\eqref{eq:ode} when $\bar{S}^2_+ - \bar{S}^2_- > 1$ and is the absolute tolerance when $\bar{S}^2_+ - \bar{S}^2_- \leq 1$. We compute $\bar{S}_\pm^2$ by solving Eq.~\eqref{eq:cubic_barred}. The default value of $\delta_\text{lin}$ is the same as the absolute tolerance of the differential equation integrator.

\section{Error bound for reducing the cubic to a quadratic}
\label{app:err_quadratic}

When $q$ is either small or close to $1$ and/or $u$ is small, the coefficient of the highest power in the cubic [Eq.~\eqref{eq:cubic_barred}] is small, and one can obtain some solutions to the cubic to a good approximation by setting that coefficient to zero and solving a quadratic. Here we derive a strict error bound for this approximation. In this case, we need only to consider the error bound for the approximation $\bar{S}^2_+ + \bar{S}^2_- \simeq -\bar{C}/\bar{B}$, since in this case $|\bar{S}^2_3|$ will be large, so $m$ will be small and thus satisfy the requirement for linearization being a good approximation given in Appendix~\ref{app:lin_err_bound}.

We only consider the case $\bar{B}, \bar{C}, \bar{D} > 0$, since this is the case where this simplification is necessary in practice and also simplifies the analysis.
We first want to obtain a lower bound on $|\bar{S}^2_3|$, for which we rewrite Eq.~\eqref{eq:cubic_barred} in terms of $z := 1/\bar{S}^2$, so it becomes
\<
P_\varepsilon(z) := \varepsilon + \bar{B} z + \bar{C} z^2 + \bar{D} z^3 = 0,
\?
where we have defined $\varepsilon := q(1-q^2)u^2 > 0$ and are interested in the case where this is small. We now want to consider the solution to $P_\varepsilon(z) = 0$ that goes to zero as $\varepsilon \searrow 0$, i.e., $z_3 := 1/\bar{S}_3^2$, and obtain a bound on how close to zero it is. We do this by finding an interval near zero on which $P_\varepsilon$ changes sign and thus has a root. Since $\varepsilon  > 0$, we have $P_\varepsilon(0) = \varepsilon > 0$, and since we assume $\bar{B}, \bar{C}, \bar{D} > 0$, we have $P_\varepsilon(z_\varepsilon) = \bar{D} z_\varepsilon^3 < 0$, where
\<
z_\varepsilon := \frac{-\bar{B} + \sqrt{\bar{B}^2 - 4\bar{C}\varepsilon}}{2\bar{C}} < 0
\?
is obtained by solving $\varepsilon + \bar{B} z_\varepsilon + \bar{C} z_\varepsilon^2 = 0$ and choosing the root that goes to zero as $\varepsilon \searrow 0$. We thus know that $z_3 \in (z_\varepsilon, 0)$, so $|\bar{S}^2_3| \geq 1/|z_\varepsilon|$.

Now, we want to use this to obtain a bound on $|\bar{S}^2_+ + \bar{S}^2_- + \bar{C}/\bar{B}|$. To do this, we note that Vieta's formulas applied to $P_\varepsilon(z)$ give
\begin{subequations}
\begin{align}
\frac{1}{\bar{S}_+^2} + \frac{1}{\bar{S}_-^2} + \frac{1}{\bar{S}_3^2} &= -\frac{\bar{C}}{\bar{D}},\\
\frac{1}{\bar{S}_3^2}\left(\frac{1}{\bar{S}_+^2} + \frac{1}{\bar{S}_-^2}\right) + \frac{1}{\bar{S}_+^2\bar{S}_-^2} &= \frac{\bar{B}}{\bar{D}},
\end{align}
\end{subequations}
which yield [noting that $1/\bar{S}_+^2 + 1/\bar{S}_-^2 = (\bar{S}^2_+ + \bar{S}^2_-)/\bar{S}_+^2\bar{S}_-^2$, so we solve the above equations for $1/\bar{S}_+^2 + 1/\bar{S}_-^2$ and $1/\bar{S}_+^2\bar{S}_-^2$]
\<
\bar{S}^2_+ + \bar{S}^2_- = -\frac{\bar{C} + \bar{D}z_3}{\bar{B} + \bar{C}z_3 + \bar{D}z_3^2}.
\?
Thus, we have an error bound of
\<
\begin{split}
\left|\bar{S}^2_+ + \bar{S}^2_- + \frac{\bar{C}}{\bar{B}}\right| &= \left|\frac{\bar{C}}{\bar{B}} -\frac{\bar{C} + \bar{D}z_3}{\bar{B} + \bar{C}z_3 + \bar{D}z_3^2}\right|\\
&= |z_3|\left|\frac{\bar{C}^2 - \bar{B}\bar{D} + \bar{C}\bar{D}z_3}{\bar{B}(\bar{B} + \bar{C}z_3 + \bar{D}z_3^2)}\right|\\
&< |z_\varepsilon|\frac{|\bar{C}^2 - \bar{B}\bar{D}| + \bar{C}\bar{D}|z_\varepsilon|}{\bar{B}(\bar{B} - \bar{C}|z_\varepsilon|)}\\
&=: \cE_\varepsilon^\text{sum}.
\end{split}
\?
To obtain the inequality, we have recalled that $z_\varepsilon < z_3 < 0$ and $\bar{B}, \bar{C}, \bar{D} > 0$, so we have $\bar{B} + \bar{C}z_3 + \bar{D}z_3^2 > \bar{B} + \bar{C}z_3 > \bar{B} + \bar{C}z_\varepsilon = \bar{B} - \bar{C}|z_\varepsilon|$. Since $\bar{B} + \bar{C}z_\varepsilon = -\varepsilon/z_\varepsilon > 0$ from the defining equation for $z_\varepsilon$, the right-hand side of the previous inequality is positive, so it also holds with absolute values added. We also applied the triangle inequality to the numerator.

We also need to bound $m$, for which we note that $\bar{S}^2_\star < 0$ in this case, since the coefficients of $P_\varepsilon$ are all positive, so all its roots are negative (assuming that they are real).\footnote{Of course, while the $\bar{S}_\pm^2$ are negative in this case, the associated unbarred versions [obtained from Eq.~\eqref{eq:Sbar_def}] are positive, as expected.} We assume that $|\bar{S}_3^2|$ is large enough so that $|\bar{S}_3^2| > |\bar{S}^2_+ + \bar{S}^2_-|$, and in fact assume that $|z_\varepsilon|(\bar{C}/\bar{B} + \cE_\varepsilon^\text{sum}) < 1$, which implies the previous inequality. Thus, we have
\<
\begin{split}
m &= \frac{\bar{S}^2_+ - \bar{S}^2_-}{\bar{S}^2_+ - \bar{S}^2_3}\\
&\leq \frac{|\bar{S}^2_+ + \bar{S}^2_-|}{\bar{S}^2_+ + \bar{S}^2_- - \bar{S}^2_3}\\
&< |z_\varepsilon|\frac{\bar{C}/\bar{B} + \cE_\varepsilon^\text{sum}}{1 - |z_\varepsilon|(\bar{C}/\bar{B} + \cE_\varepsilon^\text{sum})},
\end{split}
\?
recalling that $\bar{S}^2_3 \leq 1/z_\varepsilon < 0$, so $\bar{S}^2_+ + \bar{S}^2_- - \bar{S}^2_3 \geq 1/|z_\varepsilon| - |\bar{S}^2_+ + \bar{S}^2_-|$.
This bound on $m$ is not sharp, but suffices for our purposes.

In this case, we incur errors both due to linearizing in $m$ and approximating $\bar{S}^2_+ + \bar{S}^2_-$ by $-\bar{C}/\bar{B}$, so we apply this simplification if the bound on $m$ satisfies Eq.~\eqref{eq:m_bound} with $\delta_\text{lin} \to \delta_\text{lin}/2$ [recalling that $\cE_\text{ratio}(m)$ is an increasing function of $m$] and $\cE_\varepsilon^\text{sum} \leq \delta_\text{lin}$ [since we get a factor of $1/2$ in Eq.~\eqref{eq:m_lin}], choosing to divide the total allowed error equally between the two cases. We also replace $1/(\bar{S}_+^2 - \bar{S}_-^2) \to 1/(\bar{C}/\bar{B} + \cE_\varepsilon^\text{sum}) \leq 1/|\bar{S}^2_+ + \bar{S}^2_-| \leq 1/(\bar{S}_+^2 - \bar{S}_-^2)$ in Eq.~\eqref{eq:m_bound}.

\section{Details about the internal checks in the precession-averaged evolution}
\label{app:prec_avg_internal_checks}

The precession-averaged code implements a number of internal checks to make sure that the evolution is going well. Specifically, it checks that the imaginary parts of the solutions to the cubic are smaller than a certain absolute tolerance, by default $10^{-6}$. It also checks that the unbarred $S^2_\pm$ are nonnegative, as they must be physically [as the maximum and minimum values of $S^2$; see the discussion below Eq.~(21) in~\cite{Chatziioannou:2017tdw}]. Additionally, it checks that the evolution reached the desired final value of $u$ (with a tolerance of $0.75\Delta u$). If any of these checks is triggered, the code prints a message. For the checks that are indicative of a serious problem with the evolution, the code then terminates the evolution and initiates the fallback evolution if it is selected (as it is by default). If the fallback evolution (and second fallback evolution, if enabled) also fails, the code prints a message and returns None for the tilts by default, but it can also be set to return \verb,numpy.nan, for the tilts or to raise an error.

The code also prints a message if there are two roots with a nonzero imaginary part whose real parts are the same, since this is likely an indication of complex conjugate roots. (Real roots that are equal to numerical precision occurs for close-to-aligned-spin cases.) Similarly, it prints a message if the residual of the cubic equation does not change sign over an interval of a specified size around the root, as detailed below. By default this check does not stop the evolution, but it can be set to do so as an option. The code returns the initial tilts if both of a pair of roots are zero, so that the binary is nonprecessing to numerical precision. 

For the check on the accuracy of the roots of the cubic, the interval over which the cubic should change sign is chosen to have endpoints $\bar{S}_\star^2 \pm \max(|\bar{S}_\star^2|, 1)\delta_\text{r}$, where $\delta_\text{r}$ is the associated tolerance parameter, whose default value is $10^{-6}$. To account for cases where $\bar{S}^2_+ - \bar{S}^2_-$ or $\bar{S}^2_- - \bar{S}^2_3$ is very small, the code sets $\delta_\text{r} \to \min(\delta_\text{r}^{+,-}, \delta_\text{r}^{-,3})$ when $\bar{S}^2_A - \bar{S}^2_B < 0.55\max(|\bar{S}^2_A + \bar{S}^2_B|, 2)\delta_\text{r}$ is true for either of $(A,B) \in \{(+,-), (-,3)\}$. Here
$\delta_\text{r}^{A,B} := (\bar{S}^2_A - \bar{S}^2_B)/\max(|\bar{S}^2_A + \bar{S}^2_B|, 2)$ when the previous inequality is satisfied and $\delta_r$ otherwise. For cases when there is a double root (i.e., when $m = 0$ or~$1$), we check that the derivative of the cubic changes sign for the double root, with the same logic as above, though just considering the distance between the double root and the other root.

\section{Example code usage}
\label{app:examples}

\begingroup
\setlength{\tabcolsep}{6pt}
\renewcommand{\arraystretch}{1.25}
\begin{table*}[]
\caption{\label{tab:examples}
Examples of using the code introduced here to compute tilts at infinity or bounds on tilts at a finite separation in an interactive Python session.
}
\begin{tabular}{l}
\hline \hline
\verb|# Setup|\\
\verb|>>> from lalsimulation.tilts_at_infinity import prec_avg_tilt_comp,|\\
\verb|    calc_tilts_at_infty_hybrid_evolve|\\
\verb|>>> from lal import MSUN_SI|\\
\verb|>>> m1, m2 = 50., 45. # solar masses|\\
\verb|>>> chi1, chi2 = 0.8, 0.6|\\
\verb|>>> tilt1, tilt2, phi12 = 1.3, 0.4, 2.1 # rad|\\
\verb|>>> f0 = 20. # Hz|\\
\\
\verb|# Calculate tilts at infinity|\\
\\
\verb|# For only precession-average evolution|\\
\verb|>>> prec_avg_tilt_comp(m1*MSUN_SI, m2*MSUN_SI, chi1, chi2, tilt1, tilt2, phi12, f0)|\\
\verb|{'tilt1_inf': 1.1390970822856608, 'tilt2_inf': 0.8136099686362966}|\\
\\
\verb|# The same through the hybrid evolution interface (so with LPNspins=False)|\\
\verb|>>> calc_tilts_at_infty_hybrid_evolve(m1*MSUN_SI, m2*MSUN_SI, chi1, chi2, tilt1, tilt2, phi12,|\\
\verb|    f0, prec_only=True)|\\
\verb|{'tilt1_inf': 1.141893652039253, 'tilt2_inf': 0.7936508631342601,|\\
\verb| 'tilt1_transition': None, 'tilt2_transition': None,|\\
\verb| 'phi12_transition': None, 'f_transition': None}|\\
\\
\verb|# For hybrid evolution|\\
\verb|>>> calc_tilts_at_infty_hybrid_evolve(m1*MSUN_SI, m2*MSUN_SI, chi1, chi2, tilt1, tilt2,|\\
\verb|    phi12, f0, version='v2')|\\
\verb|{'tilt1_inf': 1.1584177372277964, 'tilt2_inf': 0.7637854727141383,|\\
\verb| 'tilt1_transition': 1.2227660639594675, 'tilt2_transition': 0.6253846448618053,|\\
\verb| 'phi12_transition': 6.102198158202932, 'f_transition': 0.005044066737021535}|\\
\\
\verb|# Calculate bounds on tilts at finite separation with both methods|\\
\verb|>>> Lf = 100. # dimensionless|\\
\verb|>>> prec_avg_tilt_comp(m1*MSUN_SI, m2*MSUN_SI, chi1, chi2, tilt1, tilt2, phi12, f0,|\\
\verb|    Lf=Lf)|\\
\verb|{'tilt1_sep_min': 1.1292402468578164, 'tilt1_sep_max': 1.1487850778054958|\\
\verb| 'tilt1_sep_avg': 1.1390338016598198, 'tilt2_sep_min': 0.7954772415249947,|\\
\verb| 'tilt2_sep_max': 0.8316644029539157, 'tilt2_sep_avg': 0.8137271199148113}|\\
\verb|>>> calc_tilts_at_infty_hybrid_evolve(m1*MSUN_SI, m2*MSUN_SI, chi1, chi2, tilt1, tilt2,|\\
\verb|    phi12, f0, Lf=Lf, version='v2')|\\
\verb|{'tilt1_sep_min': 1.1490357804932505, 'tilt1_sep_max': 1.1676210227670598,|\\
\verb| 'tilt1_sep_avg': 1.1583465084488291, 'tilt2_sep_min': 0.7455138665464005,|\\
\verb| 'tilt2_sep_max': 0.7819862871558729, 'tilt2_sep_avg': 0.7639252370956654, |\\
\verb| 'tilt1_transition': 1.2227660639594675, 'tilt2_transition': 0.6253846448618053, |\\
\verb| 'phi12_transition': 6.102198158202932, 'f_transition': 0.005044066737021535}|\\
\hline \hline
\end{tabular}
\end{table*}
\endgroup

In Table~\ref{tab:examples}, we give examples of the use of the code to compute the tilts at infinity and bounds on the tilts at a finite separation using both purely precession-averaged and hybrid evolution. We do this for a binary with masses $m_1 = 50M_\odot$, $m_2 = 45M_\odot$, dimensionless spin magnitudes $\chi_1 = 0.8$, $\chi_2 = 0.6$, and spin angles $\theta_1 = 1.3$~rad, $\theta_2 = 0.4$~rad, $\phi_{12} = 2.1$~rad at a reference frequency of $f_0 = 20$~Hz. The finite separation corresponds to an orbital angular momentum with magnitude $L_f = 100M^2$. One will obtain slightly different values for the higher decimal places with different numpy and scipy versions. These results were obtained with numpy \edit{1.21.2} and scipy \edit{1.7.1}.

\bibliography{../tilts_at_infty}

\begin{thebibliography}{91}%
\makeatletter
\providecommand \@ifxundefined [1]{%
 \@ifx{#1\undefined}
}%
\providecommand \@ifnum [1]{%
 \ifnum #1\expandafter \@firstoftwo
 \else \expandafter \@secondoftwo
 \fi
}%
\providecommand \@ifx [1]{%
 \ifx #1\expandafter \@firstoftwo
 \else \expandafter \@secondoftwo
 \fi
}%
\providecommand \natexlab [1]{#1}%
\providecommand \enquote  [1]{``#1''}%
\providecommand \bibnamefont  [1]{#1}%
\providecommand \bibfnamefont [1]{#1}%
\providecommand \citenamefont [1]{#1}%
\providecommand \href@noop [0]{\@secondoftwo}%
\providecommand \href [0]{\begingroup \@sanitize@url \@href}%
\providecommand \@href[1]{\@@startlink{#1}\@@href}%
\providecommand \@@href[1]{\endgroup#1\@@endlink}%
\providecommand \@sanitize@url [0]{\catcode `\\12\catcode `\$12\catcode
  `\&12\catcode `\#12\catcode `\^12\catcode `\_12\catcode `\%12\relax}%
\providecommand \@@startlink[1]{}%
\providecommand \@@endlink[0]{}%
\providecommand \url  [0]{\begingroup\@sanitize@url \@url }%
\providecommand \@url [1]{\endgroup\@href {#1}{\urlprefix }}%
\providecommand \urlprefix  [0]{URL }%
\providecommand \Eprint [0]{\href }%
\providecommand \doibase [0]{http://dx.doi.org/}%
\providecommand \selectlanguage [0]{\@gobble}%
\providecommand \bibinfo  [0]{\@secondoftwo}%
\providecommand \bibfield  [0]{\@secondoftwo}%
\providecommand \translation [1]{[#1]}%
\providecommand \BibitemOpen [0]{}%
\providecommand \bibitemStop [0]{}%
\providecommand \bibitemNoStop [0]{.\EOS\space}%
\providecommand \EOS [0]{\spacefactor3000\relax}%
\providecommand \BibitemShut  [1]{\csname bibitem#1\endcsname}%
\let\auto@bib@innerbib\@empty
\bibitem [{\citenamefont {Abbott}\ \emph
  {et~al.}(2020{\natexlab{a}})\citenamefont {Abbott} \emph
  {et~al.}}]{GW190412}%
  \BibitemOpen
  \bibfield  {author} {\bibinfo {author} {\bibfnamefont {R.}~\bibnamefont
  {Abbott}} \emph {et~al.} (\bibinfo {collaboration} {LIGO Scientific
  Collaboration and Virgo Collaboration}),\ }\href {\doibase
  10.1103/PhysRevD.102.043015} {\bibfield  {journal} {\bibinfo  {journal}
  {Phys. Rev. D}\ }\textbf {\bibinfo {volume} {102}},\ \bibinfo {pages}
  {043015} (\bibinfo {year} {2020}{\natexlab{a}})},\ \Eprint
  {http://arxiv.org/abs/2004.08342} {arXiv:2004.08342 [astro-ph.HE]}
  \BibitemShut {NoStop}%
\bibitem [{\citenamefont {Abbott}\ \emph
  {et~al.}(2020{\natexlab{b}})\citenamefont {Abbott} \emph
  {et~al.}}]{GW190521_discovery}%
  \BibitemOpen
  \bibfield  {author} {\bibinfo {author} {\bibfnamefont {R.}~\bibnamefont
  {Abbott}} \emph {et~al.} (\bibinfo {collaboration} {LIGO Scientific
  Collaboration and Virgo Collaboration}),\ }\href {\doibase
  10.1103/PhysRevLett.125.101102} {\bibfield  {journal} {\bibinfo  {journal}
  {Phys. Rev. Lett.}\ }\textbf {\bibinfo {volume} {125}},\ \bibinfo {pages}
  {101102} (\bibinfo {year} {2020}{\natexlab{b}})},\ \Eprint
  {http://arxiv.org/abs/2009.01075} {arXiv:2009.01075 [gr-qc]} \BibitemShut
  {NoStop}%
\bibitem [{\citenamefont {Abbott}\ \emph
  {et~al.}(2020{\natexlab{c}})\citenamefont {Abbott} \emph
  {et~al.}}]{GW190521_implications}%
  \BibitemOpen
  \bibfield  {author} {\bibinfo {author} {\bibfnamefont {R.}~\bibnamefont
  {Abbott}} \emph {et~al.} (\bibinfo {collaboration} {LIGO Scientific
  Collaboration and Virgo Collaboration}),\ }\href {\doibase
  10.3847/2041-8213/aba493} {\bibfield  {journal} {\bibinfo  {journal}
  {Astrophys. J. Lett.}\ }\textbf {\bibinfo {volume} {900}},\ \bibinfo {pages}
  {L13} (\bibinfo {year} {2020}{\natexlab{c}})},\ \Eprint
  {http://arxiv.org/abs/2009.01190} {arXiv:2009.01190 [astro-ph.HE]}
  \BibitemShut {NoStop}%
\bibitem [{\citenamefont {Hannam}\ \emph {et~al.}(2021)\citenamefont {Hannam},
  \citenamefont {Hoy}, \citenamefont {Thompson}, \citenamefont {Fairhurst},
  \citenamefont {Raymond} \emph {et~al.}}]{Hannam:2021pit}%
  \BibitemOpen
  \bibfield  {author} {\bibinfo {author} {\bibfnamefont {M.}~\bibnamefont
  {Hannam}}, \bibinfo {author} {\bibfnamefont {C.}~\bibnamefont {Hoy}},
  \bibinfo {author} {\bibfnamefont {J.~E.}\ \bibnamefont {Thompson}}, \bibinfo
  {author} {\bibfnamefont {S.}~\bibnamefont {Fairhurst}}, \bibinfo {author}
  {\bibfnamefont {V.}~\bibnamefont {Raymond}},  \emph {et~al.},\ }\href@noop {}
  {\  (\bibinfo {year} {2021})},\ \Eprint {http://arxiv.org/abs/2112.11300}
  {arXiv:2112.11300 [gr-qc]} \BibitemShut {NoStop}%
\bibitem [{\citenamefont {Abbott}\ \emph
  {et~al.}(2021{\natexlab{a}})\citenamefont {Abbott} \emph {et~al.}}]{O3b_pop}%
  \BibitemOpen
  \bibfield  {author} {\bibinfo {author} {\bibfnamefont {R.}~\bibnamefont
  {Abbott}} \emph {et~al.} (\bibinfo {collaboration} {LIGO Scientific
  Collaboration, Virgo Collaboration, and KAGRA Collaboration}),\ }\href@noop
  {} {\  (\bibinfo {year} {2021}{\natexlab{a}})},\ \Eprint
  {http://arxiv.org/abs/2111.03634} {arXiv:2111.03634 [astro-ph.HE]}
  \BibitemShut {NoStop}%
\bibitem [{\citenamefont {Abbott}\ \emph
  {et~al.}(2021{\natexlab{b}})\citenamefont {Abbott} \emph
  {et~al.}}]{GWTC-2_paper}%
  \BibitemOpen
  \bibfield  {author} {\bibinfo {author} {\bibfnamefont {R.}~\bibnamefont
  {Abbott}} \emph {et~al.} (\bibinfo {collaboration} {LIGO Scientific
  Collaboration and Virgo Collaboration}),\ }\href {\doibase
  10.1103/PhysRevX.11.021053} {\bibfield  {journal} {\bibinfo  {journal} {Phys.
  Rev. X}\ }\textbf {\bibinfo {volume} {11}},\ \bibinfo {pages} {021053}
  (\bibinfo {year} {2021}{\natexlab{b}})},\ \Eprint
  {http://arxiv.org/abs/2010.14527} {arXiv:2010.14527 [gr-qc]} \BibitemShut
  {NoStop}%
\bibitem [{\citenamefont {Abbott}\ \emph
  {et~al.}(2021{\natexlab{c}})\citenamefont {Abbott} \emph
  {et~al.}}]{GWTC-3_paper}%
  \BibitemOpen
  \bibfield  {author} {\bibinfo {author} {\bibfnamefont {R.}~\bibnamefont
  {Abbott}} \emph {et~al.} (\bibinfo {collaboration} {LIGO Scientific
  Collaboration, Virgo Collaboration, and KAGRA Collaboration}),\ }\href@noop
  {} {\  (\bibinfo {year} {2021}{\natexlab{c}})},\ \Eprint
  {http://arxiv.org/abs/2111.03606} {arXiv:2111.03606 [gr-qc]} \BibitemShut
  {NoStop}%
\bibitem [{\citenamefont {Kalogera}(2000)}]{Kalogera:1999tq}%
  \BibitemOpen
  \bibfield  {author} {\bibinfo {author} {\bibfnamefont {V.}~\bibnamefont
  {Kalogera}},\ }\href {\doibase 10.1086/309400} {\bibfield  {journal}
  {\bibinfo  {journal} {Astrophys. J.}\ }\textbf {\bibinfo {volume} {541}},\
  \bibinfo {pages} {319} (\bibinfo {year} {2000})},\ \Eprint
  {http://arxiv.org/abs/astro-ph/9911417} {arXiv:astro-ph/9911417} \BibitemShut
  {NoStop}%
\bibitem [{\citenamefont {O'Shaughnessy}\ \emph {et~al.}(2017)\citenamefont
  {O'Shaughnessy}, \citenamefont {Gerosa},\ and\ \citenamefont
  {Wysocki}}]{OShaughnessy:2017eks}%
  \BibitemOpen
  \bibfield  {author} {\bibinfo {author} {\bibfnamefont {R.}~\bibnamefont
  {O'Shaughnessy}}, \bibinfo {author} {\bibfnamefont {D.}~\bibnamefont
  {Gerosa}}, \ and\ \bibinfo {author} {\bibfnamefont {D.}~\bibnamefont
  {Wysocki}},\ }\href {\doibase 10.1103/PhysRevLett.119.011101} {\bibfield
  {journal} {\bibinfo  {journal} {Phys. Rev. Lett.}\ }\textbf {\bibinfo
  {volume} {119}},\ \bibinfo {pages} {011101} (\bibinfo {year} {2017})},\
  \Eprint {http://arxiv.org/abs/1704.03879} {arXiv:1704.03879 [astro-ph.HE]}
  \BibitemShut {NoStop}%
\bibitem [{\citenamefont {Gerosa}\ \emph {et~al.}(2018)\citenamefont {Gerosa},
  \citenamefont {Berti}, \citenamefont {O'Shaughnessy}, \citenamefont
  {Belczynski}, \citenamefont {Kesden}, \citenamefont {Wysocki},\ and\
  \citenamefont {Gladysz}}]{Gerosa:2018wbw}%
  \BibitemOpen
  \bibfield  {author} {\bibinfo {author} {\bibfnamefont {D.}~\bibnamefont
  {Gerosa}}, \bibinfo {author} {\bibfnamefont {E.}~\bibnamefont {Berti}},
  \bibinfo {author} {\bibfnamefont {R.}~\bibnamefont {O'Shaughnessy}}, \bibinfo
  {author} {\bibfnamefont {K.}~\bibnamefont {Belczynski}}, \bibinfo {author}
  {\bibfnamefont {M.}~\bibnamefont {Kesden}}, \bibinfo {author} {\bibfnamefont
  {D.}~\bibnamefont {Wysocki}}, \ and\ \bibinfo {author} {\bibfnamefont
  {W.}~\bibnamefont {Gladysz}},\ }\href {\doibase 10.1103/PhysRevD.98.084036}
  {\bibfield  {journal} {\bibinfo  {journal} {Phys. Rev. D}\ }\textbf {\bibinfo
  {volume} {98}},\ \bibinfo {pages} {084036} (\bibinfo {year} {2018})},\
  \Eprint {http://arxiv.org/abs/1808.02491} {arXiv:1808.02491 [astro-ph.HE]}
  \BibitemShut {NoStop}%
\bibitem [{\citenamefont {Belczynski}\ \emph {et~al.}(2020)\citenamefont
  {Belczynski} \emph {et~al.}}]{Belczynski:2017gds}%
  \BibitemOpen
  \bibfield  {author} {\bibinfo {author} {\bibfnamefont {K.}~\bibnamefont
  {Belczynski}} \emph {et~al.},\ }\href {\doibase 10.1051/0004-6361/201936528}
  {\bibfield  {journal} {\bibinfo  {journal} {Astron. Astrophys.}\ }\textbf
  {\bibinfo {volume} {636}},\ \bibinfo {pages} {A104} (\bibinfo {year}
  {2020})},\ \bibinfo {note} {data available from
  \url{https://www.syntheticuniverse.org}},\ \Eprint
  {http://arxiv.org/abs/1706.07053} {arXiv:1706.07053 [astro-ph.HE]}
  \BibitemShut {NoStop}%
\bibitem [{\citenamefont {Mould}\ and\ \citenamefont
  {Gerosa}(2022)}]{Mould:2021xst}%
  \BibitemOpen
  \bibfield  {author} {\bibinfo {author} {\bibfnamefont {M.}~\bibnamefont
  {Mould}}\ and\ \bibinfo {author} {\bibfnamefont {D.}~\bibnamefont {Gerosa}},\
  }\href {\doibase 10.1103/PhysRevD.105.024076} {\bibfield  {journal} {\bibinfo
   {journal} {Phys. Rev. D}\ }\textbf {\bibinfo {volume} {105}},\ \bibinfo
  {pages} {024076} (\bibinfo {year} {2022})},\ \Eprint
  {http://arxiv.org/abs/2110.05507} {arXiv:2110.05507 [astro-ph.HE]}
  \BibitemShut {NoStop}%
\bibitem [{\citenamefont {Stevenson}\ \emph {et~al.}(2017)\citenamefont
  {Stevenson}, \citenamefont {Berry},\ and\ \citenamefont
  {Mandel}}]{Stevenson:2017dlk}%
  \BibitemOpen
  \bibfield  {author} {\bibinfo {author} {\bibfnamefont {S.}~\bibnamefont
  {Stevenson}}, \bibinfo {author} {\bibfnamefont {C.~P.~L.}\ \bibnamefont
  {Berry}}, \ and\ \bibinfo {author} {\bibfnamefont {I.}~\bibnamefont
  {Mandel}},\ }\href {\doibase 10.1093/mnras/stx1764} {\bibfield  {journal}
  {\bibinfo  {journal} {Mon. Not. R. Astron. Soc.}\ }\textbf {\bibinfo {volume}
  {471}},\ \bibinfo {pages} {2801} (\bibinfo {year} {2017})},\ \Eprint
  {http://arxiv.org/abs/1703.06873} {arXiv:1703.06873 [astro-ph.HE]}
  \BibitemShut {NoStop}%
\bibitem [{\citenamefont {Talbot}\ and\ \citenamefont
  {Thrane}(2017)}]{Talbot:2017yur}%
  \BibitemOpen
  \bibfield  {author} {\bibinfo {author} {\bibfnamefont {C.}~\bibnamefont
  {Talbot}}\ and\ \bibinfo {author} {\bibfnamefont {E.}~\bibnamefont
  {Thrane}},\ }\href {\doibase 10.1103/PhysRevD.96.023012} {\bibfield
  {journal} {\bibinfo  {journal} {Phys. Rev. D}\ }\textbf {\bibinfo {volume}
  {96}},\ \bibinfo {pages} {023012} (\bibinfo {year} {2017})},\ \Eprint
  {http://arxiv.org/abs/1704.08370} {arXiv:1704.08370 [astro-ph.HE]}
  \BibitemShut {NoStop}%
\bibitem [{\citenamefont {Farr}\ \emph {et~al.}(2017)\citenamefont {Farr},
  \citenamefont {Stevenson}, \citenamefont {Miller}, \citenamefont {Mandel},
  \citenamefont {Farr},\ and\ \citenamefont {Vecchio}}]{Farr:2017uvj}%
  \BibitemOpen
  \bibfield  {author} {\bibinfo {author} {\bibfnamefont {W.~M.}\ \bibnamefont
  {Farr}}, \bibinfo {author} {\bibfnamefont {S.}~\bibnamefont {Stevenson}},
  \bibinfo {author} {\bibfnamefont {M.~C.}\ \bibnamefont {Miller}}, \bibinfo
  {author} {\bibfnamefont {I.}~\bibnamefont {Mandel}}, \bibinfo {author}
  {\bibfnamefont {B.}~\bibnamefont {Farr}}, \ and\ \bibinfo {author}
  {\bibfnamefont {A.}~\bibnamefont {Vecchio}},\ }\href {\doibase
  10.1038/nature23453} {\bibfield  {journal} {\bibinfo  {journal} {Nature
  (London)}\ }\textbf {\bibinfo {volume} {548}},\ \bibinfo {pages} {426}
  (\bibinfo {year} {2017})},\ \Eprint {http://arxiv.org/abs/1706.01385}
  {arXiv:1706.01385 [astro-ph.HE]} \BibitemShut {NoStop}%
\bibitem [{\citenamefont {Tiwari}\ \emph {et~al.}(2018)\citenamefont {Tiwari},
  \citenamefont {Fairhurst},\ and\ \citenamefont {Hannam}}]{Tiwari:2018qch}%
  \BibitemOpen
  \bibfield  {author} {\bibinfo {author} {\bibfnamefont {V.}~\bibnamefont
  {Tiwari}}, \bibinfo {author} {\bibfnamefont {S.}~\bibnamefont {Fairhurst}}, \
  and\ \bibinfo {author} {\bibfnamefont {M.}~\bibnamefont {Hannam}},\ }\href
  {\doibase 10.3847/1538-4357/aae8df} {\bibfield  {journal} {\bibinfo
  {journal} {Astrophys. J.}\ }\textbf {\bibinfo {volume} {868}},\ \bibinfo
  {pages} {140} (\bibinfo {year} {2018})},\ \Eprint
  {http://arxiv.org/abs/1809.01401} {arXiv:1809.01401 [gr-qc]} \BibitemShut
  {NoStop}%
\bibitem [{\citenamefont {Knee}\ \emph {et~al.}(2022)\citenamefont {Knee},
  \citenamefont {McIver},\ and\ \citenamefont {Cabero}}]{Knee:2021noc}%
  \BibitemOpen
  \bibfield  {author} {\bibinfo {author} {\bibfnamefont {A.~M.}\ \bibnamefont
  {Knee}}, \bibinfo {author} {\bibfnamefont {J.}~\bibnamefont {McIver}}, \ and\
  \bibinfo {author} {\bibfnamefont {M.}~\bibnamefont {Cabero}},\ }\href
  {\doibase 10.3847/1538-4357/ac48f5} {\bibfield  {journal} {\bibinfo
  {journal} {The Astrophysical Journal}\ }\textbf {\bibinfo {volume} {928}},\
  \bibinfo {pages} {21} (\bibinfo {year} {2022})}\BibitemShut {NoStop}%
\bibitem [{\citenamefont {Abbott}\ \emph
  {et~al.}(2020{\natexlab{d}})\citenamefont {Abbott} \emph
  {et~al.}}]{Aasi:2013wya}%
  \BibitemOpen
  \bibfield  {author} {\bibinfo {author} {\bibfnamefont {B.~P.}\ \bibnamefont
  {Abbott}} \emph {et~al.} (\bibinfo {collaboration} {KAGRA Collaboration, LIGO
  Scientific Collaboration, and Virgo Collaboration}),\ }\href {\doibase
  10.1007/s41114-020-00026-9} {\bibfield  {journal} {\bibinfo  {journal}
  {Living Rev. Relativity}\ }\textbf {\bibinfo {volume} {23}},\ \bibinfo
  {pages} {3} (\bibinfo {year} {2020}{\natexlab{d}})},\ \Eprint
  {http://arxiv.org/abs/1304.0670} {arXiv:1304.0670 [gr-qc]} \BibitemShut
  {NoStop}%
\bibitem [{\citenamefont {Reitze}\ \emph {et~al.}(2019)\citenamefont {Reitze}
  \emph {et~al.}}]{Reitze:2019iox}%
  \BibitemOpen
  \bibfield  {author} {\bibinfo {author} {\bibfnamefont {D.}~\bibnamefont
  {Reitze}} \emph {et~al.},\ }\href@noop {} {\bibfield  {journal} {\bibinfo
  {journal} {Bull. Am. Astron. Soc.}\ }\textbf {\bibinfo {volume} {51}},\
  \bibinfo {pages} {(7), 35} (\bibinfo {year} {2019})},\ \Eprint
  {http://arxiv.org/abs/1907.04833} {arXiv:1907.04833 [astro-ph.IM]}
  \BibitemShut {NoStop}%
\bibitem [{\citenamefont {Hall}\ \emph {et~al.}(2021)\citenamefont {Hall} \emph
  {et~al.}}]{Hall:2020dps}%
  \BibitemOpen
  \bibfield  {author} {\bibinfo {author} {\bibfnamefont {E.~D.}\ \bibnamefont
  {Hall}} \emph {et~al.},\ }\href {\doibase 10.1103/PhysRevD.103.122004}
  {\bibfield  {journal} {\bibinfo  {journal} {Phys. Rev. D}\ }\textbf {\bibinfo
  {volume} {103}},\ \bibinfo {pages} {122004} (\bibinfo {year} {2021})},\
  \Eprint {http://arxiv.org/abs/2012.03608} {arXiv:2012.03608 [gr-qc]}
  \BibitemShut {NoStop}%
\bibitem [{\citenamefont {Hild}\ \emph {et~al.}(2011)\citenamefont {Hild} \emph
  {et~al.}}]{Hild:2010id}%
  \BibitemOpen
  \bibfield  {author} {\bibinfo {author} {\bibfnamefont {S.}~\bibnamefont
  {Hild}} \emph {et~al.},\ }\href {\doibase 10.1088/0264-9381/28/9/094013}
  {\bibfield  {journal} {\bibinfo  {journal} {Classical Quantum Gravity}\
  }\textbf {\bibinfo {volume} {28}},\ \bibinfo {pages} {094013} (\bibinfo
  {year} {2011})},\ \Eprint {http://arxiv.org/abs/1012.0908} {arXiv:1012.0908
  [gr-qc]} \BibitemShut {NoStop}%
\bibitem [{\citenamefont {Kesden}\ \emph {et~al.}(2015)\citenamefont {Kesden},
  \citenamefont {Gerosa}, \citenamefont {O'Shaughnessy}, \citenamefont
  {Berti},\ and\ \citenamefont {Sperhake}}]{Kesden:2014sla}%
  \BibitemOpen
  \bibfield  {author} {\bibinfo {author} {\bibfnamefont {M.}~\bibnamefont
  {Kesden}}, \bibinfo {author} {\bibfnamefont {D.}~\bibnamefont {Gerosa}},
  \bibinfo {author} {\bibfnamefont {R.}~\bibnamefont {O'Shaughnessy}}, \bibinfo
  {author} {\bibfnamefont {E.}~\bibnamefont {Berti}}, \ and\ \bibinfo {author}
  {\bibfnamefont {U.}~\bibnamefont {Sperhake}},\ }\href {\doibase
  10.1103/PhysRevLett.114.081103} {\bibfield  {journal} {\bibinfo  {journal}
  {Phys. Rev. Lett.}\ }\textbf {\bibinfo {volume} {114}},\ \bibinfo {pages}
  {081103} (\bibinfo {year} {2015})},\ \Eprint {http://arxiv.org/abs/1411.0674}
  {arXiv:1411.0674 [gr-qc]} \BibitemShut {NoStop}%
\bibitem [{\citenamefont {Gerosa}\ \emph
  {et~al.}(2015{\natexlab{a}})\citenamefont {Gerosa}, \citenamefont {Kesden},
  \citenamefont {Sperhake}, \citenamefont {Berti},\ and\ \citenamefont
  {O'Shaughnessy}}]{Gerosa:2015tea}%
  \BibitemOpen
  \bibfield  {author} {\bibinfo {author} {\bibfnamefont {D.}~\bibnamefont
  {Gerosa}}, \bibinfo {author} {\bibfnamefont {M.}~\bibnamefont {Kesden}},
  \bibinfo {author} {\bibfnamefont {U.}~\bibnamefont {Sperhake}}, \bibinfo
  {author} {\bibfnamefont {E.}~\bibnamefont {Berti}}, \ and\ \bibinfo {author}
  {\bibfnamefont {R.}~\bibnamefont {O'Shaughnessy}},\ }\href {\doibase
  10.1103/PhysRevD.92.064016} {\bibfield  {journal} {\bibinfo  {journal} {Phys.
  Rev. D}\ }\textbf {\bibinfo {volume} {92}},\ \bibinfo {pages} {064016}
  (\bibinfo {year} {2015}{\natexlab{a}})},\ \Eprint
  {http://arxiv.org/abs/1506.03492} {arXiv:1506.03492 [gr-qc]} \BibitemShut
  {NoStop}%
\bibitem [{\citenamefont {Racine}(2008)}]{Racine:2008qv}%
  \BibitemOpen
  \bibfield  {author} {\bibinfo {author} {\bibfnamefont {{\'E}.}~\bibnamefont
  {Racine}},\ }\href {\doibase 10.1103/PhysRevD.78.044021} {\bibfield
  {journal} {\bibinfo  {journal} {Phys. Rev. D}\ }\textbf {\bibinfo {volume}
  {78}},\ \bibinfo {pages} {044021} (\bibinfo {year} {2008})},\ \Eprint
  {http://arxiv.org/abs/0803.1820} {arXiv:0803.1820 [gr-qc]} \BibitemShut
  {NoStop}%
\bibitem [{\citenamefont {Chatziioannou}\ \emph {et~al.}(2017)\citenamefont
  {Chatziioannou}, \citenamefont {Klein}, \citenamefont {Yunes},\ and\
  \citenamefont {Cornish}}]{Chatziioannou:2017tdw}%
  \BibitemOpen
  \bibfield  {author} {\bibinfo {author} {\bibfnamefont {K.}~\bibnamefont
  {Chatziioannou}}, \bibinfo {author} {\bibfnamefont {A.}~\bibnamefont
  {Klein}}, \bibinfo {author} {\bibfnamefont {N.}~\bibnamefont {Yunes}}, \ and\
  \bibinfo {author} {\bibfnamefont {N.}~\bibnamefont {Cornish}},\ }\href
  {\doibase 10.1103/PhysRevD.95.104004} {\bibfield  {journal} {\bibinfo
  {journal} {Phys. Rev. D}\ }\textbf {\bibinfo {volume} {95}},\ \bibinfo
  {pages} {104004} (\bibinfo {year} {2017})},\ \Eprint
  {http://arxiv.org/abs/1703.03967} {arXiv:1703.03967 [gr-qc]} \BibitemShut
  {NoStop}%
\bibitem [{\citenamefont {Apostolatos}\ \emph {et~al.}(1994)\citenamefont
  {Apostolatos}, \citenamefont {Cutler}, \citenamefont {Sussman},\ and\
  \citenamefont {Thorne}}]{Apostolatos:1994mx}%
  \BibitemOpen
  \bibfield  {author} {\bibinfo {author} {\bibfnamefont {T.~A.}\ \bibnamefont
  {Apostolatos}}, \bibinfo {author} {\bibfnamefont {C.}~\bibnamefont {Cutler}},
  \bibinfo {author} {\bibfnamefont {G.~J.}\ \bibnamefont {Sussman}}, \ and\
  \bibinfo {author} {\bibfnamefont {K.~S.}\ \bibnamefont {Thorne}},\ }\href
  {\doibase 10.1103/PhysRevD.49.6274} {\bibfield  {journal} {\bibinfo
  {journal} {Phys. Rev. D}\ }\textbf {\bibinfo {volume} {49}},\ \bibinfo
  {pages} {6274} (\bibinfo {year} {1994})}\BibitemShut {NoStop}%
\bibitem [{\citenamefont {Kidder}(1995)}]{Kidder:1995zr}%
  \BibitemOpen
  \bibfield  {author} {\bibinfo {author} {\bibfnamefont {L.~E.}\ \bibnamefont
  {Kidder}},\ }\href {\doibase 10.1103/PhysRevD.52.821} {\bibfield  {journal}
  {\bibinfo  {journal} {Phys. Rev. D}\ }\textbf {\bibinfo {volume} {52}},\
  \bibinfo {pages} {821} (\bibinfo {year} {1995})},\ \Eprint
  {http://arxiv.org/abs/gr-qc/9506022} {arXiv:gr-qc/9506022} \BibitemShut
  {NoStop}%
\bibitem [{\citenamefont {Gerosa}\ and\ \citenamefont
  {Kesden}(2016)}]{Gerosa:2016sys}%
  \BibitemOpen
  \bibfield  {author} {\bibinfo {author} {\bibfnamefont {D.}~\bibnamefont
  {Gerosa}}\ and\ \bibinfo {author} {\bibfnamefont {M.}~\bibnamefont
  {Kesden}},\ }\href {\doibase 10.1103/PhysRevD.93.124066} {\bibfield
  {journal} {\bibinfo  {journal} {Phys. Rev. D}\ }\textbf {\bibinfo {volume}
  {93}},\ \bibinfo {pages} {124066} (\bibinfo {year} {2016})},\ \Eprint
  {http://arxiv.org/abs/1605.01067} {arXiv:1605.01067 [astro-ph.HE]}
  \BibitemShut {NoStop}%
\bibitem [{\citenamefont {Ajith}(2011)}]{Ajith:2011ec}%
  \BibitemOpen
  \bibfield  {author} {\bibinfo {author} {\bibfnamefont {P.}~\bibnamefont
  {Ajith}},\ }\href {\doibase 10.1103/PhysRevD.84.084037} {\bibfield  {journal}
  {\bibinfo  {journal} {Phys. Rev. D}\ }\textbf {\bibinfo {volume} {84}},\
  \bibinfo {pages} {084037} (\bibinfo {year} {2011})},\ \Eprint
  {http://arxiv.org/abs/1107.1267} {arXiv:1107.1267 [gr-qc]} \BibitemShut
  {NoStop}%
\bibitem [{LAL()}]{LALSuite}%
  \BibitemOpen
  \href@noop {} {}\bibinfo {note} {{LSC Algorithm Library Suite (LALSuite),
  \url{https://doi.org/10.7935/GT1W-FZ16}}}\BibitemShut {NoStop}%
\bibitem [{\citenamefont {Boh\'e}\ \emph {et~al.}(2015)\citenamefont {Boh\'e},
  \citenamefont {Faye}, \citenamefont {Marsat},\ and\ \citenamefont
  {Porter}}]{Bohe:2015ana}%
  \BibitemOpen
  \bibfield  {author} {\bibinfo {author} {\bibfnamefont {A.}~\bibnamefont
  {Boh\'e}}, \bibinfo {author} {\bibfnamefont {G.}~\bibnamefont {Faye}},
  \bibinfo {author} {\bibfnamefont {S.}~\bibnamefont {Marsat}}, \ and\ \bibinfo
  {author} {\bibfnamefont {E.~K.}\ \bibnamefont {Porter}},\ }\href {\doibase
  10.1088/0264-9381/32/19/195010} {\bibfield  {journal} {\bibinfo  {journal}
  {Classical Quantum Gravity}\ }\textbf {\bibinfo {volume} {32}},\ \bibinfo
  {pages} {195010} (\bibinfo {year} {2015})},\ \Eprint
  {http://arxiv.org/abs/1501.01529} {arXiv:1501.01529 [gr-qc]} \BibitemShut
  {NoStop}%
\bibitem [{\citenamefont {Boh{\'e}}\ \emph {et~al.}(2013)\citenamefont
  {Boh{\'e}}, \citenamefont {Marsat}, \citenamefont {Faye},\ and\ \citenamefont
  {Blanchet}}]{Bohe:2012mr}%
  \BibitemOpen
  \bibfield  {author} {\bibinfo {author} {\bibfnamefont {A.}~\bibnamefont
  {Boh{\'e}}}, \bibinfo {author} {\bibfnamefont {S.}~\bibnamefont {Marsat}},
  \bibinfo {author} {\bibfnamefont {G.}~\bibnamefont {Faye}}, \ and\ \bibinfo
  {author} {\bibfnamefont {L.}~\bibnamefont {Blanchet}},\ }\href {\doibase
  10.1088/0264-9381/30/7/075017} {\bibfield  {journal} {\bibinfo  {journal}
  {Classical Quantum Gravity}\ }\textbf {\bibinfo {volume} {30}},\ \bibinfo
  {pages} {075017} (\bibinfo {year} {2013})},\ \Eprint
  {http://arxiv.org/abs/1212.5520} {arXiv:1212.5520 [gr-qc]} \BibitemShut
  {NoStop}%
\bibitem [{\citenamefont {Sturani}(2021)}]{SpinTaylor_TechNote}%
  \BibitemOpen
  \bibfield  {author} {\bibinfo {author} {\bibfnamefont {R.}~\bibnamefont
  {Sturani}},\ }\href {https://dcc.ligo.org/T1500554/public} {\emph {\bibinfo
  {title} {Note on the derivation of the angular momentum and spin precessing
  equations in SpinTaylor codes}}},\ \bibinfo {type} {Tech. Rep.}\ \bibinfo
  {number} {{LIGO}-T1500554}\ (\bibinfo  {institution} {{LIGO} Project},\
  \bibinfo {year} {2021})\ \bibinfo {note}
  {\url{https://dcc.ligo.org/T1500554/public}}\BibitemShut {NoStop}%
\bibitem [{\citenamefont {Phukon}\ \emph {et~al.}(2019)\citenamefont {Phukon},
  \citenamefont {Gupta}, \citenamefont {Bose},\ and\ \citenamefont
  {Jain}}]{Phukon:2019gfh}%
  \BibitemOpen
  \bibfield  {author} {\bibinfo {author} {\bibfnamefont {K.~S.}\ \bibnamefont
  {Phukon}}, \bibinfo {author} {\bibfnamefont {A.}~\bibnamefont {Gupta}},
  \bibinfo {author} {\bibfnamefont {S.}~\bibnamefont {Bose}}, \ and\ \bibinfo
  {author} {\bibfnamefont {P.}~\bibnamefont {Jain}},\ }\href {\doibase
  10.1103/PhysRevD.100.124008} {\bibfield  {journal} {\bibinfo  {journal}
  {Phys. Rev. D}\ }\textbf {\bibinfo {volume} {100}},\ \bibinfo {pages}
  {124008} (\bibinfo {year} {2019})},\ \Eprint
  {http://arxiv.org/abs/1904.03985} {arXiv:1904.03985 [gr-qc]} \BibitemShut
  {NoStop}%
\bibitem [{\citenamefont {Romero-Shaw}\ \emph {et~al.}(2019)\citenamefont
  {Romero-Shaw}, \citenamefont {Lasky},\ and\ \citenamefont
  {Thrane}}]{Romero-Shaw:2019itr}%
  \BibitemOpen
  \bibfield  {author} {\bibinfo {author} {\bibfnamefont {I.~M.}\ \bibnamefont
  {Romero-Shaw}}, \bibinfo {author} {\bibfnamefont {P.~D.}\ \bibnamefont
  {Lasky}}, \ and\ \bibinfo {author} {\bibfnamefont {E.}~\bibnamefont
  {Thrane}},\ }\href {\doibase 10.1093/mnras/stz2996} {\bibfield  {journal}
  {\bibinfo  {journal} {Mon. Not. R. Astron. Soc.}\ }\textbf {\bibinfo {volume}
  {490}},\ \bibinfo {pages} {5210} (\bibinfo {year} {2019})},\ \Eprint
  {http://arxiv.org/abs/1909.05466} {arXiv:1909.05466 [astro-ph.HE]}
  \BibitemShut {NoStop}%
\bibitem [{\citenamefont {Romero-Shaw}\ \emph {et~al.}(2020)\citenamefont
  {Romero-Shaw}, \citenamefont {Lasky}, \citenamefont {Thrane},\ and\
  \citenamefont {Calder{\'o}n~Bustillo}}]{Romero-Shaw:2020thy}%
  \BibitemOpen
  \bibfield  {author} {\bibinfo {author} {\bibfnamefont {I.~M.}\ \bibnamefont
  {Romero-Shaw}}, \bibinfo {author} {\bibfnamefont {P.~D.}\ \bibnamefont
  {Lasky}}, \bibinfo {author} {\bibfnamefont {E.}~\bibnamefont {Thrane}}, \
  and\ \bibinfo {author} {\bibfnamefont {J.}~\bibnamefont
  {Calder{\'o}n~Bustillo}},\ }\href {\doibase 10.3847/2041-8213/abbe26}
  {\bibfield  {journal} {\bibinfo  {journal} {Astrophys. J. Lett.}\ }\textbf
  {\bibinfo {volume} {903}},\ \bibinfo {pages} {L5} (\bibinfo {year} {2020})},\
  \Eprint {http://arxiv.org/abs/2009.04771} {arXiv:2009.04771 [astro-ph.HE]}
  \BibitemShut {NoStop}%
\bibitem [{\citenamefont {Romero-Shaw}\ \emph {et~al.}(2021)\citenamefont
  {Romero-Shaw}, \citenamefont {Lasky},\ and\ \citenamefont
  {Thrane}}]{Romero-Shaw:2021ual}%
  \BibitemOpen
  \bibfield  {author} {\bibinfo {author} {\bibfnamefont {I.~M.}\ \bibnamefont
  {Romero-Shaw}}, \bibinfo {author} {\bibfnamefont {P.~D.}\ \bibnamefont
  {Lasky}}, \ and\ \bibinfo {author} {\bibfnamefont {E.}~\bibnamefont
  {Thrane}},\ }\href {\doibase 10.3847/2041-8213/ac3138} {\bibfield  {journal}
  {\bibinfo  {journal} {Astrophys. J. Lett.}\ }\textbf {\bibinfo {volume}
  {921}},\ \bibinfo {pages} {L31} (\bibinfo {year} {2021})},\ \Eprint
  {http://arxiv.org/abs/2108.01284} {arXiv:2108.01284 [astro-ph.HE]}
  \BibitemShut {NoStop}%
\bibitem [{\citenamefont {Wu}\ \emph {et~al.}(2020)\citenamefont {Wu},
  \citenamefont {Cao},\ and\ \citenamefont {Zhu}}]{Wu:2020zwr}%
  \BibitemOpen
  \bibfield  {author} {\bibinfo {author} {\bibfnamefont {S.}~\bibnamefont
  {Wu}}, \bibinfo {author} {\bibfnamefont {Z.}~\bibnamefont {Cao}}, \ and\
  \bibinfo {author} {\bibfnamefont {Z.-H.}\ \bibnamefont {Zhu}},\ }\href
  {\doibase 10.1093/mnras/staa1176} {\bibfield  {journal} {\bibinfo  {journal}
  {Mon. Not. R. Astron. Soc.}\ }\textbf {\bibinfo {volume} {495}},\ \bibinfo
  {pages} {466} (\bibinfo {year} {2020})},\ \Eprint
  {http://arxiv.org/abs/2002.05528} {arXiv:2002.05528 [astro-ph.IM]}
  \BibitemShut {NoStop}%
\bibitem [{\citenamefont {Reali}\ \emph {et~al.}(2020)\citenamefont {Reali},
  \citenamefont {Mould}, \citenamefont {Gerosa},\ and\ \citenamefont
  {Varma}}]{Reali:2020vkf}%
  \BibitemOpen
  \bibfield  {author} {\bibinfo {author} {\bibfnamefont {L.}~\bibnamefont
  {Reali}}, \bibinfo {author} {\bibfnamefont {M.}~\bibnamefont {Mould}},
  \bibinfo {author} {\bibfnamefont {D.}~\bibnamefont {Gerosa}}, \ and\ \bibinfo
  {author} {\bibfnamefont {V.}~\bibnamefont {Varma}},\ }\href {\doibase
  10.1088/1361-6382/abb639} {\bibfield  {journal} {\bibinfo  {journal}
  {Classical Quantum Gravity}\ }\textbf {\bibinfo {volume} {37}},\ \bibinfo
  {pages} {225005} (\bibinfo {year} {2020})},\ \Eprint
  {http://arxiv.org/abs/2005.01747} {arXiv:2005.01747 [gr-qc]} \BibitemShut
  {NoStop}%
\bibitem [{\citenamefont {Gerosa}\ \emph {et~al.}(2017)\citenamefont {Gerosa},
  \citenamefont {Sperhake},\ and\ \citenamefont
  {Vo\v{s}mera}}]{Gerosa:2016aus}%
  \BibitemOpen
  \bibfield  {author} {\bibinfo {author} {\bibfnamefont {D.}~\bibnamefont
  {Gerosa}}, \bibinfo {author} {\bibfnamefont {U.}~\bibnamefont {Sperhake}}, \
  and\ \bibinfo {author} {\bibfnamefont {J.}~\bibnamefont {Vo\v{s}mera}},\
  }\href {\doibase 10.1088/1361-6382/aa5e58} {\bibfield  {journal} {\bibinfo
  {journal} {Classical Quantum Gravity}\ }\textbf {\bibinfo {volume} {34}},\
  \bibinfo {pages} {064004} (\bibinfo {year} {2017})},\ \Eprint
  {http://arxiv.org/abs/1612.05263} {arXiv:1612.05263 [gr-qc]} \BibitemShut
  {NoStop}%
\bibitem [{\citenamefont {Klein}(2021)}]{Klein:2021jtd}%
  \BibitemOpen
  \bibfield  {author} {\bibinfo {author} {\bibfnamefont {A.}~\bibnamefont
  {Klein}},\ }\href@noop {} {\  (\bibinfo {year} {2021})},\ \Eprint
  {http://arxiv.org/abs/2106.10291} {arXiv:2106.10291 [gr-qc]} \BibitemShut
  {NoStop}%
\bibitem [{til()}]{tilts_at_infinity}%
  \BibitemOpen
  \href@noop {} {}\bibinfo {note} {{LALSuite tilts\_at\_infinity module,
  \url{https://git.ligo.org/lscsoft/lalsuite/-/tree/master/lalsimulation/python/lalsimulation/tilts_at_infinity}}}\BibitemShut
  {NoStop}%
\bibitem [{GWT()}]{GWTC-2_PE}%
  \BibitemOpen
  \href@noop {} {}\bibinfo {note} {{GWTC-2 O3a posterior samples,
  \url{https://dcc.ligo.org/LIGO-P2000223/public/}}}\BibitemShut {NoStop}%
\bibitem [{\citenamefont {Yu}\ \emph {et~al.}(2020)\citenamefont {Yu},
  \citenamefont {Ma}, \citenamefont {Giesler},\ and\ \citenamefont
  {Chen}}]{Yu:2020iqj}%
  \BibitemOpen
  \bibfield  {author} {\bibinfo {author} {\bibfnamefont {H.}~\bibnamefont
  {Yu}}, \bibinfo {author} {\bibfnamefont {S.}~\bibnamefont {Ma}}, \bibinfo
  {author} {\bibfnamefont {M.}~\bibnamefont {Giesler}}, \ and\ \bibinfo
  {author} {\bibfnamefont {Y.}~\bibnamefont {Chen}},\ }\href {\doibase
  10.1103/PhysRevD.102.123009} {\bibfield  {journal} {\bibinfo  {journal}
  {Phys. Rev. D}\ }\textbf {\bibinfo {volume} {102}},\ \bibinfo {pages}
  {123009} (\bibinfo {year} {2020})},\ \Eprint
  {http://arxiv.org/abs/2007.12978} {arXiv:2007.12978 [gr-qc]} \BibitemShut
  {NoStop}%
\bibitem [{\citenamefont {Peters}(1964)}]{Peters:1964zz}%
  \BibitemOpen
  \bibfield  {author} {\bibinfo {author} {\bibfnamefont {P.~C.}\ \bibnamefont
  {Peters}},\ }\href {\doibase 10.1103/PhysRev.136.B1224} {\bibfield  {journal}
  {\bibinfo  {journal} {Phys. Rev.}\ }\textbf {\bibinfo {volume} {136}},\
  \bibinfo {pages} {B1224} (\bibinfo {year} {1964})}\BibitemShut {NoStop}%
\bibitem [{\citenamefont {Farr}\ \emph {et~al.}(2014)\citenamefont {Farr},
  \citenamefont {Ochsner}, \citenamefont {Farr},\ and\ \citenamefont
  {O'Shaughnessy}}]{Farr:2014qka}%
  \BibitemOpen
  \bibfield  {author} {\bibinfo {author} {\bibfnamefont {B.}~\bibnamefont
  {Farr}}, \bibinfo {author} {\bibfnamefont {E.}~\bibnamefont {Ochsner}},
  \bibinfo {author} {\bibfnamefont {W.~M.}\ \bibnamefont {Farr}}, \ and\
  \bibinfo {author} {\bibfnamefont {R.}~\bibnamefont {O'Shaughnessy}},\ }\href
  {\doibase 10.1103/PhysRevD.90.024018} {\bibfield  {journal} {\bibinfo
  {journal} {Phys. Rev. D}\ }\textbf {\bibinfo {volume} {90}},\ \bibinfo
  {pages} {024018} (\bibinfo {year} {2014})},\ \Eprint
  {http://arxiv.org/abs/1404.7070} {arXiv:1404.7070 [gr-qc]} \BibitemShut
  {NoStop}%
\bibitem [{\citenamefont {Mould}\ and\ \citenamefont
  {Gerosa}(2020)}]{Mould:2020cgc}%
  \BibitemOpen
  \bibfield  {author} {\bibinfo {author} {\bibfnamefont {M.}~\bibnamefont
  {Mould}}\ and\ \bibinfo {author} {\bibfnamefont {D.}~\bibnamefont {Gerosa}},\
  }\href {\doibase 10.1103/PhysRevD.101.124037} {\bibfield  {journal} {\bibinfo
   {journal} {Phys. Rev. D}\ }\textbf {\bibinfo {volume} {101}},\ \bibinfo
  {pages} {124037} (\bibinfo {year} {2020})},\ \Eprint
  {http://arxiv.org/abs/2003.02281} {arXiv:2003.02281 [gr-qc]} \BibitemShut
  {NoStop}%
\bibitem [{\citenamefont {Gerosa}\ \emph
  {et~al.}(2015{\natexlab{b}})\citenamefont {Gerosa}, \citenamefont {Kesden},
  \citenamefont {O'Shaughnessy}, \citenamefont {Klein}, \citenamefont {Berti},
  \citenamefont {Sperhake},\ and\ \citenamefont {Trifir\`o}}]{Gerosa:2015hba}%
  \BibitemOpen
  \bibfield  {author} {\bibinfo {author} {\bibfnamefont {D.}~\bibnamefont
  {Gerosa}}, \bibinfo {author} {\bibfnamefont {M.}~\bibnamefont {Kesden}},
  \bibinfo {author} {\bibfnamefont {R.}~\bibnamefont {O'Shaughnessy}}, \bibinfo
  {author} {\bibfnamefont {A.}~\bibnamefont {Klein}}, \bibinfo {author}
  {\bibfnamefont {E.}~\bibnamefont {Berti}}, \bibinfo {author} {\bibfnamefont
  {U.}~\bibnamefont {Sperhake}}, \ and\ \bibinfo {author} {\bibfnamefont
  {D.}~\bibnamefont {Trifir\`o}},\ }\href {\doibase
  10.1103/PhysRevLett.115.141102} {\bibfield  {journal} {\bibinfo  {journal}
  {Phys. Rev. Lett.}\ }\textbf {\bibinfo {volume} {115}},\ \bibinfo {pages}
  {141102} (\bibinfo {year} {2015}{\natexlab{b}})},\ \Eprint
  {http://arxiv.org/abs/1506.09116} {arXiv:1506.09116 [gr-qc]} \BibitemShut
  {NoStop}%
\bibitem [{PRE()}]{PRECESSION_GitHub_dev}%
  \BibitemOpen
  \href@noop {} {}\bibinfo {note} {D. Gerosa, M. Mould, and D. Gangardt,
  development version of PRECESSION,
  \url{https://github.com/dgerosa/precession/tree/dev}, commit
  a752d0c88ac0b2f10e85c6c271b047c2dc0c3dc1}\BibitemShut {NoStop}%
\bibitem [{\citenamefont {Harris}\ \emph {et~al.}(2020)\citenamefont {Harris}
  \emph {et~al.}}]{Harris:2020xlr}%
  \BibitemOpen
  \bibfield  {author} {\bibinfo {author} {\bibfnamefont {C.~R.}\ \bibnamefont
  {Harris}} \emph {et~al.},\ }\href {\doibase 10.1038/s41586-020-2649-2}
  {\bibfield  {journal} {\bibinfo  {journal} {Nature (London)}\ }\textbf
  {\bibinfo {volume} {585}},\ \bibinfo {pages} {357} (\bibinfo {year}
  {2020})},\ \Eprint {http://arxiv.org/abs/2006.10256} {arXiv:2006.10256
  [cs.MS]} \BibitemShut {NoStop}%
\bibitem [{\citenamefont {Virtanen}\ \emph {et~al.}(2020)\citenamefont
  {Virtanen} \emph {et~al.}}]{Virtanen:2019joe}%
  \BibitemOpen
  \bibfield  {author} {\bibinfo {author} {\bibfnamefont {P.}~\bibnamefont
  {Virtanen}} \emph {et~al.},\ }\href {\doibase 10.1038/s41592-019-0686-2}
  {\bibfield  {journal} {\bibinfo  {journal} {Nat. Methods}\ }\textbf {\bibinfo
  {volume} {17}},\ \bibinfo {pages} {261} (\bibinfo {year} {2020})},\ \Eprint
  {http://arxiv.org/abs/1907.10121} {arXiv:1907.10121 [cs.MS]} \BibitemShut
  {NoStop}%
\bibitem [{\citenamefont {Johansson}\ \emph {et~al.}(2021)\citenamefont
  {Johansson} \emph {et~al.}}]{mpmath}%
  \BibitemOpen
  \bibfield  {author} {\bibinfo {author} {\bibfnamefont {F.}~\bibnamefont
  {Johansson}} \emph {et~al.},\ }\href@noop {} {\emph {\bibinfo {title}
  {mpmath: a {P}ython library for arbitrary-precision floating-point arithmetic
  (version 1.2.1)}}} (\bibinfo {year} {2021}),\ \bibinfo {note}
  {\url{http://mpmath.org/}}\BibitemShut {NoStop}%
\bibitem [{\citenamefont {Hindmarsh}(1983)}]{ODEPACK}%
  \BibitemOpen
  \bibfield  {author} {\bibinfo {author} {\bibfnamefont {A.~C.}\ \bibnamefont
  {Hindmarsh}},\ }in\ \href@noop {} {\emph {\bibinfo {booktitle} {Scientific
  Computing}}},\ \bibinfo {series} {IMACS Transactions on Scientific
  Computation}, Vol.~\bibinfo {volume} {1},\ \bibinfo {editor} {edited by\
  \bibinfo {editor} {\bibfnamefont {R.~S.}\ \bibnamefont {Stepleman}} \emph
  {et~al.}}\ (\bibinfo  {publisher} {North-Holland},\ \bibinfo {address}
  {Amsterdam},\ \bibinfo {year} {1983})\ pp.\ \bibinfo {pages} {55--64},\
  \bibinfo {note}
  {\url{https://computing.llnl.gov/casc/nsde/pubs/u88007.pdf}}\BibitemShut
  {NoStop}%
\bibitem [{\citenamefont {Petzold}(1983)}]{Petzold}%
  \BibitemOpen
  \bibfield  {author} {\bibinfo {author} {\bibfnamefont {L.~R.}\ \bibnamefont
  {Petzold}},\ }\href {\doibase 10.1137/0904010} {\bibfield  {journal}
  {\bibinfo  {journal} {SIAM J. Sci. Stat. Comput.}\ }\textbf {\bibinfo
  {volume} {4}},\ \bibinfo {pages} {136} (\bibinfo {year} {1983})}\BibitemShut
  {NoStop}%
\bibitem [{\citenamefont {Mora}\ and\ \citenamefont
  {Will}(2004)}]{Mora:2003wt}%
  \BibitemOpen
  \bibfield  {author} {\bibinfo {author} {\bibfnamefont {T.}~\bibnamefont
  {Mora}}\ and\ \bibinfo {author} {\bibfnamefont {C.~M.}\ \bibnamefont
  {Will}},\ }\href {\doibase 10.1103/PhysRevD.69.104021} {\bibfield  {journal}
  {\bibinfo  {journal} {Phys. Rev. D}\ }\textbf {\bibinfo {volume} {69}},\
  \bibinfo {pages} {104021} (\bibinfo {year} {2004})},\ \bibinfo {note}
  {\href{http://doi.org/10.1103/PhysRevD.71.129901}{{\bf{71}}, 129901(E)
  (2005)}},\ \Eprint {http://arxiv.org/abs/gr-qc/0312082} {arXiv:gr-qc/0312082}
  \BibitemShut {NoStop}%
\bibitem [{\citenamefont {Memmesheimer}\ \emph {et~al.}(2004)\citenamefont
  {Memmesheimer}, \citenamefont {Gopakumar},\ and\ \citenamefont
  {Sch{\"a}fer}}]{Memmesheimer:2004cv}%
  \BibitemOpen
  \bibfield  {author} {\bibinfo {author} {\bibfnamefont {R.-M.}\ \bibnamefont
  {Memmesheimer}}, \bibinfo {author} {\bibfnamefont {A.}~\bibnamefont
  {Gopakumar}}, \ and\ \bibinfo {author} {\bibfnamefont {G.}~\bibnamefont
  {Sch{\"a}fer}},\ }\href {\doibase 10.1103/PhysRevD.70.104011} {\bibfield
  {journal} {\bibinfo  {journal} {Phys. Rev. D}\ }\textbf {\bibinfo {volume}
  {70}},\ \bibinfo {pages} {104011} (\bibinfo {year} {2004})},\ \Eprint
  {http://arxiv.org/abs/gr-qc/0407049} {arXiv:gr-qc/0407049} \BibitemShut
  {NoStop}%
\bibitem [{\citenamefont {Ossokine}\ \emph {et~al.}(2015)\citenamefont
  {Ossokine}, \citenamefont {Boyle}, \citenamefont {Kidder}, \citenamefont
  {Pfeiffer}, \citenamefont {Scheel},\ and\ \citenamefont
  {Szil\'agyi}}]{Ossokine:2015vda}%
  \BibitemOpen
  \bibfield  {author} {\bibinfo {author} {\bibfnamefont {S.}~\bibnamefont
  {Ossokine}}, \bibinfo {author} {\bibfnamefont {M.}~\bibnamefont {Boyle}},
  \bibinfo {author} {\bibfnamefont {L.~E.}\ \bibnamefont {Kidder}}, \bibinfo
  {author} {\bibfnamefont {H.~P.}\ \bibnamefont {Pfeiffer}}, \bibinfo {author}
  {\bibfnamefont {M.~A.}\ \bibnamefont {Scheel}}, \ and\ \bibinfo {author}
  {\bibfnamefont {B.}~\bibnamefont {Szil\'agyi}},\ }\href {\doibase
  10.1103/PhysRevD.92.104028} {\bibfield  {journal} {\bibinfo  {journal} {Phys.
  Rev. D}\ }\textbf {\bibinfo {volume} {92}},\ \bibinfo {pages} {104028}
  (\bibinfo {year} {2015})},\ \Eprint {http://arxiv.org/abs/1502.01747}
  {arXiv:1502.01747 [gr-qc]} \BibitemShut {NoStop}%
\bibitem [{\citenamefont {Belczynski}\ \emph {et~al.}(2016)\citenamefont
  {Belczynski}, \citenamefont {Holz}, \citenamefont {Bulik},\ and\
  \citenamefont {O'Shaughnessy}}]{Belczynski:2016obo}%
  \BibitemOpen
  \bibfield  {author} {\bibinfo {author} {\bibfnamefont {K.}~\bibnamefont
  {Belczynski}}, \bibinfo {author} {\bibfnamefont {D.~E.}\ \bibnamefont
  {Holz}}, \bibinfo {author} {\bibfnamefont {T.}~\bibnamefont {Bulik}}, \ and\
  \bibinfo {author} {\bibfnamefont {R.}~\bibnamefont {O'Shaughnessy}},\ }\href
  {\doibase 10.1038/nature18322} {\bibfield  {journal} {\bibinfo  {journal}
  {Nature (London)}\ }\textbf {\bibinfo {volume} {534}},\ \bibinfo {pages}
  {512} (\bibinfo {year} {2016})},\ \Eprint {http://arxiv.org/abs/1602.04531}
  {arXiv:1602.04531 [astro-ph.HE]} \BibitemShut {NoStop}%
\bibitem [{\citenamefont {Rodriguez}\ \emph {et~al.}(2016)\citenamefont
  {Rodriguez}, \citenamefont {Haster}, \citenamefont {Chatterjee},
  \citenamefont {Kalogera},\ and\ \citenamefont {Rasio}}]{Rodriguez:2016avt}%
  \BibitemOpen
  \bibfield  {author} {\bibinfo {author} {\bibfnamefont {C.~L.}\ \bibnamefont
  {Rodriguez}}, \bibinfo {author} {\bibfnamefont {C.-J.}\ \bibnamefont
  {Haster}}, \bibinfo {author} {\bibfnamefont {S.}~\bibnamefont {Chatterjee}},
  \bibinfo {author} {\bibfnamefont {V.}~\bibnamefont {Kalogera}}, \ and\
  \bibinfo {author} {\bibfnamefont {F.~A.}\ \bibnamefont {Rasio}},\ }\href
  {\doibase 10.3847/2041-8205/824/1/L8} {\bibfield  {journal} {\bibinfo
  {journal} {Astrophys. J. Lett.}\ }\textbf {\bibinfo {volume} {824}},\
  \bibinfo {pages} {L8} (\bibinfo {year} {2016})},\ \Eprint
  {http://arxiv.org/abs/1604.04254} {arXiv:1604.04254 [astro-ph.HE]}
  \BibitemShut {NoStop}%
\bibitem [{\citenamefont {Olejak}\ \emph {et~al.}(2020)\citenamefont {Olejak},
  \citenamefont {Fishbach}, \citenamefont {Belczynski}, \citenamefont {Holz},
  \citenamefont {Lasota}, \citenamefont {Miller},\ and\ \citenamefont
  {Bulik}}]{Olejak:2020oel}%
  \BibitemOpen
  \bibfield  {author} {\bibinfo {author} {\bibfnamefont {A.}~\bibnamefont
  {Olejak}}, \bibinfo {author} {\bibfnamefont {M.}~\bibnamefont {Fishbach}},
  \bibinfo {author} {\bibfnamefont {K.}~\bibnamefont {Belczynski}}, \bibinfo
  {author} {\bibfnamefont {D.~E.}\ \bibnamefont {Holz}}, \bibinfo {author}
  {\bibfnamefont {J.-P.}\ \bibnamefont {Lasota}}, \bibinfo {author}
  {\bibfnamefont {M.~C.}\ \bibnamefont {Miller}}, \ and\ \bibinfo {author}
  {\bibfnamefont {T.}~\bibnamefont {Bulik}},\ }\href {\doibase
  10.3847/2041-8213/abb5b5} {\bibfield  {journal} {\bibinfo  {journal}
  {Astrophys. J. Lett.}\ }\textbf {\bibinfo {volume} {901}},\ \bibinfo {pages}
  {L39} (\bibinfo {year} {2020})},\ \Eprint {http://arxiv.org/abs/2004.11866}
  {arXiv:2004.11866 [astro-ph.HE]} \BibitemShut {NoStop}%
\bibitem [{\citenamefont {Belczynski}(2020)}]{Belczynski:2020bca}%
  \BibitemOpen
  \bibfield  {author} {\bibinfo {author} {\bibfnamefont {K.}~\bibnamefont
  {Belczynski}},\ }\href {\doibase 10.3847/2041-8213/abcbf1} {\bibfield
  {journal} {\bibinfo  {journal} {Astrophys. J. Lett.}\ }\textbf {\bibinfo
  {volume} {905}},\ \bibinfo {pages} {L15} (\bibinfo {year} {2020})},\ \Eprint
  {http://arxiv.org/abs/2009.13526} {arXiv:2009.13526 [astro-ph.HE]}
  \BibitemShut {NoStop}%
\bibitem [{\citenamefont {Arca~Sedda}(2021)}]{ArcaSedda:2021zmm}%
  \BibitemOpen
  \bibfield  {author} {\bibinfo {author} {\bibfnamefont {M.}~\bibnamefont
  {Arca~Sedda}},\ }\href {\doibase 10.3847/2041-8213/abdfcd} {\bibfield
  {journal} {\bibinfo  {journal} {Astrophys. J. Lett.}\ }\textbf {\bibinfo
  {volume} {908}},\ \bibinfo {pages} {L38} (\bibinfo {year} {2021})},\ \Eprint
  {http://arxiv.org/abs/2102.03364} {arXiv:2102.03364 [astro-ph.HE]}
  \BibitemShut {NoStop}%
\bibitem [{\citenamefont {Lu}\ \emph {et~al.}(2020)\citenamefont {Lu},
  \citenamefont {Beniamini},\ and\ \citenamefont {Bonnerot}}]{Lu:2020gfh}%
  \BibitemOpen
  \bibfield  {author} {\bibinfo {author} {\bibfnamefont {W.}~\bibnamefont
  {Lu}}, \bibinfo {author} {\bibfnamefont {P.}~\bibnamefont {Beniamini}}, \
  and\ \bibinfo {author} {\bibfnamefont {C.}~\bibnamefont {Bonnerot}},\ }\href
  {\doibase 10.1093/mnras/staa3372} {\bibfield  {journal} {\bibinfo  {journal}
  {Mon. Not. R. Astron. Soc.}\ }\textbf {\bibinfo {volume} {500}},\ \bibinfo
  {pages} {1817} (\bibinfo {year} {2020})},\ \Eprint
  {http://arxiv.org/abs/2009.10082} {arXiv:2009.10082 [astro-ph.HE]}
  \BibitemShut {NoStop}%
\bibitem [{\citenamefont {Bartos}\ \emph {et~al.}(2017)\citenamefont {Bartos},
  \citenamefont {Kocsis}, \citenamefont {Haiman},\ and\ \citenamefont
  {M\'arka}}]{Bartos:2016dgn}%
  \BibitemOpen
  \bibfield  {author} {\bibinfo {author} {\bibfnamefont {I.}~\bibnamefont
  {Bartos}}, \bibinfo {author} {\bibfnamefont {B.}~\bibnamefont {Kocsis}},
  \bibinfo {author} {\bibfnamefont {Z.}~\bibnamefont {Haiman}}, \ and\ \bibinfo
  {author} {\bibfnamefont {S.}~\bibnamefont {M\'arka}},\ }\href {\doibase
  10.3847/1538-4357/835/2/165} {\bibfield  {journal} {\bibinfo  {journal}
  {Astrophys. J.}\ }\textbf {\bibinfo {volume} {835}},\ \bibinfo {pages} {165}
  (\bibinfo {year} {2017})},\ \Eprint {http://arxiv.org/abs/1602.03831}
  {arXiv:1602.03831 [astro-ph.HE]} \BibitemShut {NoStop}%
\bibitem [{\citenamefont {Stone}\ \emph {et~al.}(2017)\citenamefont {Stone},
  \citenamefont {Metzger},\ and\ \citenamefont {Haiman}}]{Stone:2016wzz}%
  \BibitemOpen
  \bibfield  {author} {\bibinfo {author} {\bibfnamefont {N.~C.}\ \bibnamefont
  {Stone}}, \bibinfo {author} {\bibfnamefont {B.~D.}\ \bibnamefont {Metzger}},
  \ and\ \bibinfo {author} {\bibfnamefont {Z.}~\bibnamefont {Haiman}},\ }\href
  {\doibase 10.1093/mnras/stw2260} {\bibfield  {journal} {\bibinfo  {journal}
  {Mon. Not. R. Astron. Soc.}\ }\textbf {\bibinfo {volume} {464}},\ \bibinfo
  {pages} {946} (\bibinfo {year} {2017})},\ \Eprint
  {http://arxiv.org/abs/1602.04226} {arXiv:1602.04226 [astro-ph.GA]}
  \BibitemShut {NoStop}%
\bibitem [{\citenamefont {McKernan}\ \emph {et~al.}(2018)\citenamefont
  {McKernan} \emph {et~al.}}]{McKernan:2017umu}%
  \BibitemOpen
  \bibfield  {author} {\bibinfo {author} {\bibfnamefont {B.}~\bibnamefont
  {McKernan}} \emph {et~al.},\ }\href {\doibase 10.3847/1538-4357/aadae5}
  {\bibfield  {journal} {\bibinfo  {journal} {Astrophys. J.}\ }\textbf
  {\bibinfo {volume} {866}},\ \bibinfo {pages} {66} (\bibinfo {year} {2018})},\
  \Eprint {http://arxiv.org/abs/1702.07818} {arXiv:1702.07818 [astro-ph.HE]}
  \BibitemShut {NoStop}%
\bibitem [{\citenamefont {Gr\"obner}\ \emph {et~al.}(2020)\citenamefont
  {Gr\"obner}, \citenamefont {Ishibashi}, \citenamefont {Tiwari}, \citenamefont
  {Haney},\ and\ \citenamefont {Jetzer}}]{Grobner:2020drr}%
  \BibitemOpen
  \bibfield  {author} {\bibinfo {author} {\bibfnamefont {M.}~\bibnamefont
  {Gr\"obner}}, \bibinfo {author} {\bibfnamefont {W.}~\bibnamefont
  {Ishibashi}}, \bibinfo {author} {\bibfnamefont {S.}~\bibnamefont {Tiwari}},
  \bibinfo {author} {\bibfnamefont {M.}~\bibnamefont {Haney}}, \ and\ \bibinfo
  {author} {\bibfnamefont {P.}~\bibnamefont {Jetzer}},\ }\href {\doibase
  10.1051/0004-6361/202037681} {\bibfield  {journal} {\bibinfo  {journal}
  {Astron. Astrophys.}\ }\textbf {\bibinfo {volume} {638}},\ \bibinfo {pages}
  {A119} (\bibinfo {year} {2020})},\ \Eprint {http://arxiv.org/abs/2005.03571}
  {arXiv:2005.03571 [astro-ph.GA]} \BibitemShut {NoStop}%
\bibitem [{PE_()}]{PE_GWTC-3}%
  \BibitemOpen
  \href@noop {} {}\bibinfo {note} {{GWTC-3 O3b posterior samples,
  \url{https://zenodo.org/record/5546662}}}\BibitemShut {NoStop}%
\bibitem [{\citenamefont {Ossokine}\ \emph {et~al.}(2020)\citenamefont
  {Ossokine} \emph {et~al.}}]{Ossokine:2020kjp}%
  \BibitemOpen
  \bibfield  {author} {\bibinfo {author} {\bibfnamefont {S.}~\bibnamefont
  {Ossokine}} \emph {et~al.},\ }\href {\doibase 10.1103/PhysRevD.102.044055}
  {\bibfield  {journal} {\bibinfo  {journal} {Phys. Rev. D}\ }\textbf {\bibinfo
  {volume} {102}},\ \bibinfo {pages} {044055} (\bibinfo {year} {2020})},\
  \Eprint {http://arxiv.org/abs/2004.09442} {arXiv:2004.09442 [gr-qc]}
  \BibitemShut {NoStop}%
\bibitem [{\citenamefont {Babak}\ \emph {et~al.}(2017)\citenamefont {Babak},
  \citenamefont {Taracchini},\ and\ \citenamefont {Buonanno}}]{Babak:2016tgq}%
  \BibitemOpen
  \bibfield  {author} {\bibinfo {author} {\bibfnamefont {S.}~\bibnamefont
  {Babak}}, \bibinfo {author} {\bibfnamefont {A.}~\bibnamefont {Taracchini}}, \
  and\ \bibinfo {author} {\bibfnamefont {A.}~\bibnamefont {Buonanno}},\ }\href
  {\doibase 10.1103/PhysRevD.95.024010} {\bibfield  {journal} {\bibinfo
  {journal} {Phys. Rev. D}\ }\textbf {\bibinfo {volume} {95}},\ \bibinfo
  {pages} {024010} (\bibinfo {year} {2017})},\ \Eprint
  {http://arxiv.org/abs/1607.05661} {arXiv:1607.05661 [gr-qc]} \BibitemShut
  {NoStop}%
\bibitem [{\citenamefont {Pan}\ \emph {et~al.}(2014)\citenamefont {Pan},
  \citenamefont {Buonanno}, \citenamefont {Taracchini}, \citenamefont {Kidder},
  \citenamefont {Mrou\'e}, \citenamefont {Pfeiffer}, \citenamefont {Scheel},\
  and\ \citenamefont {Szil\'agyi}}]{Pan:2013rra}%
  \BibitemOpen
  \bibfield  {author} {\bibinfo {author} {\bibfnamefont {Y.}~\bibnamefont
  {Pan}}, \bibinfo {author} {\bibfnamefont {A.}~\bibnamefont {Buonanno}},
  \bibinfo {author} {\bibfnamefont {A.}~\bibnamefont {Taracchini}}, \bibinfo
  {author} {\bibfnamefont {L.~E.}\ \bibnamefont {Kidder}}, \bibinfo {author}
  {\bibfnamefont {A.~H.}\ \bibnamefont {Mrou\'e}}, \bibinfo {author}
  {\bibfnamefont {H.~P.}\ \bibnamefont {Pfeiffer}}, \bibinfo {author}
  {\bibfnamefont {M.~A.}\ \bibnamefont {Scheel}}, \ and\ \bibinfo {author}
  {\bibfnamefont {B.}~\bibnamefont {Szil\'agyi}},\ }\href {\doibase
  10.1103/PhysRevD.89.084006} {\bibfield  {journal} {\bibinfo  {journal} {Phys.
  Rev. D}\ }\textbf {\bibinfo {volume} {89}},\ \bibinfo {pages} {084006}
  (\bibinfo {year} {2014})},\ \Eprint {http://arxiv.org/abs/1307.6232}
  {arXiv:1307.6232 [gr-qc]} \BibitemShut {NoStop}%
\bibitem [{\citenamefont {Pratten}\ \emph {et~al.}(2021)\citenamefont {Pratten}
  \emph {et~al.}}]{Pratten:2020ceb}%
  \BibitemOpen
  \bibfield  {author} {\bibinfo {author} {\bibfnamefont {G.}~\bibnamefont
  {Pratten}} \emph {et~al.},\ }\href {\doibase 10.1103/PhysRevD.103.104056}
  {\bibfield  {journal} {\bibinfo  {journal} {Phys. Rev. D}\ }\textbf {\bibinfo
  {volume} {103}},\ \bibinfo {pages} {104056} (\bibinfo {year} {2021})},\
  \Eprint {http://arxiv.org/abs/2004.06503} {arXiv:2004.06503 [gr-qc]}
  \BibitemShut {NoStop}%
\bibitem [{\citenamefont {Varma}\ \emph
  {et~al.}(2019{\natexlab{a}})\citenamefont {Varma}, \citenamefont {Field},
  \citenamefont {Scheel}, \citenamefont {Blackman}, \citenamefont {Gerosa},
  \citenamefont {Stein}, \citenamefont {Kidder},\ and\ \citenamefont
  {Pfeiffer}}]{Varma:2019csw}%
  \BibitemOpen
  \bibfield  {author} {\bibinfo {author} {\bibfnamefont {V.}~\bibnamefont
  {Varma}}, \bibinfo {author} {\bibfnamefont {S.~E.}\ \bibnamefont {Field}},
  \bibinfo {author} {\bibfnamefont {M.~A.}\ \bibnamefont {Scheel}}, \bibinfo
  {author} {\bibfnamefont {J.}~\bibnamefont {Blackman}}, \bibinfo {author}
  {\bibfnamefont {D.}~\bibnamefont {Gerosa}}, \bibinfo {author} {\bibfnamefont
  {L.~C.}\ \bibnamefont {Stein}}, \bibinfo {author} {\bibfnamefont {L.~E.}\
  \bibnamefont {Kidder}}, \ and\ \bibinfo {author} {\bibfnamefont {H.~P.}\
  \bibnamefont {Pfeiffer}},\ }\href {\doibase 10.1103/PhysRevResearch.1.033015}
  {\bibfield  {journal} {\bibinfo  {journal} {Phys. Rev. Research.}\ }\textbf
  {\bibinfo {volume} {1}},\ \bibinfo {pages} {033015} (\bibinfo {year}
  {2019}{\natexlab{a}})},\ \Eprint {http://arxiv.org/abs/1905.09300}
  {arXiv:1905.09300 [gr-qc]} \BibitemShut {NoStop}%
\bibitem [{\citenamefont {Varma}\ \emph {et~al.}(2022)\citenamefont {Varma},
  \citenamefont {Biscoveanu}, \citenamefont {Islam}, \citenamefont {Shaik},
  \citenamefont {Haster}, \citenamefont {Isi}, \citenamefont {Farr},
  \citenamefont {Field},\ and\ \citenamefont {Vitale}}]{Varma:2022pld}%
  \BibitemOpen
  \bibfield  {author} {\bibinfo {author} {\bibfnamefont {V.}~\bibnamefont
  {Varma}}, \bibinfo {author} {\bibfnamefont {S.}~\bibnamefont {Biscoveanu}},
  \bibinfo {author} {\bibfnamefont {T.}~\bibnamefont {Islam}}, \bibinfo
  {author} {\bibfnamefont {F.~H.}\ \bibnamefont {Shaik}}, \bibinfo {author}
  {\bibfnamefont {C.-J.}\ \bibnamefont {Haster}}, \bibinfo {author}
  {\bibfnamefont {M.}~\bibnamefont {Isi}}, \bibinfo {author} {\bibfnamefont
  {W.~M.}\ \bibnamefont {Farr}}, \bibinfo {author} {\bibfnamefont {S.~E.}\
  \bibnamefont {Field}}, \ and\ \bibinfo {author} {\bibfnamefont
  {S.}~\bibnamefont {Vitale}},\ }\href {\doibase
  10.1103/PhysRevLett.128.191102} {\bibfield  {journal} {\bibinfo  {journal}
  {Phys. Rev. Lett.}\ }\textbf {\bibinfo {volume} {128}},\ \bibinfo {pages}
  {191102} (\bibinfo {year} {2022})}\BibitemShut {NoStop}%
\bibitem [{\citenamefont {Khan}\ \emph {et~al.}(2020)\citenamefont {Khan},
  \citenamefont {Ohme}, \citenamefont {Chatziioannou},\ and\ \citenamefont
  {Hannam}}]{Khan:2019kot}%
  \BibitemOpen
  \bibfield  {author} {\bibinfo {author} {\bibfnamefont {S.}~\bibnamefont
  {Khan}}, \bibinfo {author} {\bibfnamefont {F.}~\bibnamefont {Ohme}}, \bibinfo
  {author} {\bibfnamefont {K.}~\bibnamefont {Chatziioannou}}, \ and\ \bibinfo
  {author} {\bibfnamefont {M.}~\bibnamefont {Hannam}},\ }\href {\doibase
  10.1103/PhysRevD.101.024056} {\bibfield  {journal} {\bibinfo  {journal}
  {Phys. Rev. D}\ }\textbf {\bibinfo {volume} {101}},\ \bibinfo {pages}
  {024056} (\bibinfo {year} {2020})},\ \Eprint
  {http://arxiv.org/abs/1911.06050} {arXiv:1911.06050 [gr-qc]} \BibitemShut
  {NoStop}%
\bibitem [{\citenamefont {Estell\'es}\ \emph {et~al.}(2022)\citenamefont
  {Estell\'es} \emph {et~al.}}]{Estelles:2021jnz}%
  \BibitemOpen
  \bibfield  {author} {\bibinfo {author} {\bibfnamefont {H.}~\bibnamefont
  {Estell\'es}} \emph {et~al.},\ }\href {\doibase 10.3847/1538-4357/ac33a0}
  {\bibfield  {journal} {\bibinfo  {journal} {Astrophys. J.}\ }\textbf
  {\bibinfo {volume} {924}},\ \bibinfo {pages} {79} (\bibinfo {year} {2022})},\
  \Eprint {http://arxiv.org/abs/2105.06360} {arXiv:2105.06360 [gr-qc]}
  \BibitemShut {NoStop}%
\bibitem [{\citenamefont {Estell\'es}\ \emph
  {et~al.}(2021{\natexlab{a}})\citenamefont {Estell\'es}, \citenamefont
  {Ramos-Buades}, \citenamefont {Husa}, \citenamefont {Garc\'\i{}a-Quir\'os},
  \citenamefont {Colleoni}, \citenamefont {Haegel},\ and\ \citenamefont
  {Jaume}}]{Estelles:2020osj}%
  \BibitemOpen
  \bibfield  {author} {\bibinfo {author} {\bibfnamefont {H.}~\bibnamefont
  {Estell\'es}}, \bibinfo {author} {\bibfnamefont {A.}~\bibnamefont
  {Ramos-Buades}}, \bibinfo {author} {\bibfnamefont {S.}~\bibnamefont {Husa}},
  \bibinfo {author} {\bibfnamefont {C.}~\bibnamefont {Garc\'\i{}a-Quir\'os}},
  \bibinfo {author} {\bibfnamefont {M.}~\bibnamefont {Colleoni}}, \bibinfo
  {author} {\bibfnamefont {L.}~\bibnamefont {Haegel}}, \ and\ \bibinfo {author}
  {\bibfnamefont {R.}~\bibnamefont {Jaume}},\ }\href {\doibase
  10.1103/PhysRevD.103.124060} {\bibfield  {journal} {\bibinfo  {journal}
  {Phys. Rev. D}\ }\textbf {\bibinfo {volume} {103}},\ \bibinfo {pages}
  {124060} (\bibinfo {year} {2021}{\natexlab{a}})},\ \Eprint
  {http://arxiv.org/abs/2004.08302} {arXiv:2004.08302 [gr-qc]} \BibitemShut
  {NoStop}%
\bibitem [{\citenamefont {Estell\'es}\ \emph {et~al.}(2020)\citenamefont
  {Estell\'es}, \citenamefont {Husa}, \citenamefont {Colleoni}, \citenamefont
  {Keitel}, \citenamefont {Mateu-Lucena}, \citenamefont {Garc\'\i{}a-Quir\'os},
  \citenamefont {Ramos-Buades},\ and\ \citenamefont
  {Borchers}}]{Estelles:2020twz}%
  \BibitemOpen
  \bibfield  {author} {\bibinfo {author} {\bibfnamefont {H.}~\bibnamefont
  {Estell\'es}}, \bibinfo {author} {\bibfnamefont {S.}~\bibnamefont {Husa}},
  \bibinfo {author} {\bibfnamefont {M.}~\bibnamefont {Colleoni}}, \bibinfo
  {author} {\bibfnamefont {D.}~\bibnamefont {Keitel}}, \bibinfo {author}
  {\bibfnamefont {M.}~\bibnamefont {Mateu-Lucena}}, \bibinfo {author}
  {\bibfnamefont {C.}~\bibnamefont {Garc\'\i{}a-Quir\'os}}, \bibinfo {author}
  {\bibfnamefont {A.}~\bibnamefont {Ramos-Buades}}, \ and\ \bibinfo {author}
  {\bibfnamefont {A.}~\bibnamefont {Borchers}},\ }\href@noop {} {\  (\bibinfo
  {year} {2020})},\ \Eprint {http://arxiv.org/abs/2012.11923} {arXiv:2012.11923
  [gr-qc]} \BibitemShut {NoStop}%
\bibitem [{\citenamefont {Estell\'es}\ \emph
  {et~al.}(2021{\natexlab{b}})\citenamefont {Estell\'es}, \citenamefont
  {Colleoni}, \citenamefont {Garc\'\i{}a-Quir\'os}, \citenamefont {Husa},
  \citenamefont {Keitel}, \citenamefont {Mateu-Lucena}, \citenamefont
  {de~Lluc~Planas},\ and\ \citenamefont {Ramos-Buades}}]{Estelles:2021gvs}%
  \BibitemOpen
  \bibfield  {author} {\bibinfo {author} {\bibfnamefont {H.}~\bibnamefont
  {Estell\'es}}, \bibinfo {author} {\bibfnamefont {M.}~\bibnamefont
  {Colleoni}}, \bibinfo {author} {\bibfnamefont {C.}~\bibnamefont
  {Garc\'\i{}a-Quir\'os}}, \bibinfo {author} {\bibfnamefont {S.}~\bibnamefont
  {Husa}}, \bibinfo {author} {\bibfnamefont {D.}~\bibnamefont {Keitel}},
  \bibinfo {author} {\bibfnamefont {M.}~\bibnamefont {Mateu-Lucena}}, \bibinfo
  {author} {\bibfnamefont {M.}~\bibnamefont {de~Lluc~Planas}}, \ and\ \bibinfo
  {author} {\bibfnamefont {A.}~\bibnamefont {Ramos-Buades}},\ }\href@noop {} {\
   (\bibinfo {year} {2021}{\natexlab{b}})},\ \Eprint
  {http://arxiv.org/abs/2105.05872} {arXiv:2105.05872 [gr-qc]} \BibitemShut
  {NoStop}%
\bibitem [{\citenamefont {Nitz}\ and\ \citenamefont
  {Capano}(2021)}]{Nitz:2020mga}%
  \BibitemOpen
  \bibfield  {author} {\bibinfo {author} {\bibfnamefont {A.~H.}\ \bibnamefont
  {Nitz}}\ and\ \bibinfo {author} {\bibfnamefont {C.~D.}\ \bibnamefont
  {Capano}},\ }\href {\doibase 10.3847/2041-8213/abccc5} {\bibfield  {journal}
  {\bibinfo  {journal} {Astrophys. J. Lett.}\ }\textbf {\bibinfo {volume}
  {907}},\ \bibinfo {pages} {L9} (\bibinfo {year} {2021})},\ \Eprint
  {http://arxiv.org/abs/2010.12558} {arXiv:2010.12558 [astro-ph.HE]}
  \BibitemShut {NoStop}%
\bibitem [{\citenamefont {Hannam}\ \emph {et~al.}(2014)\citenamefont {Hannam},
  \citenamefont {Schmidt}, \citenamefont {Boh\'e}, \citenamefont {Haegel},
  \citenamefont {Husa}, \citenamefont {Ohme}, \citenamefont {Pratten},\ and\
  \citenamefont {P\"urrer}}]{Hannam:2013oca}%
  \BibitemOpen
  \bibfield  {author} {\bibinfo {author} {\bibfnamefont {M.}~\bibnamefont
  {Hannam}}, \bibinfo {author} {\bibfnamefont {P.}~\bibnamefont {Schmidt}},
  \bibinfo {author} {\bibfnamefont {A.}~\bibnamefont {Boh\'e}}, \bibinfo
  {author} {\bibfnamefont {L.}~\bibnamefont {Haegel}}, \bibinfo {author}
  {\bibfnamefont {S.}~\bibnamefont {Husa}}, \bibinfo {author} {\bibfnamefont
  {F.}~\bibnamefont {Ohme}}, \bibinfo {author} {\bibfnamefont {G.}~\bibnamefont
  {Pratten}}, \ and\ \bibinfo {author} {\bibfnamefont {M.}~\bibnamefont
  {P\"urrer}},\ }\href {\doibase 10.1103/PhysRevLett.113.151101} {\bibfield
  {journal} {\bibinfo  {journal} {Phys. Rev. Lett.}\ }\textbf {\bibinfo
  {volume} {113}},\ \bibinfo {pages} {151101} (\bibinfo {year} {2014})},\
  \Eprint {http://arxiv.org/abs/1308.3271} {arXiv:1308.3271 [gr-qc]}
  \BibitemShut {NoStop}%
\bibitem [{\citenamefont {Schmidt}\ \emph {et~al.}(2015)\citenamefont
  {Schmidt}, \citenamefont {Ohme},\ and\ \citenamefont
  {Hannam}}]{Schmidt:2014iyl}%
  \BibitemOpen
  \bibfield  {author} {\bibinfo {author} {\bibfnamefont {P.}~\bibnamefont
  {Schmidt}}, \bibinfo {author} {\bibfnamefont {F.}~\bibnamefont {Ohme}}, \
  and\ \bibinfo {author} {\bibfnamefont {M.}~\bibnamefont {Hannam}},\ }\href
  {\doibase 10.1103/PhysRevD.91.024043} {\bibfield  {journal} {\bibinfo
  {journal} {Phys. Rev. D}\ }\textbf {\bibinfo {volume} {91}},\ \bibinfo
  {pages} {024043} (\bibinfo {year} {2015})},\ \Eprint
  {http://arxiv.org/abs/1408.1810} {arXiv:1408.1810 [gr-qc]} \BibitemShut
  {NoStop}%
\bibitem [{\citenamefont {Abbott}\ \emph
  {et~al.}(2021{\natexlab{d}})\citenamefont {Abbott} \emph {et~al.}}]{O3a_pop}%
  \BibitemOpen
  \bibfield  {author} {\bibinfo {author} {\bibfnamefont {R.}~\bibnamefont
  {Abbott}} \emph {et~al.} (\bibinfo {collaboration} {LIGO Scientific
  Collaboration and Virgo Collaboration}),\ }\href {\doibase
  10.3847/2041-8213/abe949} {\bibfield  {journal} {\bibinfo  {journal}
  {Astrophys. J. Lett.}\ }\textbf {\bibinfo {volume} {913}},\ \bibinfo {pages}
  {L7} (\bibinfo {year} {2021}{\natexlab{d}})},\ \Eprint
  {http://arxiv.org/abs/2010.14533} {arXiv:2010.14533 [astro-ph.HE]}
  \BibitemShut {NoStop}%
\bibitem [{\citenamefont {Gerosa}\ \emph {et~al.}(2021)\citenamefont {Gerosa},
  \citenamefont {Mould}, \citenamefont {Gangardt}, \citenamefont {Schmidt},
  \citenamefont {Pratten},\ and\ \citenamefont {Thomas}}]{Gerosa:2020aiw}%
  \BibitemOpen
  \bibfield  {author} {\bibinfo {author} {\bibfnamefont {D.}~\bibnamefont
  {Gerosa}}, \bibinfo {author} {\bibfnamefont {M.}~\bibnamefont {Mould}},
  \bibinfo {author} {\bibfnamefont {D.}~\bibnamefont {Gangardt}}, \bibinfo
  {author} {\bibfnamefont {P.}~\bibnamefont {Schmidt}}, \bibinfo {author}
  {\bibfnamefont {G.}~\bibnamefont {Pratten}}, \ and\ \bibinfo {author}
  {\bibfnamefont {L.~M.}\ \bibnamefont {Thomas}},\ }\href {\doibase
  10.1103/PhysRevD.103.064067} {\bibfield  {journal} {\bibinfo  {journal}
  {Phys. Rev. D}\ }\textbf {\bibinfo {volume} {103}},\ \bibinfo {pages}
  {064067} (\bibinfo {year} {2021})},\ \Eprint
  {http://arxiv.org/abs/2011.11948} {arXiv:2011.11948 [gr-qc]} \BibitemShut
  {NoStop}%
\bibitem [{\citenamefont {Bogdanovic}\ \emph {et~al.}(2007)\citenamefont
  {Bogdanovic}, \citenamefont {Reynolds},\ and\ \citenamefont
  {Miller}}]{Bogdanovic:2007hp}%
  \BibitemOpen
  \bibfield  {author} {\bibinfo {author} {\bibfnamefont {T.}~\bibnamefont
  {Bogdanovic}}, \bibinfo {author} {\bibfnamefont {C.~S.}\ \bibnamefont
  {Reynolds}}, \ and\ \bibinfo {author} {\bibfnamefont {M.~C.}\ \bibnamefont
  {Miller}},\ }\href {\doibase 10.1086/518769} {\bibfield  {journal} {\bibinfo
  {journal} {Astrophys. J. Lett.}\ }\textbf {\bibinfo {volume} {661}},\
  \bibinfo {pages} {L147} (\bibinfo {year} {2007})},\ \Eprint
  {http://arxiv.org/abs/astro-ph/0703054} {arXiv:astro-ph/0703054} \BibitemShut
  {NoStop}%
\bibitem [{\citenamefont {Kullback}\ and\ \citenamefont
  {Leibler}(1951)}]{Kullback:1951zyt}%
  \BibitemOpen
  \bibfield  {author} {\bibinfo {author} {\bibfnamefont {S.}~\bibnamefont
  {Kullback}}\ and\ \bibinfo {author} {\bibfnamefont {R.~A.}\ \bibnamefont
  {Leibler}},\ }\href {\doibase 10.1214/aoms/1177729694} {\bibfield  {journal}
  {\bibinfo  {journal} {Ann. Math. Stat.}\ }\textbf {\bibinfo {volume} {22}},\
  \bibinfo {pages} {79} (\bibinfo {year} {1951})}\BibitemShut {NoStop}%
\bibitem [{\citenamefont {Hoy}\ and\ \citenamefont
  {Raymond}(2021)}]{Hoy:2020vys}%
  \BibitemOpen
  \bibfield  {author} {\bibinfo {author} {\bibfnamefont {C.}~\bibnamefont
  {Hoy}}\ and\ \bibinfo {author} {\bibfnamefont {V.}~\bibnamefont {Raymond}},\
  }\href {\doibase 10.1016/j.softx.2021.100765} {\bibfield  {journal} {\bibinfo
   {journal} {SoftwareX}\ }\textbf {\bibinfo {volume} {15}},\ \bibinfo {pages}
  {100765} (\bibinfo {year} {2021})},\ \Eprint
  {http://arxiv.org/abs/2006.06639} {arXiv:2006.06639 [astro-ph.IM]}
  \BibitemShut {NoStop}%
\bibitem [{\citenamefont {Yang}\ and\ \citenamefont
  {Leibovich}(2019)}]{Yang:2019oqm}%
  \BibitemOpen
  \bibfield  {author} {\bibinfo {author} {\bibfnamefont {Z.}~\bibnamefont
  {Yang}}\ and\ \bibinfo {author} {\bibfnamefont {A.~K.}\ \bibnamefont
  {Leibovich}},\ }\href {\doibase 10.1103/PhysRevD.100.084021} {\bibfield
  {journal} {\bibinfo  {journal} {Phys. Rev. D}\ }\textbf {\bibinfo {volume}
  {100}},\ \bibinfo {pages} {084021} (\bibinfo {year} {2019})},\ \Eprint
  {http://arxiv.org/abs/1908.05688} {arXiv:1908.05688 [gr-qc]} \BibitemShut
  {NoStop}%
\bibitem [{\citenamefont {Varma}\ \emph
  {et~al.}(2019{\natexlab{b}})\citenamefont {Varma}, \citenamefont {Gerosa},
  \citenamefont {Stein}, \citenamefont {H\'ebert},\ and\ \citenamefont
  {Zhang}}]{Varma:2018aht}%
  \BibitemOpen
  \bibfield  {author} {\bibinfo {author} {\bibfnamefont {V.}~\bibnamefont
  {Varma}}, \bibinfo {author} {\bibfnamefont {D.}~\bibnamefont {Gerosa}},
  \bibinfo {author} {\bibfnamefont {L.~C.}\ \bibnamefont {Stein}}, \bibinfo
  {author} {\bibfnamefont {F.}~\bibnamefont {H\'ebert}}, \ and\ \bibinfo
  {author} {\bibfnamefont {H.}~\bibnamefont {Zhang}},\ }\href {\doibase
  10.1103/PhysRevLett.122.011101} {\bibfield  {journal} {\bibinfo  {journal}
  {Phys. Rev. Lett.}\ }\textbf {\bibinfo {volume} {122}},\ \bibinfo {pages}
  {011101} (\bibinfo {year} {2019}{\natexlab{b}})},\ \Eprint
  {http://arxiv.org/abs/1809.09125} {arXiv:1809.09125 [gr-qc]} \BibitemShut
  {NoStop}%
\bibitem [{\citenamefont {Hunter}(2007)}]{Hunter:2007ouj}%
  \BibitemOpen
  \bibfield  {author} {\bibinfo {author} {\bibfnamefont {J.~D.}\ \bibnamefont
  {Hunter}},\ }\href {\doibase 10.1109/MCSE.2007.55} {\bibfield  {journal}
  {\bibinfo  {journal} {Comput. Sci. Eng.}\ }\textbf {\bibinfo {volume} {9}},\
  \bibinfo {pages} {90} (\bibinfo {year} {2007})}\BibitemShut {NoStop}%
\bibitem [{\citenamefont {Abramowitz}\ and\ \citenamefont {Stegun}(1964)}]{AS}%
  \BibitemOpen
  \bibinfo {editor} {\bibfnamefont {M.}~\bibnamefont {Abramowitz}}\ and\
  \bibinfo {editor} {\bibfnamefont {I.~A.}\ \bibnamefont {Stegun}},\ eds.,\
  \href@noop {} {\emph {\bibinfo {title} {Handbook of Mathematical Functions
  With Formulas, Graphs, and Mathematical Tables}}},\ \bibinfo {series}
  {Applied Mathematics Series}\ No.~\bibinfo {number} {55}\ (\bibinfo
  {publisher} {National Bureau of Standards},\ \bibinfo {address} {Washington,
  D.C.},\ \bibinfo {year} {1964})\BibitemShut {NoStop}%
\end{thebibliography}%

\end{document}